\documentclass[12pt]{article}
\usepackage{amsmath} 
\usepackage{color}
\input{epsf} 
\def\ltap{\raisebox{-.4ex}{\rlap{$\,\sim\,$}} \raisebox{.4ex}{$\,<\,$}} 
\def\gtap{\raisebox{-.4ex}{\rlap{$\,\sim\,$}} \raisebox{.4ex}{$\,>\,$}}

\newcommand\as{\alpha_{\mathrm{S}}} 
\newcommand\f[2]{\frac{#1}{#2}} 
\def\dO{{\cal D}_{0}} 
\def\dl{{\cal D}_{1}} 
\def\dll{{\cal D}_{2}} 
\def\dlll{{\cal D}_{3}} 
 
\def\la{\lambda} 
\def\LN{\ln N} 
\def\beq{\begin{equation}} 
\def\eeq{\end{equation}} 

\def\to{\rightarrow} 
\def\nn{\nonumber}

\def\ms{${\overline {\rm MS}}$} 
 
\def\asp{{\alpha_s}\over{\pi}} 
\def\b0{b_0}
\def\bone{b_1}
\def\btwo{b_2}

\newcommand{\ccaption}[2]{
    \begin{center}
    \parbox{0.85\textwidth}{
      \caption[#1]{\small{{#2}}}
      }
    \end{center}
    }

\newcommand\mlbl[1]{{\mbox{\footnotesize #1}}}

\begin{document} 

\begin{titlepage}
\renewcommand{\thefootnote}{\fnsymbol{footnote}}
\begin{flushright}
Bicocca-FT-03-12 \\
CERN--TH/2003-117   \\ hep-ph/0306211
     \end{flushright}
\par \vspace{10mm}

\begin{center}
{\Large \bf
Soft-gluon resummation for Higgs boson production\\
\vskip .2cm
at hadron colliders}
\end{center}
\par \vspace{2mm}
\begin{center}
{\bf Stefano Catani${}^{(a)}$, Daniel de Florian${}^{(b)}$\footnote{Partially
supported by Fundaci\'on Antorchas and CONICET.},
Massimiliano Grazzini${}^{(c)}$ and Paolo Nason${}^{(d)}$}\\

\vspace{5mm}

${}^{(a)}$INFN, Sezione di Firenze, I-50019 Sesto Fiorentino, Florence, Italy\\

${}^{(b)}$Departamento de F\'\i sica, FCEYN, Universidad de Buenos Aires,
(1428) Pabell\'on 1\\ Ciudad Universitaria, Capital Federal, Argentina\\

${}^{(c)}$Theory Division, CERN, CH-1211 Geneva 23, Switzerland \\

${}^{(d)}$INFN, Sezione di Milano, I-20136 Milan, Italy\\

\vspace{5mm}

\end{center}

\par \vspace{2mm}
\begin{center} {\large \bf Abstract} \end{center}
\begin{quote}
\pretolerance 10000

We consider QCD corrections to Higgs boson production through gluon--gluon fusion in hadron collisions. We compute the cross section, performing the all-order
resummation  of
multiple soft-gluon emission at next-to-next-to-leading logarithmic level.
Known fixed-order results  (up to
next-to-next-to-leading order) are consistently included in our
calculation.
We give phenomenological predictions for Higgs boson production at the
Tevatron and at the LHC.
We estimate the residual theoretical uncertainty from perturbative QCD
contributions.
We also quantify the differences obtained by using the 
presently available sets of
parton distributions.

\end{quote}

\vspace*{\fill}
\begin{flushleft}
     CERN--TH/2003-117 \\ June 2003

\end{flushleft}
\end{titlepage}

\setcounter{footnote}{1}
\renewcommand{\thefootnote}{\fnsymbol{footnote}}

\section{Introduction}
\label{sec:intro}

Higgs boson production is nowadays
a topic of central importance
in hadron collider physics \cite{Carena:2000yx,atlascms}. 
The Tevatron will be actively searching for a Higgs boson
signal in the near future. The discovery of the Higgs boson is also a primary physics goal of the
LHC.

The main Higgs production mechanism at hadron colliders is the gluon
fusion process
\cite{Georgi:1978gs},
an essentially strong-interaction process, that has attracted a
large amount of theoretical work in recent years.
Indeed, limits on the Higgs mass will rely upon QCD
calculations of the cross sections. Conversely, if a Higgs boson is
discovered, discrepancies of its measured cross section from QCD
calculations may signal deviations of the Yukawa couplings from
the Standard Model (SM) predictions. It is thus important to provide an accurate
calculation of the Higgs production cross section, together with a reliable
estimate of the associated theoretical error.

The QCD computation of the gluon fusion production cross section
was carried out at next-to-leading order (NLO) in Refs.~\cite{Dawson:1991zj,
Djouadi:1991tk} in the heavy-top limit, and in Ref.~\cite{Spira:1995rr}
with the full dependence on the top-quark mass. Perturbative corrections at the NLO
level were found to be quite large (of the order of 100\%), thus
casting doubts upon the reliability of the perturbative calculation.

Next-to-next-to-leading order (NNLO) corrections have been computed in the
heavy-top limit.
The virtual contributions were evaluated in Ref.~\cite{Harlander:2000mg}.
The soft contributions were computed in
Refs.~\cite{Catani:2001ic,Harlander:2001is}.
The remaining hard real contributions were included in Ref.~\cite{Harlander:2002wh},
with a semi-numerical approach.
More recently,
fully analytical results for the
real contributions have been obtained
\cite{Anastasiou:2002yz,Kilgore:2002sk,Ravindran:2003um}.

Numerically, it is found that NNLO corrections are moderate in size.
There is thus good hope that the NNLO calculation gives a reasonable
estimate of the cross section. Nevertheless, a better understanding of the
pattern of radiative corrections would prove useful both to assess the
reliability of the available calculations and to include higher-order 
corrections.

In the present work, we include the dominant
effect of the uncalculated higher-order terms,
by exploiting  the resummation of soft-gluon emission.
The possibility of performing such an improvement relies
upon the observation that an accurate use of
the soft-gluon approximation provides the bulk of the NLO term and a reliable
estimate of the NNLO effects \cite{Catani:2001ic}.
Now that the NNLO corrections are fully known, and confirm that observation,
it makes 
sense
to add the higher-order terms that can be obtained in the
soft-gluon approximation, in order to give a more precise prediction.
Since the soft approximation proves reliable for the NLO and NNLO
contributions, we make the reasonable assumption that it maintains
the same reliability for higher-order terms.

The paper is organized as follows. In Sect.~\ref{sec:fixedorder}
we give our notation for the QCD cross section and the fixed-order
radiative corrections. In Sect.~\ref{sec:res} we present the formalism
of soft-gluon resummation to all logarithmic orders, and derive the explicit
resummation formulae up to the next-to-next-to-leading logarithmic (NNLL)
level.

The resummation formalism correctly predicts the structure and the coefficients
of the singular terms
in the exact NLO and NNLO calculations. The quantitative reliability of the soft-gluon
approximation can be tested by comparing the truncation of the resummation
formalism at the NLO and NNLO levels with the exact result.
This comparison is performed in Sect.~\ref{sec:SV}. Several variants
of the soft-gluon approximation are considered there, in order to justify
the validity of our approach.
Furthermore, in Sect.~\ref{sec:benchmarks}, the order of magnitude
of the soft-gluon effects is studied, by simply considering the value
of the short-distance partonic cross section in $N$-moment space.

In Sect.~\ref{sec:predictions}, full numerical predictions for the
Higgs production cross section, including soft-gluon resummation at NNLL
accuracy and
the exact NLO and NNLO contributions,
are given, both at the Tevatron and at the LHC. We also perform a study of the 
remaining theoretical uncertainty.
Incidentally, we note that the calculation presented here is the first
calculation of a QCD cross section at such nominal theoretical accuracy,
namely NNLO+NNLL accuracy.

Finally, in Sect.~\ref{sec:conclusions} we give our conclusions.

In Appendix~\ref{appa} we derive a simple prescription to evaluate the
large-$N$ Mellin moments of the soft-gluon contributions at an
arbitrary logarithmic accuracy. The method is a generalization
of the prescription $z^{N-1}-1\to -\theta(1-z-N_0/N)$.
In Appendix~\ref{appb}, we present a formal proof of the
equivalence of two different formulations of the $N$-space resummation
formulae. This equivalence clarifies the distinction between the all-order 
logarithmic terms in the soft-gluon resummation formalism and
the large-order perturbative behaviour due to infrared renormalons.
In Appendix~\ref{appc} we
show how to compute the logarithmic functions that control soft-gluon 
resummation.
In Appendix~\ref{appd} we check the numerical convergence of the fixed-order
soft-gluon expansion, and in Appendix~\ref{appe}
we provide the soft-gluon approximation of the N$^3$LO contribution to the
partonic cross section.

Preliminary results of this work have been presented in 
Ref.~\cite{Giele:2002hx}.

\section{Notation and QCD cross section at fixed orders}
\label{sec:fixedorder}
We consider the collision of two hadrons $h_1$ and $h_2$ with 
centre-of-mass energy ${\sqrt s}$. The inclusive cross section for the 
production of the SM Higgs boson  can be written 
as
\begin{align}
\label{had}
\sigma(s,M_H^2) =& 
\sum_{a,b} \int_0^1 dx_1 \;dx_2 \; f_{a/h_1}(x_1,\mu_F^2) 
\;f_{b/h_2}(x_2,\mu_F^2) \int_0^1 dz \;\delta\!\left(z -
\frac{\tau_H}{x_1x_2}\right) \nn \\
& \cdot \sigma^{(0)}\,z\;G_{ab}(z;\as(\mu_R^2), M_H^2/\mu_R^2;M_H^2/\mu_F^2) \;,
\end{align}
where $M_H$ is the Higgs boson mass, $\tau_H=M_H^2/s$,
and $\mu_F$ and $\mu_R$ are factorization and 
renormalization scales, respectively. 
The parton densities of the colliding hadrons are denoted by 
$f_{a/h}(x,\mu_F^2)$ and the subscript $a$ labels the type
of massless partons ($a=g,q_f,{\bar q}_f$,
with $N_f$ different flavours of light quarks). 
We use parton densities as defined in the \ms\ factorization scheme.

From Eq.~(\ref{had}) the cross section ${\hat \sigma}_{ab}$ for the partonic 
subprocess $ab \to H + X$ at the centre-of-mass energy 
${\hat s}=x_1x_2s=M_H^2/z$ is
\begin{equation}
{\hat \sigma}_{ab}({\hat s},M_H^2) = \frac{1}{\hat s} 
\;\sigma^{(0)} M_H^2 \;G_{ab}(z) = \sigma^{(0)} \;z \;\;G_{ab}(z) \;,
\end{equation}
where the term $1/{\hat s}$ corresponds to the flux factor and leads to 
an overall $z$ factor. The  Born-level cross section $\sigma^{(0)}$ and 
the hard coefficient function $G_{ab}$ arise 
from the phase-space integral of the matrix elements squared. 

The incoming partons $a,b$ couple to the Higgs boson
through heavy-quark loops and, therefore,  $\sigma^{(0)}$ and $G_{ab}$ also depend on the 
masses $M_Q$ of the heavy quarks. The Born-level contribution $\sigma^{(0)}$
is~\cite{Georgi:1978gs}
\vspace*{ -.2cm}
\begin{equation}
\label{borncs}
\sigma^{(0)}=\f{G_F}{288\pi\sqrt{2}} 
\;\left| \sum_Q A_Q\!\left(\frac{4M_Q^2}{M_H^2}\right) \right|^2 \;\;,
\end{equation}
where $G_F=1.16639 \times 10^{-5}$ GeV$^{-2}$ is the Fermi constant,
and the amplitude $A_Q$ is given by
\begin{align}
A_Q(x)  &= \f{3}{2} x \Big[ 1+(1-x) f(x) 
\Big] \;,\nonumber \\
f(x) & = \left\{ \begin{array}{ll}
\displaystyle \arcsin^2 \frac{1}{\sqrt{x}} \;, & x \ge 1 \\
\displaystyle - \frac{1}{4} \left[ \ln \frac{1+\sqrt{1-x}}
{1-\sqrt{1-x}} - i\pi \right]^2 \;, & x < 1
\end{array} \right. \;.
\end{align}
In the following, 
$M_Q=M_t$ or $M_b$
denotes the on-shell pole mass of the top quark or bottom quark.

The coefficient function $G_{ab}$ in Eq.~(\ref{had}) is computable in QCD
perturbation theory according to the expansion
\begin{align}
\label{alloexpansion}
&G_{ab}(z;\as(\mu_R^2), M_H^2/\mu_R^2;M_H^2/\mu_F^2) =
\as^2(\mu_R^2) \sum_{n=0}^{+\infty} \left(\f{\as(\mu_R^2)}{\pi}\right)^n
\;G_{ab}^{(n)}(z;M_H^2/\mu_R^2;M_H^2/\mu_F^2)  \\
\label{expansion}
&= \as^2(\mu_R^2)
G_{ab}^{(0)}(z) + \f{\as^3(\mu_R^2)}{\pi} \;
G_{ab}^{(1)}\left(z;\frac{M_H^2}{\mu_R^2};\frac{M_H^2}{\mu_F^2}\right) 
+ \f{\as^4(\mu_R^2)}{\pi^2} \;
G_{ab}^{(2)}\left(z;\frac{M_H^2}{\mu_R^2};\frac{M_H^2}{\mu_F^2}\right) 
+ {\cal O}(\as^5) \;,
\end{align}
where the (scale-independent) LO contribution is
\begin{equation}
G_{ab}^{(0)}(z) = \delta_{ag} \; \delta_{bg} \;\delta(1-z) \;.
\end{equation}
The terms $G_{ab}^{(1)}$ and $G_{ab}^{(2)}$ give the NLO and NNLO
contributions, respectively.

The NLO coefficients $G_{ab}^{(1)}$ are known. Their calculation with the
exact dependence on $M_t$ (and $M_b$) was performed in Ref.~\cite{Spira:1995rr}, where it
was also observed that the NLO Higgs boson cross section is well approximated
by considering its limit $M_t \gg M_H$ \cite{Dawson:1991zj,Djouadi:1991tk}.
Therefore, throughout the paper we work in the framework of the large-$M_t$
approximation: we consider the case of a single heavy quark, the top quark,
and $N_f=5$ light-quark flavours, and we 
neglect all the contributions to $G_{ab}^{(n)}$ that
vanish when $M_H/M_t \to 0$. However, unless otherwise stated, we include in
$\sigma^{(0)}$ the full dependence on $M_t$ and $M_b$.
At NLO this approximation
\cite{Spira:1995rr,Kramer:1998iq} turns out to be very good when $M_H \leq
2M_t$, and it is still accurate\footnote{The accuracy of this approximation
when $M_H \ltap 2M_t$ is not accidental. In fact, as pointed out in
Refs.~\cite{Catani:2001ic,Catani:2001cr} and discussed below, the main part of
the QCD corrections to direct Higgs production is due to parton radiation at
relatively low transverse momenta. Such radiation is weakly sensitive to the
mass of the heavy quark in the loop.}  to better than $10\%$ when $M_H \ltap
1$~TeV.

The use of the large-$M_t$ expansion considerably simplifies the calculation
of the QCD radiative corrections, since one
can exploit the effective-lagrangian approach
\cite{efflag,Chetyrkin:1997iv,Kramer:1998iq}
to embody the heavy-quark loop in an effective point-like vertex.
The virtual \cite{Harlander:2000mg} and soft 
\cite{Bern:1998sc,Catani:2000pi,Campbell:1998hg,Catani:2000ss} contributions
(i.e. the contributions that are singular when $z \to 1$)
to the NNLO coefficients $G_{ab}^{(2)}(z)$ were independently computed in 
Refs.~\cite{Catani:2001ic} and \cite{Harlander:2001is}. 
The hard contributions were
considered in Ref.~\cite{Harlander:2002wh}, 
by expanding $G_{ab}^{(2)}(z)$ in powers of
$(1-z)$ and evaluating the coefficients of the expansion explicitly up to 
order $(1-z)^{16}$. An independent NNLO calculation, which includes all the
hard contributions in closed analytic form, was carried out
in Ref.~\cite{Anastasiou:2002yz}. This result has recently been confirmed
\cite{Ravindran:2003um} by using a different method of calculation.
In the high-energy limit, the results of 
Refs.~\cite{Anastasiou:2002yz,Ravindran:2003um} agree with the calculation of 
Ref.~\cite{Hautmann:2002tu}, based on $k_\perp$-factorization \cite{kperpfact}.

The NLO coefficient functions $G_{ab}^{(1)}$
in the large-$M_t$ limit (i.e. neglecting corrections that vanish
when  $M_H/M_t \to 0$) are \cite{Dawson:1991zj, Djouadi:1991tk}
\begin{align}
\label{gg1}
G_{gg}^{(1)}(z;M_H^2/\mu_R^2;&M_H^2/\mu_F^2)= \delta(1-z) \left( \f{11}{2} + 6
\zeta(2) + \f{33-2N_f}{6}\ln\f{\mu_R^2}{\mu_F^2}\right)
+12\,\dl(z)\nn\\
&+ 6\,\dO(z) \,\ln\f{M_H^2}{\mu_F^2}\, 
+P_{gg}^{{\rm reg}}(z)\,\ln\f{(1-z)^2M_H^2}{z\mu_F^2}-6\f{\ln z}{1-z}
-\f{11}{2}\f{(1-z)^3}{z} \;,
\end{align}
\begin{equation}
\label{gq1}
G_{gq}^{(1)}(z;M_H^2/\mu_R^2;M_H^2/\mu^2_F)=\f{1}{2}P_{gq}(z)
\ln \f{(1-z)^2 M_H^2}{z\mu^2_F}+\f{2}{3}\, z-\f{(1-z)^2}{z} \;,
\end{equation}
\begin{equation}
\label{qq}
G_{q{\bar q}}^{(1)}(z;M_H^2/\mu_R^2;M_H^2/\mu_F^2)=\f{32}{27}\,\f{(1-z)^3}{z}
\;, \quad G_{qq}^{(1)}(z;M_H^2/\mu_R^2;M_H^2/\mu_F^2) = 0 \;,
\end{equation}
where $\zeta(n)$ is the Riemann zeta-function ($\zeta(2)=\pi^2/6=1.645\dots$,
$\zeta(3)=1.202\dots$), and we have defined
\begin{equation}
\label{singd}
{\cal D}_i(z) \equiv \left[ \f{\ln^i(1-z)}{1-z}\right]_+\,\, .
\end{equation}
The kernels $P_{ab}(z)$ are the LO Altarelli--Parisi splitting functions
for real emission,
\begin{equation}
\label{apreg}
P_{gg}(z)=6\left[\f{1}{1-z}+\f{1}{z}-2+z(1-z)\right] \;,~~~~~
P_{gq}(z)=\f{4}{3}\,\f{1+(1-z)^2}{z} \;,
\end{equation}
and $P_{gg}^{{\rm reg}}(z)$ is the regular (when $z \to 1$) part of $P_{gg}(z)$:
\begin{equation}
P_{gg}^{{\rm reg}}(z)=P_{gg}(z)-\f{6}{1-z}\;.
\end{equation}
The analytic formulae for the NNLO coefficient functions
$G_{ab}^{(2)}$ are given in
Refs.~\cite{Anastasiou:2002yz,Ravindran:2003um} 
($\eta_{ab}^{(2)}(z)= z G_{ab}^{(2)}(z)$,
according to the notation of Ref.~\cite{Anastasiou:2002yz}).

For the purpose of the discussion in the following sections, we note that
we can identify three kinds of contributions
in Eqs.~(\ref{gg1})--(\ref{qq}) and in the analytic formulae for $G_{ab}^{(2)}$:
\begin{itemize}
\item Soft and virtual corrections, which involve only the $gg$ channel and
 give rise to the ${\cal D}_i$ and $\delta(1-z)$ terms (see Eq.~(\ref{gg1})).
 These are the most singular terms when $z\to 1$.
\item Purely collinear logarithmic contributions, which are controlled by 
the regular part of the Altarelli--Parisi splitting kernels 
(see Eqs.~(\ref{gg1}), (\ref{gq1})).
The argument of the collinear logarithm corresponds to the maximum value
$( q^2_{T \, {\rm max}} \sim (1-z)^2M_H^2/z)$ of the
transverse momentum $q_T$ of the Higgs boson. These contributions give the
next-to-dominant singular terms when $z\to 1$.
\item Hard contributions, which are present in all partonic channels and
lead to finite corrections in the limit $z\to 1$ .
\end{itemize}

In this work we are mainly interested in studying the effect of soft-gluon
contributions to all perturbative orders.
Soft-gluon resummation
has to be carried out in the Mellin (or $N$-moment) space 
\cite{Sterman:1986aj,Catani:ne}. We thus
introduce our notation in the $N$-space.

We consider the Mellin transform $\sigma_N(M_H^2)$ of the hadronic cross
section $\sigma(s,M_H^2)$. The $N$-moments with respect to $\tau_H=M_H^2/s$
at fixed $M_H$ are thus defined as follows:
\begin{equation}
\label{sigman}
\sigma_N(M_H^2) \equiv \int_0^1 \;d\tau_H \;\tau_H^{N-1} \;\sigma(s,M_H^2) 
\;\;.
\end{equation} 
In $N$-moment space, Eq.~(\ref{had}) takes a simple factorized form
\begin{equation}
\label{hadn}
\sigma_{N-1}(M_H^2) = \sigma^{(0)} \;\sum_{a,b}
\; f_{a/h_1, \, N}(\mu_F^2) \; f_{b/h_2\, N}(\mu_F^2) 
\; {G}_{ab,\, N}(\as(\mu_R^2), M_H^2/\mu_R^2;M_H^2/\mu_F^2) \;,
\end{equation}
where we have introduced the customary $N$-moments  of the
parton distributions ($f_{a/h, \, N}$) and of the hard coefficient function
(${G}_{ab,\, N}$):
\begin{align} 
\label{pdfn}
f_{a/h, \, N}(\mu_F^2) &= \int_0^1 \;dx \;x^{N-1} \;
f_{a/h}(x,\mu_F^2) \;, \\
\label{gndef}
G_{ab,\, N} &= \int_0^1 dz \;z^{N-1} \;G_{ab}(z) \;\;.
\end{align}
Once these $N$-moments are known,
the physical cross section in $x$-space can be obtained by Mellin inversion:
\begin{align}
\sigma(s,M_H^2) = \sigma^{(0)} \;\sum_{a,b} & 
\;\int_{C_{MP}-i\infty}^{C_{MP}+i\infty}
\;\frac{dN}{2\pi i} \;\left( \frac{M_H^2}{s} \right)^{-N+1} \;
f_{a/h_1, \, N}(\mu_F^2) \; f_{b/h_2\, N}(\mu_F^2) \nonumber \\
\label{invmt}
& \times
\; {G}_{ab,\, N}(\as(\mu_R^2), M_H^2/\mu_R^2;M_H^2/\mu_F^2) \;,
\end{align} 
where the constant $C_{MP}$ that defines the integration contour in the $N$-plane
is on the right of all the possible singularities of the $N$-moments.

Note that the evaluation of $G_{ab}(z)$ in the limit $z \to 1$ corresponds to 
the evaluation of its $N$-moments $G_{ab,N}$ in the limit $N \to \infty$.
In particular, the soft, virtual and collinear contributions to 
$G_{ab}^{(n)}(z)$ lead to $\ln N$-enhanced contributions in $N$-space
according to the following correspondence (see Appendix~\ref{appa}):
\begin{align}
\label{lnsoft}
\int_0^1 \;dz \;z^{N-1} \;{\cal D}_k(z) &= \frac{(-1)^{k+1}}{k+1} \ln^{k+1} N 
+ {\cal O}(\ln^k N)\;\;,\\
\label{lnvir}
\int_0^1 \;dz \;z^{N-1} \;\delta(1-z) &= 1 \;\;,\\
\label{lncol}
\int_0^1 \;dz \;z^{N-1} \;\ln^k (1-z) &= \frac{(-1)^{k}}{N} \ln^{k} N
+ {\cal O}\left( \frac{1}{N} \ln^{k-1} N \right) \;\;.
\end{align}

\section{Soft-gluon resummation}
\label{sec:res}

\subsection{Summation to all logarithmic orders}
\label{sec:resall}

In this section we consider the all-order perturbative summation of enhanced
threshold (soft and virtual) contributions to the partonic cross section 
for Higgs boson production. The threshold region $z \to 1$ corresponds to the
limit $N \to \infty$ in $N$-moment space. 
We are thus interested in evaluating the hard coefficient function
$G_{ab, \,N}$, by keeping all the terms that are not vanishing when 
$N \to \infty$. To this purpose, we first note that in this limit only the
$gg$ partonic channel is not suppressed. In other words, we have:
\begin{equation}
\label{pcscaling}
G_{ab, \,N}(\as(\mu_R^2), M_H^2/\mu_R^2;M_H^2/\mu_F^2) = {\cal O}(1/N) \;\;
\quad (ab \neq gg) \;\;,
\end{equation}
where the notation ${\cal O}(1/N)$ means that the right-hand side vanishes
at least as a single power of $1/N$ (modulo $\ln N$ corrections) when
$N \to \infty$. When the partonic channel is  $ab \neq q{\bar q}$, 
Eq.~(\ref{pcscaling}) simply follows from power counting: the final state $X$
in the partonic subprocess $ab \to {H+X}$ contains at least a fermion, and 
the corresponding cross section thus vanishes in the soft limit.
When $ab=q{\bar q}$, the threshold limit selects the exclusive subprocess
$q{\bar q} \to H$ that vanishes since the
gluon-mediated production of a spin 0 particle
through $q{\bar q}$ annihilation is forbidden in the massless quark case.
As a matter of fact, gluonic interactions conserve helicity, so that the total spin
projection along the incoming $q\bar{q}$ direction is $\pm 1$,
which is incompatible with the production of a spin 0 state.

Observe that the large-$N$ behaviour of the hadronic cross section
$\sigma_N(M_H^2)$
also depends upon the large-$N$ behaviour of the parton densities,
according to Eq.~(\ref{hadn}). Thus, the ${\cal O}(1/N)$ relative suppression of
the partonic cross section in the $qg$ channel relative to the $gg$
channel may be compensated by the enhancement of the quark with
respect to the gluon density.  Under the typical assumption that the
gluon density is softer than the (valence) quark density at large $x$, it is
possible to show (see Sect.~2.4 in Ref.~\cite{Catani:1999hs})
that the two parton channels contribute with the same power behaviour in
$N$ to the total hadronic cross section. In the present work, however,
we are not considering the large-$N$ limit of the hadronic cross section
$\sigma_N(M_H^2)$ for Higgs boson production.
We are rather using the soft-gluon approximation to find a good
approximation of the full {\em partonic} cross section, to be convoluted with 
the parton densities and used in kinematical regimes\footnote{More discussion
about this point can be found in Sect.~\ref{sec:SV}.}
that are far from the {\em hadronic} large-$N$ limit. In this context,
Eq.~(\ref{pcscaling}) implies 
that the $gg$ channel strongly prevails over the
other channels in the evaluation of the cross section for Higgs boson 
production.
 
We are thus led to consider the $gg$ partonic channel.
The formalism to systematically perform soft-gluon resummation for 
hadronic processes,
in which a colourless massive particle is produced by $q{\bar q}$ annihilation 
or $gg$ fusion, was set up in 
Refs.~\cite{Sterman:1986aj,Catani:ne,Catani:1990rp}. In the case of Higgs
boson production, we have
\begin{equation}
\label{ggscaling}
G_{gg, \,N} = \as^2 \left\{ 1 + \sum_{n=1}^{+\infty} \as^n 
\sum_{m=0}^{2n} G_H^{(n,m)} \ln^mN \right\}+ {\cal O}(1/N)
= {G}_{gg,\, N}^{{\rm (res)}} + {\cal O}(1/N) \;\;,
\end{equation}
where the non-vanishing (singular and constant) contributions in the large-$N$
limit can be organized in the following {\em all-order} resummation formula:
\begin{align} 
\label{resformula} 
{G}_{gg,\, N}^{{\rm (res)}}(\as(\mu_R^2), M_H^2/\mu_R^2;M_H^2/\mu_F^2) 
&=\as^2(\mu_R^2)\, C_{gg}(\as(\mu^2_R),M_H^2/\mu^2_R;M_H^2/\mu_F^2) \nn \\ 
&\cdot \exp \{ {\cal G}_H(\as(\mu^2_R), \ln N;M_H^2/\mu^2_R,M_H^2/\mu_F^2)\}
\; . 
\end{align}

The function $C_{gg}(\as)$ contains all the contributions that are 
constant in the large-$N$ limit. They are produced by the hard virtual 
contributions and non-logarithmic soft corrections, and  
can be computed as a power series expansions in $\as$: 
\begin{equation}
\label{Cfun}
C_{gg}(\as(\mu^2_R),M_H^2/\mu^2_R;M_H^2/\mu_F^2) =  
1 + \sum_{n=1}^{+\infty} \;  
\left( \frac{\as(\mu^2_R)}{\pi} \right)^n \; 
C_{gg}^{(n)}(M_H^2/\mu^2_R;M_H^2/\mu_F^2) \;\;.
\end{equation}

All the large logarithmic terms $\as^n \ln^mN$ (with $1 \leq m \leq 2n$), which
are due to soft-gluon radiation,
are included in the exponential factor $\exp{\cal G}_H$. It can be expanded
as
\begin{align} 
\label{calgnnll} 
~\vspace{-.5cm} {\cal G}_H\!\left(\as(\mu^2_R),\ln N;\frac{M_H^2}{\mu^2_R}, 
\frac{M_H^2}{\mu_F^2}\right) &= \sum_{n=1}^{+\infty} \as^n
\sum_{m=1}^{n+1} {\cal G}_H^{(n,m)} \ln^mN  \\
&= \ln N \; g_H^{(1)}(\b0 \as(\mu^2_R) \ln N) + 
g_H^{(2)}(\b0 \as(\mu^2_R) \ln N, M_H^2/\mu^2_R;M_H^2/\mu_F^2 )  \nonumber \\ 
&+ \as(\mu^2_R) 
\;g_H^{(3)}(\b0 \as(\mu^2_R)\ln N,M_H^2/\mu^2_R;M_H^2/\mu_F^2 ) 
\nonumber \\
\label{gexpan}
&+ \sum_{n=4}^{+\infty} \left[ \as(\mu^2_R)\right]^{n-2}
\; g_H^{(n)}(\b0 \as(\mu^2_R)\ln N,M_H^2/\mu^2_R;M_H^2/\mu_F^2 )\;, 
\end{align} 
where, for later convenience, we have introduced the first coefficient, $\b0$,
of the QCD $\beta$-function. The functions $g_H^{(n)}$ are defined such that
$g_H^{(n)}(\b0 \as \ln N)=0$ when $\as=0$.

Note that the {\em exponentiation} in Eqs.~(\ref{resformula}) and 
(\ref{calgnnll}) is not trivial \cite{Sterman:1986aj,Catani:ne}. 
The sum over $m$ in Eq.~(\ref{ggscaling})
extends up to $m=2n$, while in Eq.~(\ref{calgnnll}) the maximum value for $m$ is
smaller, $ m \leq n+1$. In particular, this means that all the double
logarithmic (DL) terms $\as^n G_H^{(n,2n)} \ln^{2n}N$ in Eq.~(\ref{ggscaling})
are taken into account by simply exponentiating the lowest-order contribution
$\as G_H^{(1,2)} \ln^{2}N$. Then, the exponentiation allows us to define the
resummed perturbative expansion in Eq.~(\ref{gexpan}). The function
$\ln N \; g_H^{(1)}$ resums all the {\em leading} logarithmic (LL) contributions
$\as^n \ln^{n+1}N$, $g_H^{(2)}$ contains the {\em next-to-leading} logarithmic 
(NLL) terms $\as^n \ln^{n}N$, $\as g_H^{(3)}$ collects
the {\em next-to-next-to-leading} logarithmic (NNLL) terms 
$\as^{n+1} \ln^{n}N$, and so forth.
Note that in the context of soft-gluon resummation, the parameter $\as \ln N$
is formally considered as being of order unity. Thus, the ratio of two
successive terms in the expansion (\ref{gexpan}) is formally of ${\cal O}(\as)$
(with no $\ln N$ enhancement). In this respect, 
the resummed logarithmic expansion in Eq.~(\ref{gexpan}) is as systematic 
as any customary fixed-order expansion in powers of $\as$.

The purpose of the soft-gluon resummation program is to explicitly evaluate
the logarithmic functions $g^{(n)}$ of Eq.~(\ref{gexpan}) in terms of few
coefficients that are perturbatively computable.
In the case of Higgs boson production, this goal is achieved by showing that 
the all-order resummation formula (\ref{resformula}) can be recast in the 
following form \cite{Catani:2001ic}:
\begin{align}
\label{resfdelta}
{G}_{gg,\, N}^{{\rm (res)}}(\as(\mu_R^2), M_H^2/\mu_R^2;M_H^2/\mu_F^2) 
&= \as^2(\mu_R^2)\, 
{\overline C}_{gg}(\as(\mu^2_R),M_H^2/\mu^2_R;M_H^2/\mu_F^2) \nn \\ 
&\cdot  \Delta_{N}^{H}(\as(\mu^2_R),M_H^2/\mu^2_R;M_H^2/\mu_F^2) +
{\cal O}(1/N)\; . 
\end{align}
The factor ${\overline C}_{gg}(\as)$ in Eq.~(\ref{resfdelta}) is completely 
analogous to the factor ${C}_{gg}(\as)$ in Eq.~(\ref{resformula}). The
difference between ${C}_{gg}$ and ${\overline C}_{gg}$ is simply due to 
the fact that some constant terms at large $N$ have been moved from 
${C}_{gg}$ to $\Delta_{N}^{H}$. The Sudakov radiative factor
$\Delta_{N}^{H}$ has the following integral representation:
\begin{align}
\Delta_N^{H}\!\left(\as(\mu^2_R),\f{M_H^2}{\mu^2_R};\f{M_H^2}{\mu_F^2}\right) 
= \exp &\left\{ \int_0^1 \;dz \; \frac{z^{N-1} - 1}{1-z} \right. \nonumber \\
\label{deltarep}
&\times \left. \left[ 2 \int_{\mu_F^2}^{(1-z)^2M_H^2}
\; \frac{dq^2}{q^2} \; A(\as(q^2)) + D(\as((1-z)^2M_H^2))
\right] \right\} \;\;,
\end{align}
where $A(\as)$ and $D(\as)$ are perturbative functions
\begin{align}
\label{Afun}
A(\as) &= \sum_{n=1}^{+\infty} \left(\asp \right)^n A^{(n)} 
={\asp} A^{(1)}+\left(\asp \right)^2 A^{(2)} 
+\left(\asp \right)^3 A^{(3)}+ {\cal O}(\as^4)
\;\;, \\
\label{Dfun}
D(\as) &= \sum_{n=2}^{+\infty} \left(\asp \right)^n D^{(n)} 
= \left(\asp \right)^2 D^{(2)} + {\cal O}(\as^3)
\;\;.
\end{align} 
The coefficients $A^{(n)}$ and $D^{(n)}$ are perturbatively computable.
For example, they can be extracted from the calculation of
${G}_{gg,\, N}$ at N$^n$LO.

By inspection of $z$ and $q^2$ integrations in Eq.~(\ref{deltarep}),
it is evident that the radiative factor
leads to the logarithmic structure of Eq.~(\ref{gexpan}), plus
corrections of ${\cal O}(1/N)$ that vanish when $N\to \infty$. 
The functions $g_H^{(n)}$ depend on the coefficients in Eqs.~(\ref{Afun}) and
(\ref{Dfun}), and the functional dependence is completely specified by
Eq.~(\ref{deltarep}). More precisely (see Eqs.~(\ref{g1fun})--(\ref{g3fun})),
the LL function $g_H^{(1)}$ depends on $A^{(1)}$, the NLL function 
$g_H^{(2)}$ depends also on $A^{(2)}$, the NNLL function $g_H^{(3)}$ depends 
also on $A^{(3)}$ and $D^{(2)}$, and so forth. In Appendix~\ref{appa}, we
describe in detail a method to obtain the functions $g_H^{(n)}$ 
(for arbitrary values of $n$) from the integral representation 
in Eq.~(\ref{deltarep}).

The structure of Eq.~(\ref{deltarep}) is completely analogous to that of the
radiative factor of the Drell--Yan (DY) process. The derivation of 
this result
up to NLL accuracy (i.e. keeping only the coefficients $A^{(1)}$ and 
$A^{(2)}$) was discussed in Refs.~\cite{Sterman:1986aj,Catani:ne,Catani:1990rr}.
The discussion of Refs.~\cite{Sterman:1986aj,Catani:ne,Catani:1990rr}
can be extended to any logarithmic accuracy by taking into account
the following
two main points.

First, beyond ${\cal O}(\as^2)$ the Sudakov radiative factor acquires an
additional contribution 
\cite{Catani:1998tm,Laenen:1998qw,Kidonakis:1997gm,Bonciani:1998vc}
due to final-state soft partons emitted
at large angles with respect to the direction of the colliding gluons
(or of the colliding $q{\bar q}$ pair, in the case of the DY process).
In Eq.~(\ref{deltarep}) this contribution\footnote{This contribution
is denoted by $\Delta_N^{({\rm int})}$ in 
Refs.~\cite{Catani:2001ic,Catani:1998tm}.}
is included in the term that depends on the function $D(\as((1-z)^2M_H^2))$.

Second, the term proportional to the function $A(\as(q^2))$ in 
Eq.~(\ref{deltarep}) embodies the effect of soft-parton radiation emitted
collinearly to the initial-state partons; it therefore depends on both
the factorization scheme and the factorization scale $\mu_F$ of the gluon
parton distributions in Eq.~(\ref{hadn}). The invariance of the hadronic
cross section with respect to $\mu_F$ variations thus implies that 
the function $A(\as)$ gets higher-order ($n \geq 3$) contributions 
(given by the coefficients $A^{(n)}$ in Eq.~(\ref{Afun})) to compensate
the factorization-scale dependence of the parton distributions, as given by the
Altarelli--Parisi evolution equations
\begin{equation}
\label{apeqs}
\f{d \,f_{a/h, \,N}(\mu_F^2)}{d \ln \mu_F^2} = \sum_{b} 
\gamma_{ab, \,N}(\as(\mu_F^2)) \;f_{b/h, \,N}(\mu_F^2) \;\;.
\end{equation}
It is straightforward to show that the function $A(\as)$ in 
Eqs.~(\ref{deltarep}) and (\ref{Afun}) coincides at any perturbative order
with the function that controls the large-$N$ behaviour 
\cite{Korchemsky:1988si} of the gluon 
anomalous dimensions $\gamma_{gg, \,N}$ in the \ms\ factorization scheme:
\begin{equation}
\label{gammagg}
\gamma_{gg, \,N}(\as) = - A(\as) \,\ln N + {\cal O}(1) \quad\quad
(N \to \infty) \;\;.
\end{equation}

The Sudakov radiative factor in Eq.~(\ref{deltarep})
can also be expressed
by using an alternative integral representation:
\begin{align}
\label{deltarepn0}
\Delta_N^{H}\!\left(\as(\mu^2_R),\f{M_H^2}{\mu^2_R};\f{M_H^2}{\mu_F^2}\right) 
&=
\exp \left \{ - \int_{N_0/N}^1 \;\frac{dy}{y}  \left[ 2 \int_{\mu_F^2}^{y^2M_H^2}
\; \frac{dq^2}{q^2} \; A(\as(q^2)) + {\widetilde D}(\as(y^2M_H^2))
\right] \right\} \\
&\times {\widetilde C}_{gg}(\as(\mu^2_R),M_H^2/\mu^2_R) + {\cal O}(1/N)\, , \nn 
\end{align}
where $N_0=e^{-\gamma_E}$ ($\gamma_E=0.5772\dots$ is the Euler number), 
and the new perturbative functions
${\widetilde C}_{gg}(\as)$ and ${\widetilde D}(\as)$ 
are completely analogous to the functions $C_{gg}(\as)$ and $D(\as)$
in Eqs.~(\ref{Cfun}) (unlike $C_{gg}$, ${\widetilde C}_{gg}$ does not depend
on $\mu_F^2$) and (\ref{Dfun}), respectively. Note that the
$\ln N$ dependence of $\Delta_N^{H}$ is fully included in the exponential factor
on the right-hand side of Eq.~(\ref{deltarepn0}).
The representation in 
Eq.~(\ref{deltarepn0}) was first introduced in Ref.~\cite{Catani:ne} 
up to NLL accuracy.
The equivalence between Eq.~(\ref{deltarep}) and (\ref{deltarepn0}) 
to any logarithmic accuracy is proved in Appendix~\ref{appb}, where we also 
derive the formulae that explicitly relate the functions $A(\as)$ and 
$D(\as)$ to the functions ${\widetilde C}_{gg}(\as)$ and ${\widetilde D}(\as)$.
Using Eq.~(\ref{deltarepn0}), it is straightforward to carry out the 
$z$ and $q^2$ integrations and to obtain the logarithmic functions $g_H^{(n)}$
in Eq.~(\ref{gexpan}) (see Sect.~\ref{sec:resNNLL} and Appendix~\ref{appc}).

These results on soft-gluon resummation at any logarithmic accuracy deserve
some comments on the large-order perturbative behaviour. As is well known
(see \cite{Beneke:1998ui} and references therein), any perturbative QCD 
expansion, such as Eq.~(\ref{alloexpansion}), 
has to be regarded as an asymptotic rather than a
convergent series. In fact, the perturbative coefficients 
(e.g. $G_{ab}^{(n)}$) are expected to diverge as $n!$ 
when the perturbative order $n$ becomes very large.
The ambiguities related to the definition of the asymptotic series are then
interpreted as perturbative evidence of non-perturbative power corrections.
These features apply to the fixed-order perturbative expansion in
Eq.~(\ref{alloexpansion}) as well as to the resummed logarithmic expansion
in Eq.~(\ref{gexpan}).
In other words, the functions $g_H^{(n)}$ are expected to
diverge as $n!$ at large $n$.

Infrared renormalons \cite{Beneke:1998ui}
are a known source of factorially divergent terms
in perturbative QCD. They arise from the behaviour of the running coupling 
$\as(q^2)$ in the infrared region and, more precisely, from integrating the
coupling down to momentum scales $q$ below the Landau pole, as set by the QCD
scale $\Lambda_{QCD}$. The representation in Eq.~(\ref{deltarep}) may lead to 
infrared renormalons, since it involves
$z$ and $q^2$ integrals that extend in the infrared region
{\em independently} of the value of $N$. 
In Ref.~\cite{Korchemsky:1994is}, it was observed that the ensuing
factorially divergent behaviour corresponds to a power correction 
$\Lambda_{QCD}/M_H$, which is linear in $1/M_H$. This observation was based on
the truncation of the perturbative function $A(\as)$ at a fixed perturbative
order. However, as pointed out by Beneke and Braun \cite{Beneke:1995pq}, this
simplifying assumption is not sufficient to draw conclusions on power
corrections. Factorial divergences can arise both from the $z$ and $q^2$ 
integrals and from the large-order behaviour of the functions $A(\as)$
and $D(\as)$: both effects have to be taken into account \cite{Beneke:1995pq}.
This feature is evident by comparing the integral representations in
Eqs.~(\ref{deltarep}) and (\ref{deltarepn0}). The two all-order
representations are fully equivalent, but the integrals 
in Eq.~(\ref{deltarepn0}) are perfectly 
convergent as long as $N \leq N_0M_H/\Lambda_{QCD} \sim M_H/\Lambda_{QCD}$. 
In Eq.~(\ref{deltarepn0})
factorial divergences can appear only through the large-order behaviour of the
functions $A(\as)$ and ${\widetilde D}(\as)$ (see also Appendix~\ref{appb}). 
The detailed study of Ref.~\cite{Beneke:1995pq} shows that the functions
$D(\as)$ and ${\widetilde D}(\as)$ (due to large-angle soft-gluon radiation)
are indeed factorially divergent. In particular, $D(\as)$ has a factorial
divergence that corresponds to a linear power correction and cancels
the linear power correction
found in Ref.~\cite{Korchemsky:1994is}. Renormalon calculations, 
based on the explicit evaluation of the dominant terms
at large $N_f$, lead to power corrections of the type $\Lambda_{QCD}^2/M_H^2$
\cite{Beneke:1995pq} (or, more generally, integer powers of
$N^2\Lambda_{QCD}^2/M_H^2$ \cite{Gardi:2001di}).

Note that the $z$ integral in Eq.~(\ref{deltarepn0}) is not regular when
$N > N_0M_H/\Lambda_{QCD} \sim M_H/\Lambda_{QCD}$, 
but this does not lead to factorial divergences
\cite{Catani:1996yz}. We postpone further comments on this point to 
Sect.~\ref{sec:resxs}.

The all-order resummation formulae presented in this section are valid
for Higgs boson production, independently of the use of the large-$M_t$
approximation. In particular, the exponential factor $\exp{\cal G}_H$
in Eq.~(\ref{resformula}) and the Sudakov radiative factor $\Delta_N^{H}$ 
in Eqs.~(\ref{deltarep}) and (\ref{deltarepn0}) do not depend on $M_t$.
The full dependence on $M_t$ of the resummation formula (\ref{resformula})
is embodied in the $N$-independent function $C_{gg}(\as)$, namely, in its
perturbative coefficients $C_{gg}^{(n)}$ (see Eq.~(\ref{Cfun})).
As shown below in Eqs.~(\ref{cgg1})--(\ref{deltaG2}),
in the large-$M_t$ limit, the coefficient $C_{gg}^{(1)}$ becomes
independent of $M_t$, while $C_{gg}^{(2)}$ depends logarithmically on
$M_H/M_t$.

\subsection{Soft-gluon resummation at NNLL accuracy}
\label{sec:resNNLL}

In the following we are interested in a quantitative study of soft-gluon
resummation effects up to NNLL accuracy. We thus need the Higgs bosons 
coefficients
$A^{(1)}, A^{(2)}, A^{(3)}$ and $D^{(2)}$ in Eqs.~(\ref{Afun}) and 
(\ref{Dfun}). Note that these coefficients are related\footnote{This relation
follows from the general structure \cite{Catani:2000pi,Catani:2000ss}
of the soft-gluon factorization formulae
at ${\cal O}(\as^2)$.} to the analogous
coefficients of the DY process \cite{Catani:ne,Matsuura:sm,vogtresum}
by a simple overall factor $C_a$, which is proportional to the 
colour charges of the colliding partons in the different processes.
Therefore we explicitly introduce in the following expressions
the factor $C_a$, where $C_a=C_A=N_c=3$ in Higgs boson production
and $C_a=C_F=(N_c^2-1)/2N_c=4/3$ in DY production.

The LL and NLL coefficients $A^{(1)}$ and $A^{(2)}$ are well 
known \cite{KT,Catani:vd}:
\begin{equation} 
\label{A12coef} 
A^{(1)}= C_a\;,\;\;\;\; A^{(2)}=\frac{1}{2} \; C_a K \;, 
\end{equation} 
with
\begin{equation} 
\label{kcoef} 
K = C_A \left( \frac{67}{18} - \frac{\pi^2}{6} \right)  
- \frac{5}{9} N_f \;. 
\end{equation}
The NNLL coefficient $D^{(2)}$ 
was evaluated in Refs.~\cite{Catani:2001ic,Harlander:2001is}:
\begin{align} 
D^{(2)}= C_a \left[ C_A \left( -\f{101}{27} + \f{11}{3} \,\zeta(2)+ \f{7}{2}
 \,\zeta(3) \right) +  N_f \left( \f{14}{27} -\f{2}{3} \,\zeta(2)\right) 
 \right]\;\;.
\end{align}
The higher-order coefficients $A^{(n)}$ with $n \geq 3$ are not fully known.
However, the contribution to $A^{(n)}$ of the term proportional to $N_f^{n-1}$
can be extracted from calculations \cite{Beneke:1995pq,Gracey} 
in the large-$N_f$ limit. Moreover, by exploiting the relation 
(\ref{gammagg}) between the anomalous dimensions and $A(\as)$, 
the available approximation \cite{vnvogt}
of the NNLO anomalous dimensions can be used to obtain a corresponding
numerical estimate \cite{vogtresum} of the NNLL coefficient $A^{(3)}$:
\begin{align} 
\label{A3coef}
A^{(3)}&= C_a \left\{ (13.81 \pm 0.13) 
- \frac{1}{2} \left[ C_F \left( \frac{55}{48} - \zeta(3) \right) 
+ C_A \left( \frac{209}{216} - \frac{5\pi^2}{54}+ \frac{7 \zeta(3)}{6} 
\right) \right] N_f
-\f{1}{108} N_f^2 \right\} \\
&= C_a \left[ 
 (13.81 \pm 0.13) -2.1467\dots \;N_f -\f{1}{108} N_f^2 \right] \;. \nonumber
\end{align}
where we have used the recent analytical computation 
\cite{Moch:2002sn,Berger:2002sv}
of the term proportional to $N_f$,
which agrees with the
approximate numerical calculation in Ref.~\cite{vnvogt}.

The LL, NLL and NNLL functions $g_H^{(1)}, g_H^{(2)}$ and $g_H^{(3)}$ in 
Eq.~(\ref{gexpan}) have the following explicit expressions 
(see Appendix~\ref{appc}):
\begin{align} 
\label{g1fun}
g_H^{(1)}(\la) =&+ \f{A^{(1)}}{\pi \b0 \la} 
\left[ 2 \la+(1-2 \la)\ln(1-2\la)\right] \;,\\ 
g_H^{(2)}(\la,M_H^2/\mu^2_R;M_H^2/\mu_F^2) 
=&-\f{A^{(2)}}{\pi^2 \b0^2 } \left[ 2 \la+\ln(1-2\la)\right] - 
\f{2 A^{(1)} \gamma_E}{\pi \b0 } \ln(1-2\la)\nn \\ 
&+ \f{A^{(1)} \bone}{\pi \b0^3} 
\left[2 \la+\ln(1-2\la)+\f{1}{2} \ln^2(1-2\la)\right]\nn \\ 
\label{g2fun}
&+ \f{A^{(1)}}{\pi \b0}\left[2 \la+\ln(1-2\la) \right]  
\ln\f{M_H^2}{\mu^2_R}-\f{2 A^{(1)}}{\pi \b0} \,\la \ln\f{M_H^2}{\mu^2_F} \;,  
\end{align} 

\begin{align} 
g_H^{(3)}(\la,M_H^2/\mu^2_R;M_H^2/\mu_F^2) = 
&+ \f{4 A^{(1)}}{\pi} (\zeta(2)+\gamma_E^2) \f{\la}{1-2\la} 
-\f{2 A^{(1)} \gamma_E \bone}{\pi \b0^2 (1-2\la)} \left[2\la+\ln(1-2\la) \right]
\nn \\ 
& + \f{A^{(1)}  \bone^2}{\pi \b0^4 (1-2\la)} 
\left[2\la^2+ 2\la \ln(1-2\la)+\f{1}{2} \ln^2(1-2\la) \right] \nn \\ 
& + \f{A^{(1)}  \btwo}{\pi\b0^3} 
\left[2\la+\ln(1-2\la) +\f{2\la^2}{1-2\la} \right] 
+ \f{2 A^{(3)}}{\pi^3 \b0^2} \f{\la^2}{1-2\la}
-  \f{ D^{(2)}}{\pi^2 \b0} \f{\la}{1-2\la} \nn \\ 
&+ \f{4 \gamma_E A^{(2)}}{\pi^2 \b0} \f{\la}{1-2\la}  
- \f{A^{(2)} \bone }{\pi^2 \b0^3} \f{1}{1-2\la} 
 \left[2\la+\ln(1-2\la) +2 \la^2 \right]  
 \nn \\ 
&- \f{2 A^{(2)}}{\pi^2 \b0} \,\la \ln\f{M_H^2}{\mu^2_F} 
- \f{A^{(1)}}{\pi} \, \la  \ln^2\f{M_H^2}{\mu^2_F} 
+ \f{2 A^{(1)}}{\pi} \la \ln\f{M_H^2}{\mu^2_R} \ln\f{M_H^2}{\mu^2_F} \nn \\ 
&+ \f{1}{1-2\la} \Big( 
 \f{ A^{(1)} \bone}{\pi \b0^2} \left[ 2\la+\ln(1-2\la) \right] - 
\f{4 A^{(1)} \gamma_E}{\pi} \la  - \f{4 A^{(2)}}{\pi^2 \b0}  \la^2 \Big)
\ln\f{M_H^2}{\mu^2_R} \nn \\ 
\label{g3fun}
&+ \f{2A^{(1)}}{\pi} \f{\la^2}{1-2\la}\ln^2\f{M_H^2}{\mu^2_R}  \;\;,
\end{align} 
where
\begin{equation}
\lambda= \b0 \as(\mu^2_R) \ln N \;\;,
\end{equation}
and
$\b0, \bone, \btwo$ are the first three coefficients of the QCD 
$\beta$-function \cite{beta2}:
\begin{align}
\b0 &= \frac{1}{12 \pi} \left( 11 C_A - 2 N_f \right) \;\;,
\quad\quad \bone=  \frac{1}{24 \pi^2} 
\left( 17 C_A^2 - 5 C_A N_f - 3 C_F N_f \right) \;\;,
\nonumber \\
\label{bcoef}
\btwo &= \frac{1}{(4 \pi)^3} \left( \f{2857}{54} C_A^3
- \f{1415}{54} C_A^2 N_f - \f{205}{18} C_A C_F N_f + C_F^2 N_f
+ \f{79}{54} C_A N_f^2 + \f{11}{9} C_F N_f^2 \right) \;\;.
\end{align}
The functions $g_H^{(1)}$ and $g_H^{(2)}$ are well known 
(see e.g. Ref.~\cite{Catani:1998tm}).
The NNLL function $g_H^{(3)}$ was first evaluated in Ref.~\cite{vogtresum}. 
Our result in Eq.~(\ref{g3fun}) is obtained by using a different method,
and confirms the result of Ref.~\cite{vogtresum}.

To fully exploit the content of the resummation formula (\ref{resformula})
up to NLL (and NNLL) accuracy we need the constant coefficient 
$C_{gg}^{(1)}$ (and $C_{gg}^{(2)}$) in Eq.~(\ref{Cfun}). 
The coefficients $C_{gg}^{(1)}$ and $C_{gg}^{(2)}$ read
\begin{align} 
\label{cgg1}
 C_{gg}^{(1)}&= \delta G_{gg}^{(1)} + 6\gamma_E^2 + \pi^2 - 6\gamma_E
 \ln\f{M_H^2}{\mu_F^2} \;,
\\
 C_{gg}^{(2)}&= \delta G_{gg}^{(2)} + \gamma_E \left(\f{101}{3}-\f{14}{9} N_f
 -\f{63}{2}\zeta(3)\right)+ \gamma_E^2 \left( \frac{133}{2} - \frac{5 N_f}{3} +
 \frac{21\,{\pi }^2}{2} \right)+ \gamma_E^3 \left(11 - \frac{2 {N_f} }{3} 
 \right) + 18 \gamma_E^4 \nn\\ 
&+ \frac{133\,{\pi }^2}{12} - \frac{5\,{N_f}\,{\pi }^2}{18}
 + \frac{29\,{\pi }^4}{20} + 22\,\zeta(3) - \frac{4\,{N_f}\,\zeta(3)}{3}
 +\ln^2\f{M_H^2}{\mu_F^2} \left( -\f{165}{4} \gamma_E + 18 \gamma_E^2 + \f{5}{2}
 N_f \gamma_E +3 \pi^2 \right) \nn \\ 
&+ \f{3}{2}\gamma_E (33-2 N_f)
 \ln\f{M_H^2}{\mu_F^2}\ln\f{M_H^2}{\mu_R^2} - \f{1}{4} (33-2
 N_f) (6 \gamma_E^2+\pi^2) \ln\f{M_H^2}{\mu_R^2} \nn \\ 
&+ \ln\f{M_H^2}{\mu_F^2} \left[-36 \gamma_E^3
 + (33-2 N_f) \gamma_E^2 + \gamma_E \left( - \frac{133}{2} +
 \frac{5\,{N_f}}{3} - \frac{21\,{\pi }^2}{2} \right) \right.\nn\\
 &\left.+ \frac{11\,{\pi }^2}{2} - \frac{{N_f}\,{\pi }^2}{3} - 72\,{\zeta}(3) 
 \right]\;,
\end{align} 
where
\begin{align}
\label{deltaG1}
\delta G_{gg}^{(1)}&=
\f{11}{2} + 6 \zeta(2) + 
\f{33-2N_f}{6}\ln\f{\mu_R^2}{\mu_F^2} \;,
\\
\delta G_{gg}^{(2)}&=
\f{11399}{144} +\f{133}{2}\zeta(2) -\f{9}{20}\zeta(2)^2
  -\f{165}{4}\zeta(3)\nn\\ 
&
+  \left(\f{19}{8}+\f{2}{3} N_f\right)
 \ln\f{M_H^2}{M_t^2}
  +N_f \left( -\f{1189}{144} -\f{5}{3} \zeta(2) +\f{5}{6}\zeta(3)\right)
 \nn \\
&
+\f{\left(33-2N_f\right)^2}{48}\ln^2\f{\mu_F^2}{\mu_R^2}- 18\,\zeta(2)
\ln^2\f{M_H^2}{\mu_F^2}\nn \\
&
+ \left( \f{169}{4}+\f{171}{2}\zeta(3)- \f{19}{6}N_f
 +\left(33-2N_f\right)\zeta(2)\right)
\ln\f{M_H^2}{\mu_F^2}\nn \\
\label{deltaG2}
&
+\left(-\f{465}{8}+\f{13}{3}N_f-\f{3}{2}\left(33-2N_f\right)\zeta(2)\right)
\ln\f{M_H^2}{\mu_R^2} \;\;.
\end{align}
The terms $\delta G_{gg}^{(1)}$ and $\delta G_{gg}^{(2)}$ are
the coefficients of the contribution proportional to $\delta(1-z)$
in the coefficient functions $G_{gg}^{(1)}(z)$ and $G_{gg}^{(2)}(z)$,
respectively. The NLO term $\delta G_{gg}^{(1)}$ can be read from 
Eq.~(\ref{gg1}). The NNLO term $\delta G_{gg}^{(2)}$ was computed in 
Refs.~\cite{Catani:2001ic,Harlander:2001is}.

As pointed out in Ref.~\cite{Kramer:1998iq},
the dominant part of the corrections of ${\cal O}(1/N)$ in Eq.~(\ref{ggscaling})
is due to collinear-parton radiation, since it produces terms that are
enhanced by powers of $\ln N$ (or $\ln (1-z)$, as shown in Eqs.~(\ref{gg1})
and (\ref{lncol})). 
These corrections can be included in the soft-gluon resummation formula
(\ref{resformula}). In particular, by implementing
the simple modification 
\begin{equation}
\label{lcollpres}
C_{gg}^{(1)}  \to C_{gg}^{(1)} + 2 A^{(1)} \;\f{\ln N}{N} \;,
\end{equation}
to the coefficient $C_{gg}^{(1)}$, Eq.~(\ref{resformula}) correctly resums
\cite{Kramer:1998iq,Catani:2001ic}
all the leading collinear contributions to $G_{gg,\, N}$, i.e. all the terms
of the type $(\as^n \ln^{2n-1}N)/N$ that appear in the large-$N$ behaviour
of $G_{gg,\, N}^{(n)}$.
In the following sections
we use the prescription in Eq.~(\ref{lcollpres}) to quantitatively estimate
the dominant corrections to soft-gluon resummation.

\section{Soft-virtual approximation}
\label{sec:SV}
\subsection{Soft-virtual approximation at NNLO}
\label{sec:fonsp}
In Sect.~\ref{sec:res} we have discussed a method
to resum soft-gluon effects to all perturbative
orders. The resummed formula
is formally justified in
the threshold limit $z \to 1$, where the expansion parameter $\ln N$
is really large, so that terms that are suppressed
by powers of $1/N$
are negligible. We wish, however, to
use the resummed formulae also away from the threshold region.
Our justification for doing so is that we expect that the
large-$N$ approximation is a good quantitative approximation to the exact
result. 

The reasons for this expectation are discussed in detail in 
Refs.~\cite{Catani:2001ic,Catani:2001cr}.
We briefly summarize the main point of that discussion.
In the evaluation of the hadronic cross section
in Eq.~(\ref{had}), the partonic cross section $\hat \sigma_{ab}({\hat
s},M_H^2)$ has to be weighted (convoluted) with the parton densities.
Owing to the strong suppression of the parton densities $f_{a/h}(x,\mu_F^2)$
at large $x$, the partonic centre-of-mass energy $\sqrt {\hat s}$ is typically
substantially smaller than $\sqrt s$ ($\langle {\hat s} \rangle = 
\langle x_1 x_2 s \rangle = \langle \tau_H \rangle s$), and the dominant values
of the variable 
$z=M_H^2/\hat s$ in the hard coefficient function 
$G_{ab}(z)$ can be close to unity also when $\sqrt s$ is not very close to
$M_H$ \cite{Contogouris:1990zq}.

At fixed $M_H$, the reliability of the large-$z$ (large-$N$)
approximation of the hard coefficient function
depends on the value of $\sqrt s$ and on the actual value of the parton
densities. Therefore, the qualitative expectation of the relevance of the 
large-$N$ approximation
has to be quantitatively tested at the level at which the
exact result is known, that is up to the NNLO level.
Since in the rest of the paper we are interested in studying
higher-order soft-gluon effects within the $N$-space resummation formalism
of Sect.~\ref{sec:res},
in this subsection we study the soft
approximations to the fixed-order perturbative coefficients in
$N$-space up to NNLO.  At the end of the subsection, we also discuss
the soft approximations in $x$-space, which were already considered in
Refs.~\cite{Catani:2001ic,Catani:2001cr}.

We begin by defining the $N$-space {\em soft-virtual} (SV-$N$) approximation 
of the hard coefficient function. We write
\begin{equation}
{G}_{gg,\, N}^{{\rm (res)}}
=\as^2(\mu_R^2) \left[ 1+ \frac{\as(\mu_R^2)}{\pi} \,G_{gg, N}^{(1) 
\,\mlbl{SV-$N$}}
+\left( \frac{\as(\mu_R^2)}{\pi}\right)^2 G_{gg, N}^{(2) \,\mlbl{SV-$N$}}
+{\cal O}(\as^3) \right]\;.
\end{equation}
The coefficients $G_{gg, N}^{(1,2) \,\mlbl{SV-$N$}}$
can be obtained
either by Mellin transformation of $G^{(1,2)}_{gg}(z)$, neglecting all terms
formally suppressed by powers of $N$, or by expanding
Eq.~(\ref{resformula}) to the fourth order in $\as$.
We get
\begin{align}
\label{ggSV-N}
G_{gg, N}^{(1) \,\mlbl{SV-$N$}}(& M_H^2/\mu_R^2; M_H^2/\mu_F^2)=
6 \ln^2 N +12 \gamma_E \LN -\,6\, \LN
\ln\f{M_H^2}{\mu_F^2} + C_{gg}^{(1)} \;\;,\\
G_{gg, N}^{(2) \,\mlbl{SV-$N$}}(& M_H^2/\mu_R^2,
M_H^2/\mu_F^2)= 18\,\ln^4 N 
+ \ln^3 N 
\left[ \f{1}{3} (33\,- 2\,N_f) + 72\,\gamma_E\, -
36\,\ln\f{M_H^2}{\mu_F^2} \right] \nn \\
+ \ln^2 N & \left[ 6\,C_{gg}^{(1)} + (33-2\, N_f)\,\gamma_E +
 72\,{\gamma_E}^2 -\frac{5}{3}\,N_f + \frac{67}{2} - \frac{3}{2}{\pi
 }^2 \right.  \nn \\
 & \left.+ 18\,{\ln^2\f{M_H^2}{\mu_F^2}} -72
 {\gamma_E} \ln\f{M_H^2}{\mu_F^2} - \f{1}{2} (33-
 2\,N_f)\,\ln\f{M_H^2}{\mu_R^2} \right] \nn \\
 +\LN &\left[ (33 - 2
 N_f) \gamma_E^2 + 12 \gamma_E\,C_{gg}^{(1)} + \gamma_E \left( 67 -
 \frac{10}{3} N_f - 3\,{\pi }^2 \right) + \frac{101}{3} - \frac{14}{9}
 N_f - \frac{63}{2}\,{\zeta}(3) \right.  \nn \\ 
& \left. -\f{1}{4}
 (33-2 N_f) \ln^2\f{M_H^2}{\mu_F^2} +\f{1}{2} (33-2 N_f)
 \ln\f{M_H^2}{\mu_F^2} \, \ln\f{M_H^2}{\mu_R^2}- (33-2 N_f) \gamma_E
 \ln\f{M_H^2}{\mu_R^2}\right.  \nn \\ 
& \left. + \ln\f{M_H^2}{\mu_F^2}
 \left( -\f{67}{2} + \f{5}{3} N_f +\f{3}{2} \pi^2 -6 \,C_{gg}^{(1)}
 \right) \right] + C_{gg}^{(2)}\;\;.
\label{gg2SV-N} 
\end{align}
The $N$-space {\em soft-virtual-collinear} (SVC-$N$)
approximation is defined as
\begin{align}
\label{ggSVC-N}
G_{gg, N}^{(1) \,\mlbl{SVC-$N$}}(M_H^2/\mu_R^2;M_H^2/\mu_F^2)&=
G_{gg, N}^{(1) \,\mlbl{SV-$N$}}(M_H^2/\mu_R^2;M_H^2/\mu_F^2) + 6 \f{\LN}{N} \;,
\\
G_{gg, N}^{(2) \,\mlbl{SVC-$N$}}(M_H^2/\mu_R^2;M_H^2/\mu_F^2)&=
G_{gg, N}^{(2) \,\mlbl{SV-$N$}}(M_H^2/\mu_R^2;M_H^2/\mu_F^2) 
+ 36 \f{\ln^3N}{N} \nn\\
\label{gg2SVC-N}
&+ 72 \gamma_E \f{\ln^2 N}{N} -36 \f{\ln^2 N}{N}
\ln\f{M_H^2}{\mu_F^2} \;\;.
\end{align}
Note that we keep
terms proportional to $\ln^2 N/N$ in the two-loop coefficient
$G_{gg, N}^{(2) \,\mlbl{SVC-$N$}}$. These subleading collinear terms\footnote{We
recall that these terms are not the complete contribution proportional 
to $\ln^2 N/N$ in $G_{gg, N}^{(2)}$. We also anticipate that the effect of
including these terms in Eq.~(\ref{gg2SVC-N}) is numerically negligible.}
appear in the expansion of the (modified) resummed formulae when
the leading collinear terms are taken into account according to 
Eq.~(\ref{lcollpres}).

We now want to compare the SV-$N$ and SVC-$N$ approximations to the exact
results, both at NLO and NNLO. 
To do this, at each order  we define the quantity
\begin{equation}
\label{eq:delta}
\Delta_{A}(\mu_F,\mu_R)=
\f{\sigma_{A}(\mu_F,\mu_R)-\sigma(\mu_F=\mu_R=M_H)}
{\sigma(\mu_F=\mu_R=M_H)} \;\;,
\end{equation}
where the subscript $A$ stands for SV, SVC, or nothing
(in the case of no approximation) at the given order.
The
approximated cross sections $\sigma_{\rm SV}$ and $\sigma_{\rm SVC}$
at NLO are defined by Eq.~(\ref{invmt}) with the replacements
\begin{eqnarray}
G_{ab,N}&\Longrightarrow &0
\quad \mbox{for $ab\neq gg$}
 \\
G_{gg,N}
&\Longrightarrow &\as^2(\mu_R^2) \left[ 1+ 
\frac{\as(\mu_R^2)}{\pi}
\,G_{gg, N}^{(1) 
\,\mlbl{A-$N$}}\right]\;, \nonumber
\end{eqnarray}
where $A$ stands for SV or SVC, and $G_{gg, N}^{(1) \,\mlbl{A-$N$}}$
is given in Eq.~(\ref{ggSV-N}) or (\ref{ggSVC-N}), respectively.
At the NNLO level, the approximated cross sections $\sigma_{A}$
are defined by the replacements
\begin{eqnarray}
G_{ab,N}&\Longrightarrow &
\as^2(\mu_R^2)\,\frac{\as(\mu_R^2)}{\pi}\, G_{ab,N}^{(1)}
\quad \mbox{for $ab\neq gg$}
 \\
G_{gg,N}
&\Longrightarrow &\as^2(\mu_R^2) \left[1+ 
\frac{\as(\mu_R^2)}{\pi}G_{gg,N}^{(1)}
+\left(\frac{\as(\mu_R^2)}{\pi}\right)^2
\,G_{gg, N}^{(2)\,\mlbl{A-$N$}}\right]\;, \nonumber
\end{eqnarray}
where $G_{gg, N}^{(2)\,\mlbl{A-$N$}}$ is given in Eqs.~(\ref{gg2SV-N}) 
and (\ref{gg2SVC-N}).
Note that at NNLO, $\sigma_{\rm SV}$ and $\sigma_{\rm SVC}$ include
the complete (i.e. without any large-$N$ approximation and including the $qg$ 
and $q{\bar q}$ channels) hard coefficient function
up to NLO.

\begin{figure}[htb]
\begin{center}
\begin{tabular}{c}
\epsfxsize=12truecm
\epsffile{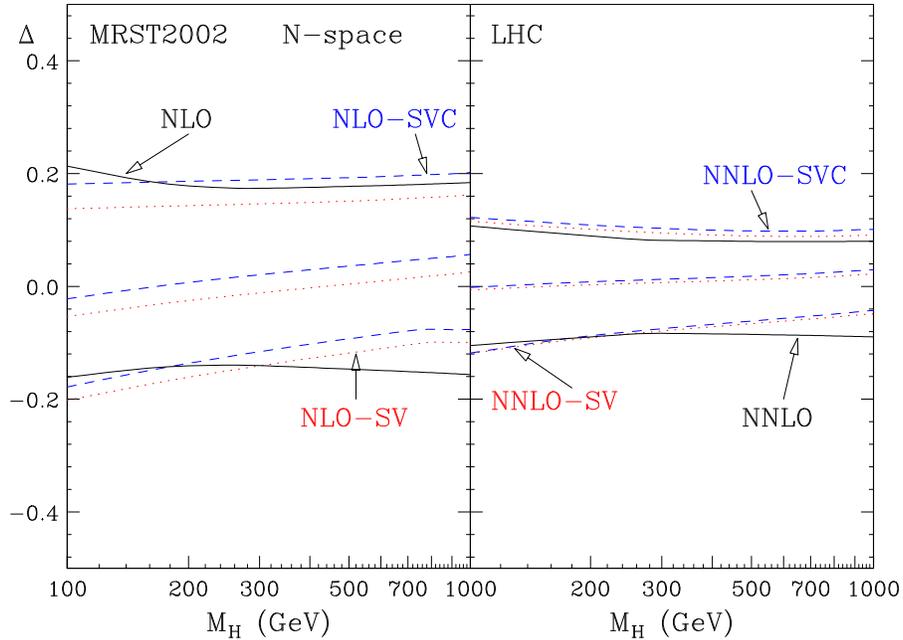}\\
\end{tabular}
\end{center}
\caption{\label{fig:deltaNlhc}
{\em 
The SV (dotted lines) and SVC (dashed lines) approximations in $N$-space
 versus the exact results (solid lines)
 at the LHC (${\sqrt s}=14$~TeV).}}
\end{figure}

\begin{figure}[htb]
\begin{center}
\begin{tabular}{c}
\epsfxsize=12truecm
\epsffile{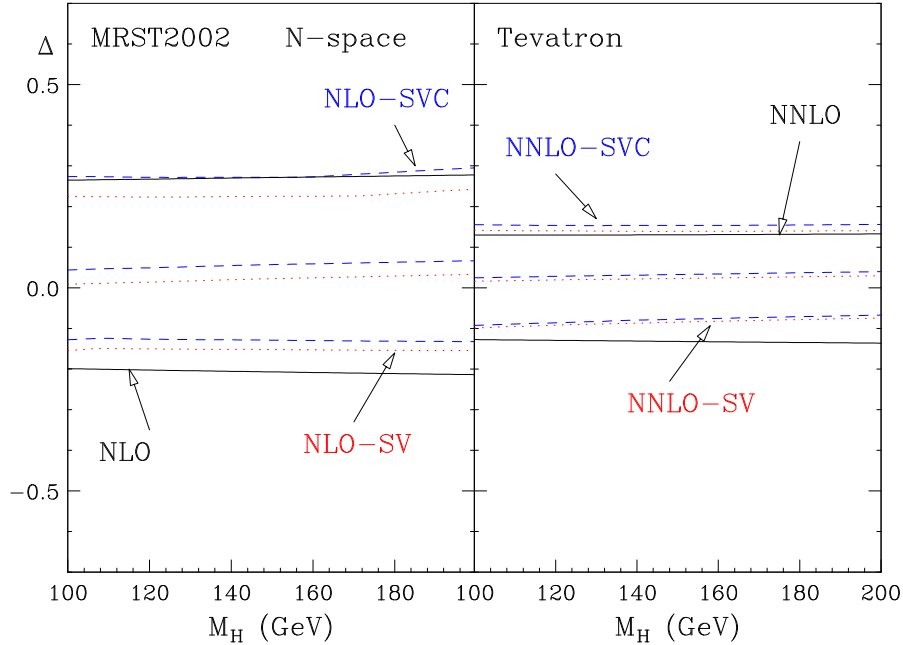}\\
\end{tabular}
\end{center}
\caption{\label{fig:deltaNtev}
{\em 
The SV (dotted lines) and SVC (dashed lines) approximations in $N$-space
 versus the exact results (solid lines)
 at the Tevatron (${\sqrt s}=1.96$~TeV).}}
\end{figure}

The quantity $\Delta$ is
plotted in
Figs.~\ref{fig:deltaNlhc} and \ref{fig:deltaNtev}
for the LHC
and for the Tevatron,
respectively.
The central curves are obtained by fixing $\mu_F=\mu_R=M_H$. 
The bands are obtained by varying $\mu_F$ and $\mu_R$
simultaneously and independently
in the range $0.5M_H\leq \mu_F,\mu_R\leq 2M_H$ with the constraint
$0.5 \leq \mu_F/\mu_R \leq 2$.

Here and in the following, we use the MRST2002 \cite{mrst2002}
set of parton distributions, which includes
(approximated \cite{vnvogt,Moch:2002sn,vermaseren}) NNLO parton distributions.
The parton densities
and QCD coupling are evaluated at each corresponding order,
by using 1-loop $\as$ at LO, 2-loop $\as$ at NLO and 3-loop $\as$ at NNLO.
The corresponding values of $\as(M_Z)$ are
0.130, 0.1197, 0.1154,
at 1-loop, 2-loop and 3-loop  order, respectively.

As can be observed from Figs.~\ref{fig:deltaNlhc} and \ref{fig:deltaNtev}, 
the SV and SVC approximations in $N$-space
agree very well with the exact NLO and NNLO
calculations. Moreover, the differences between the approximated and exact
results are substantially smaller than the effects produced by the scale
variations in the exact results at each fixed order.
A sizeable part of these small differences can be
understood from the fact that in the approximation only the $gg$
channel contribution is included, thus neglecting the {\it
negative} contribution from the $qg$ channel, which incidentally
is also well approximated by its leading collinear behaviour at large $N$.
The very small numerical difference between the SV and SVC approximations
is consistent with the fact that they formally differ by $1/N$
suppressed terms, thus again confirming the consistency of the
approximation  based on the large-$N$ expansion.

The soft approximation,
as well as the SV and SVC approximations,
can also be defined in $x$ space.
In this formulation, the large
logarithm to be considered is $\ln(1-x)$, instead of $\ln N$.
It is easy to go from one formulation to the other by Mellin transform.
For example (see e.g. Eqs.~(\ref{lnsoft})--(\ref{lncol}) and 
Appendix~\ref{appa}), 
one goes from the $x$-space formulae to
the $N$-space ones by performing a Mellin transformation and discarding
all $1/N$ suppressed terms, and all terms that are subleading (in the large-$N$
limit) with
respect to the logarithmic accuracy of the initial $x$-space
formulae. Thus, $x$-space and
$N$-space formulae generally differ by terms that are formally subleading.

It has been shown in Ref.~\cite{Catani:1996yz} that large
subleading terms arise in the $x$-space formulation of the
resummation program. These subleading terms
grow factorially with the order of the perturbative expansion.
Their factorial growth is
not related to renormalons or Landau singularities.
These terms are rather an artefact of the $x$-space approximation,
which mistreats the kinematical constraint of energy conservation,
and
they should not be present in the exact theory.
As a consequence of the factorial growth of these terms, all-order
resummation cannot systematically be defined (implemented) in $x$-space,
since the series of LL, NLL, NNLL, $\dots$ terms are separately divergent.
It is therefore interesting to compare the $x$-space and $N$-space approximations
for the first few exactly known orders.
\begin{figure}[htb]
\begin{center}
\begin{tabular}{c}
\epsfxsize=12truecm
\epsffile{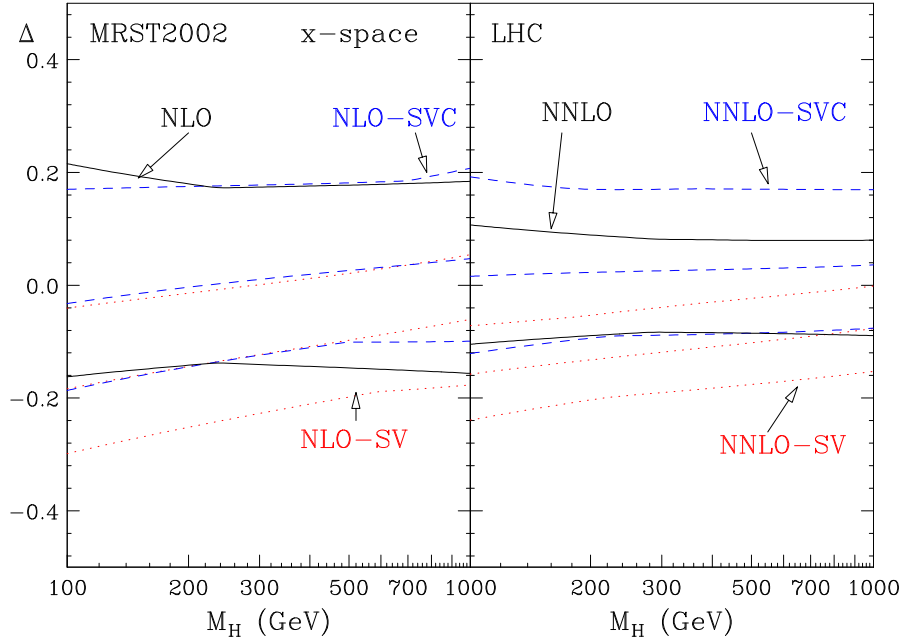}\\
\end{tabular}
\end{center}
\caption{\label{fig:deltalhc}
{\em The SV-$x$ (dotted lines) and SVC-$x$ (dashed lines) approximations
versus the exact results (solid lines)
 at the LHC (${\sqrt s}=14$~TeV).}}
\end{figure}
\begin{figure}[htb]
\begin{center}
\begin{tabular}{c}
\epsfxsize=12truecm
\epsffile{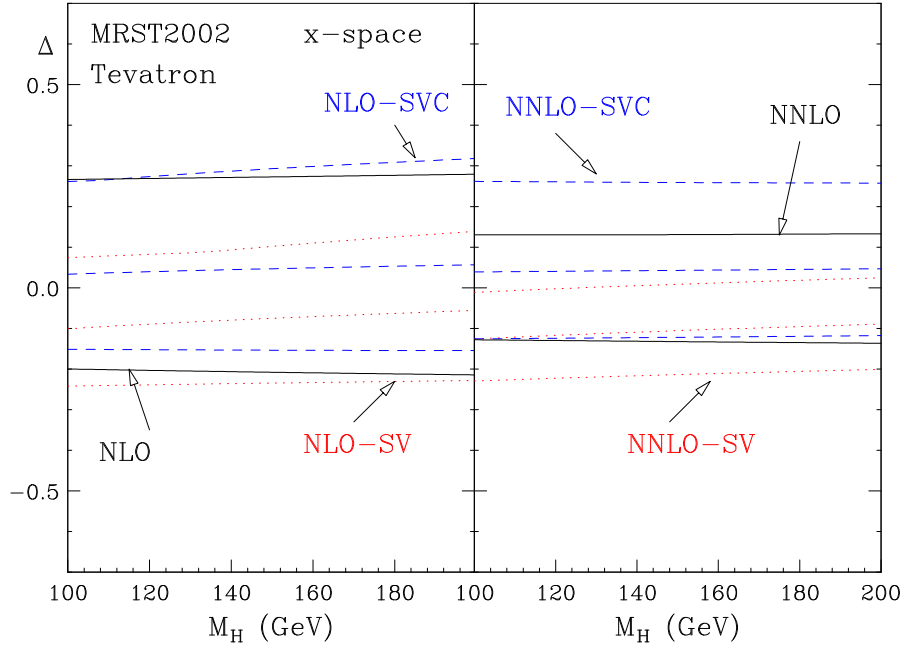}\\
\end{tabular}
\end{center}
\caption{\label{fig:deltatev}
{\em The SV-$x$ (dotted lines) and SVC-$x$ (dashed lines) approximations
versus the exact results (solid lines)
 at the Tevatron (${\sqrt s}=1.96$~TeV).}}
\end{figure}

At NLO, the $x$-space soft-virtual contribution (called SV-$x$ approximation,
here) to the gluon coefficient function $G_{gg}^{(1)}$ in Eq.~(\ref{gg1}) is
\begin{align}
\label{gg1SV}
G_{gg}^{(1) \,\mlbl{SV-$x$}}(z;M_H^2/\mu_R^2;M_H^2/\mu_F^2)&=
 \delta(1-z) \left( \f{11}{2} + 6 \zeta(2) + 
\f{33-2N_f}{6}\ln\f{\mu_R^2}{\mu_F^2}\right) \nn\\
& +\,6\,\dO(z) \, \ln\f{M_H^2}{\mu_F^2} +12\,\dl(z)  \;.
\end{align}
The same approximation can be defined at NNLO, and the corresponding
contribution to the gluon coefficient function $G_{gg}^{(2)}(z)$
has the form
\begin{equation}
\label{gg2SV}
G_{gg}^{(2) \,{\mlbl{SV-$x$}}}(z;M_H^2/\mu_R^2,M_H^2/\mu_F^2)=
 \delta(1-z)\;\delta G^{(2)}+\dO\;G_0^{(2)}
+\dl\;G_1^{(2)}+\dll\;G_2^{(2)}+\dlll\;G_3^{(2)}\;,
\end{equation}
where the coefficients $\delta G^{(2)}, G_0^{(2)},  G_1^{(2)}$ and $G_3^{(2)}$
were computed in Refs.~\cite{Catani:2001ic,Harlander:2001is}
(these coefficients can be found in Eq.~(2.13) of Ref.~\cite{Catani:2001ic}).

After including the {\it leading} collinear logarithmic
contributions, the $x$-space soft-virtual-collinear (SVC-$x$)
approximation is defined \cite{Catani:2001ic} by
\begin{equation}
\label{gg1SVC}
 G_{gg}^{(1) \,{\mlbl{SVC-$x$}}}(z;M_H^2/\mu^2_R;M_H^2/\mu^2_F) =
G_{gg}^{(1) \,{\mlbl{SV-$x$}}}(z;M_H^2/\mu^2_R;M_H^2/\mu^2_F) -12\, 
\ln(1-z) \;\;,
\end{equation}
\begin{equation}
\label{gg2SVC}
G_{gg}^{(2) \,{\mlbl{SVC-$x$}}}(z;M_H^2/\mu^2_R;M_H^2/\mu^2_F) =
G_{gg}^{(2) \,{\mlbl{SV-$x$}}}(z;M_H^2/\mu^2_R;M_H^2/\mu^2_F) - 72\, \ln^3(1-z) \;\;.
\end{equation}

Figures \ref{fig:deltalhc} and \ref{fig:deltatev} 
are obtained in the same way as Figs.~\ref{fig:deltaNlhc} and
\ref{fig:deltaNtev},
except for the use of the SV-$x$ and SVC-$x$ approximations
instead of the analogous $N$-space approximations.
These figures show that the bulk of the fixed-order radiative corrections
is given by the SV-$x$ and SVC-$x$ approximations, as observed in 
Refs.~\cite{Catani:2001ic,Catani:2001cr}.
However, comparing Figs.~\ref{fig:deltalhc} and \ref{fig:deltatev}
with Figs.~\ref{fig:deltaNlhc} and \ref{fig:deltaNtev},
we also see that the $x$-space approximations are
worse than the $N$-space ones. 
Furthermore, the difference between
the SV-$x$ and SVC-$x$ approximations (which is formally suppressed
by a power of $(1-z)$) is considerably larger than the 
difference between the SV-$N$ and SVC-$N$ approximations 
(which is formally suppressed by a power of $1/N$). 

The lower quality of the $x$-space approximations at NLO and NNLO
can have two different
origins. First, the typical value $\langle 1-z \rangle$ of the distance from the
partonic threshold, which is the parameter that
formally controls the large-$x$ expansion,
can be quantitatively larger than the typical value $\langle 1/N \rangle$
of the analogous expansion parameter in the $N$-space approach. Second,
the numerical coefficients in the $x$-space expansion formulae 
can be larger than those in the $N$-space expansion formulae, 
as is the case at very high perturbative orders, because of the 
presence of factorially-growing subleading terms in the $x$-space approach.
Of course, given the few orders available in the perturbative
expansion, it is very difficult to explicitly check for the presence
or absence of these factorial terms.
We
conclude, however, that these results
provide no justification
for estimating
terms of still higher order
through the soft-gluon approximation in the $x$-space approach.

\subsection{Numerical relevance of the resummation}
\label{sec:benchmarks}

Before moving to the phenomenological results for soft-gluon
resummation in Higgs production at the LHC and
at the Tevatron, it is interesting
to study the main features of the resummed coefficient function.
This is better done in $N$-space, since both the resummed and fixed-order
coefficient functions in $x$-space are distributions, and their numerical
impact is therefore more difficult
to assess.

\begin{figure}[htb]
\begin{center}
\begin{tabular}{c}
\epsfxsize=14truecm
\epsffile{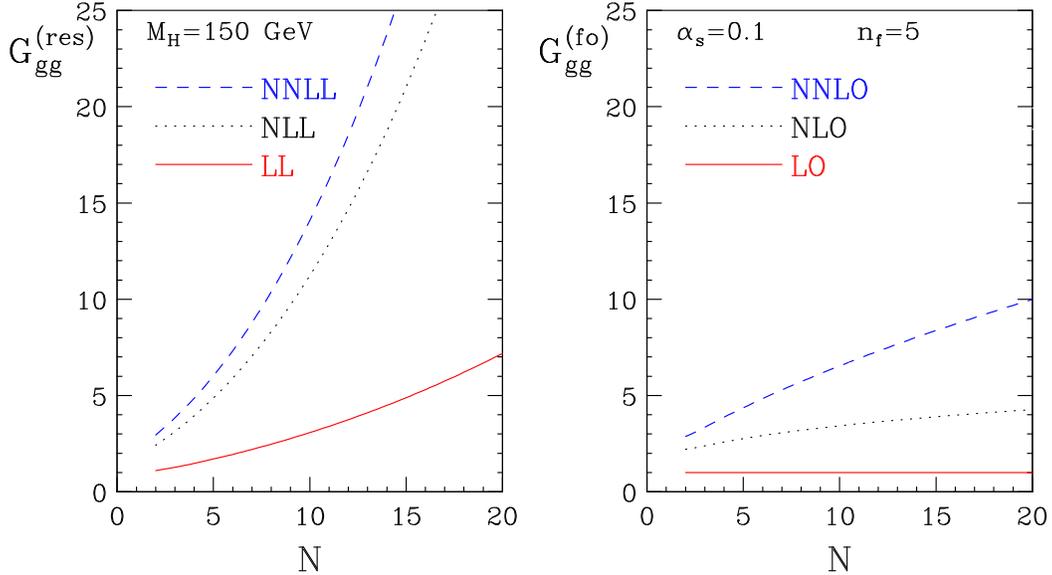}
\end{tabular}
\end{center}
\caption{\label{fig:sudakov}{\em $N$-dependence of resummed (left-hand side) 
and fixed-order (right-hand side) gluon coefficient functions 
for Higgs production 
($M_H=150$~GeV) with fixed 
coupling constant $\as=0.1$. }}
\end{figure}

To directly observe the
effect of the exponentiation, we fix the coupling constant at
$\as=0.1$, the Higgs mass at $M_H=150$~GeV, and the scales at $\mu_F=\mu_R=M_H$.
Figure~\ref{fig:sudakov} shows the $N$-moments, $G_{gg, \,N}$, of the 
resummed (left-hand side) and fixed-order (right-hand side) gluon coefficient
function. The fixed-order coefficient function (see Eq.~(\ref{expansion}))
is evaluated at LO, NLO and NNLO. The resummed coefficient function
(see Eq.~(\ref{resformula})) is evaluated at LL (i.e. including the function
$g_H^{(1)}$), NLL (i.e. including also the function $g_H^{(2)}$ and the coefficient
$C_{gg}^{(1)}$) 
and NNLL (i.e. including also the function $g_H^{(3)}$ and the coefficient
$C_{gg}^{(2)}$) order.
At large values of $N$,
there is a noticeable improvement in the convergence of the
perturbative expansion once the resummation is performed. Indeed, the 
ratio
between NNLL and NLL results is considerably smaller than
the one between NLL and LL results. On the contrary, fixed-order results
show an increasing ratio between two successive orders, which is
due to the appearance of new and 
large $\ln^{(i)}N$ terms in the fixed-order expansion.
Note, however, that we do not anticipate very large resummation effects on Higgs
boson production at the LHC and the Tevatron. In fact, we know 
\cite{Catani:2001ic}--\cite{Ravindran:2003um} that the
ratio between the NNLO and LO cross sections does not exceed a factor of about 3
at these hadron colliders.
From inspection of the right-hand side of 
Fig.~\ref{fig:sudakov}, we thus expect that the Higgs boson cross section
at LHC and Tevatron energies is
mainly sensitive to the gluon coefficient function at
moderate values of $N$.

\section{Phenomenological results}
\label{sec:predictions}

\subsection{Resummed cross section}
\label{sec:resxs}

We use soft-gluon resummation in $N$-space at the parton level 
(i.e. at the level of the
partonic coefficient function $G_{ab}$)
to introduce an improved (resummed) hadronic cross section
$\sigma^{\rm (res)}(s,M_H^2)$, which is obtained by inverse Mellin
transformation (see Eq.~(\ref{invmt})) as follows:
\begin{align}
\label{hadnres}
\sigma^{\rm (res)}(s,M_H^2) &= \sigma^{(0)} 
\;\int_{C_{MP}-i\infty}^{C_{MP}+i\infty}
\;\frac{dN}{2\pi i} \;\left( \frac{M_H^2}{s} \right)^{-N+1}
\; f_{g/h_1, \, N}(\mu_F^2) \; f_{g/h_2\, N}(\mu_F^2) \nn \\
&\times \left[ \;
{G}_{gg,\, N}^{\rm (res)}(\as(\mu_R^2), M_H^2/\mu_R^2;M_H^2/\mu_F^2)
- \left( 
{G}_{gg,\, N}^{\rm (res)}(\as(\mu_R^2), M_H^2/\mu_R^2;M_H^2/\mu_F^2)
\right)_{\rm (f.o.)} \, \right] \nn \\
&+ \sigma^{\rm (f.o.)}(s,M_H^2) \;\;,
\end{align}
where $\sigma^{\rm (f.o.)}(s,M_H^2)$ is the Higgs boson hadronic cross 
section at a given fixed order (f.o.~=~LO, NLO, NNLO),
${G}_{gg,\, N}^{\rm (res)}$ is given in Eq.~(\ref{resformula}),
and 
$\left( {G}_{gg,\, N}^{\rm (res)} \right)_{\rm (f.o.)}$
represents its perturbative truncation at the same fixed order in 
$\as(\mu_R^2)$.
Thus, because of the subtraction
in the square bracket on the right-hand side, Eq.~(\ref{hadnres}) exactly
reproduces the fixed-order results and resums soft-gluon effects beyond
those fixed orders up to a certain logarithmic accuracy.

\setcounter{footnote}{0}
In the following subsections, we present numerical results for the resummed
cross section $\sigma^{\rm (res)}(s,M_H^2)$
at LL, NLL and NNLL accuracy. The resummed coefficient function
${G}_{gg,\, N}^{\rm (res)}$ in Eq.~(\ref{hadnres}) is evaluated from the
expressions  in Eqs.~(\ref{resformula})--(\ref{gexpan}):
at LL accuracy we include the 
function $g_H^{(1)}$; at NLL accuracy we include also the function $g_H^{(2)}$ 
and the coefficient $C_{gg}^{(1)}$; at NNLL accuracy 
we include also $g_H^{(3)}$ and $C_{gg}^{(2)}$.
Although they are briefly denoted as N$^k$LL $(k=0,1,2)$, 
the resummed results are always matched to the corresponding fixed order 
(f.o.~=~N$^k$LO) according to Eq.~(\ref{hadnres}),
i.e. LL is matched to LO, NLL to NLO and NNLL to NNLO. 

Unless otherwise stated, cross sections are computed using 
sets of parton distributions, with densities and QCD coupling 
evaluated at each corresponding order, by using 1-loop $\as$ at LO
(LL), 2-loop $\as$ at NLO (NLL), and 3-loop $\as$ at NNLO (NNLL).

We recall that the hard coefficient function $G_{ab}$ is evaluated in the
large-$M_t$ approximation, 
whereas
the exact dependence on the masses $M_t$ and $M_b$ of the top and bottom quark
is included in the Born-level cross 
section $\sigma^{(0)}$. We use $M_t=176$~GeV and $M_b=4.75$~GeV.

The inverse Mellin transformation in Eq.~(\ref{hadnres}) involves an integral
in the complex $N$ plane.
When the $N$-moments ${G}_{gg,\, N}$ 
are evaluated at a fixed perturbative order,
they are analytic functions in a
right half-plane of the complex variable $N$. In this case, the constant 
$C_{MP}$
that defines the integration contour has to be chosen in
this half-plane, that is, on the right of all the possible singularities of 
the $N$-moments.

When ${G}_{gg,\, N}$ is evaluated in resummed
perturbation theory, the resummed functions $g_H^{(n)}(\lambda)$
in Eqs.~(\ref{g1fun})--(\ref{g3fun}) are singular at the point $\lambda=1/2$,
which corresponds to $N=N_L=\exp(1/2b_0\as(\mu_R^2))$ 
(i.e. $N_L \sim M_H/\Lambda_{QCD}$).
These singularities, which are related to the divergent behaviour of the 
perturbative running coupling  $\as$ near the Landau pole, signal the onset of
non-perturbative phenomena at very large values of $N$ or, equivalently,
in the region very close to threshold. We deal with these singularities
by using the {\em Minimal Prescription} introduced in  
Ref.~\cite{Catani:1996yz}. 
In the evaluation of the inverse Mellin 
transformation in Eq.~(\ref{hadnres}), 
the constant $C_{MP}$ is chosen in such a way that all
singularities in the integrand are to the left of the integration contour,
except for the Landau singularity at $N=N_L$, that should lie to the far right.
The results obtained by using this prescription converge asymptotically to the
perturbative series\footnote{An explicit check of the numerical
convergence is presented in Appendix~\ref{appd}.} 
and do not introduce (unjustified) 
power corrections\footnote{The only remaining asymptotic ambiguity
is more suppressed than any power law \cite{Catani:1996yz}.}
of non-perturbative origin.
These corrections are certainly present in physical
cross sections, but their effect is not expected to be sizeable,
as long as $M_H$ is sufficiently perturbative and $\tau_H=M_H^2/s$ is
sufficiently far from the hadronic threshold, as is the case in Higgs boson
production at the Tevatron and the LHC.

The resummed cross section in Eq.~(\ref{hadnres}) can 
equivalently be rewritten as
\begin{equation}
\sigma^{\rm (res)}(s,M_H^2) = \sigma^{\rm (SV)}(s,M_H^2)
+ \sigma^{\rm (match.)}(s,M_H^2) \;\;,
\end{equation}
where $\sigma^{\rm (SV)}$ denotes the contribution obtained by Mellin inversion
of ${G}_{gg,\, N}^{\rm (res)}$, while the matching contribution 
$\sigma^{\rm (match.)}$ denotes the fixed-order cross section minus
the corresponding fixed-order truncation of the soft-gluon resummed terms.
As can easily be argued from the numerical results in Sect.~\ref{sec:fonsp},
$\sigma^{\rm (SV)}$ gives the bulk of the QCD radiative corrections to the
Higgs boson cross section at the Tevatron and the LHC. 
The order of magnitude of the relative contribution 
from $\sigma^{\rm (match.)}$ can be estimated from the size of the ratio
$\Delta$ (see the definition in Eq.~(\ref{eq:delta})) 
in Figs.~\ref{fig:deltaNlhc} and \ref{fig:deltaNtev}: it is of ${\cal O}(10\%)$
and of ${\cal O}(1\%)$ at NLO and NNLO, respectively. Thus, the fixed-order
cross section $\sigma^{\rm (match.)}$ quantitatively behaves as naively expected
from a power series expansion whose expansion parameter is $\as \sim 0.1$.
We expect that the currently unknown (beyond NNLO) corrections to  
$\sigma^{\rm (match.)}$ have no practical quantitative impact on
the QCD predictions for Higgs boson production at the Tevatron and the LHC.

We note that the predictions we are going to present regard the production of an
on-shell Higgs boson. Therefore they are directly applicable at low values of $M_H$,
where the small-width approximation is valid.
At high values of $M_H$, corrections due to finite-width effects have to be implemented.

\subsection{LHC}

\begin{figure}[htb]
\begin{center}
\begin{tabular}{c}
\epsfxsize=15truecm
\epsffile{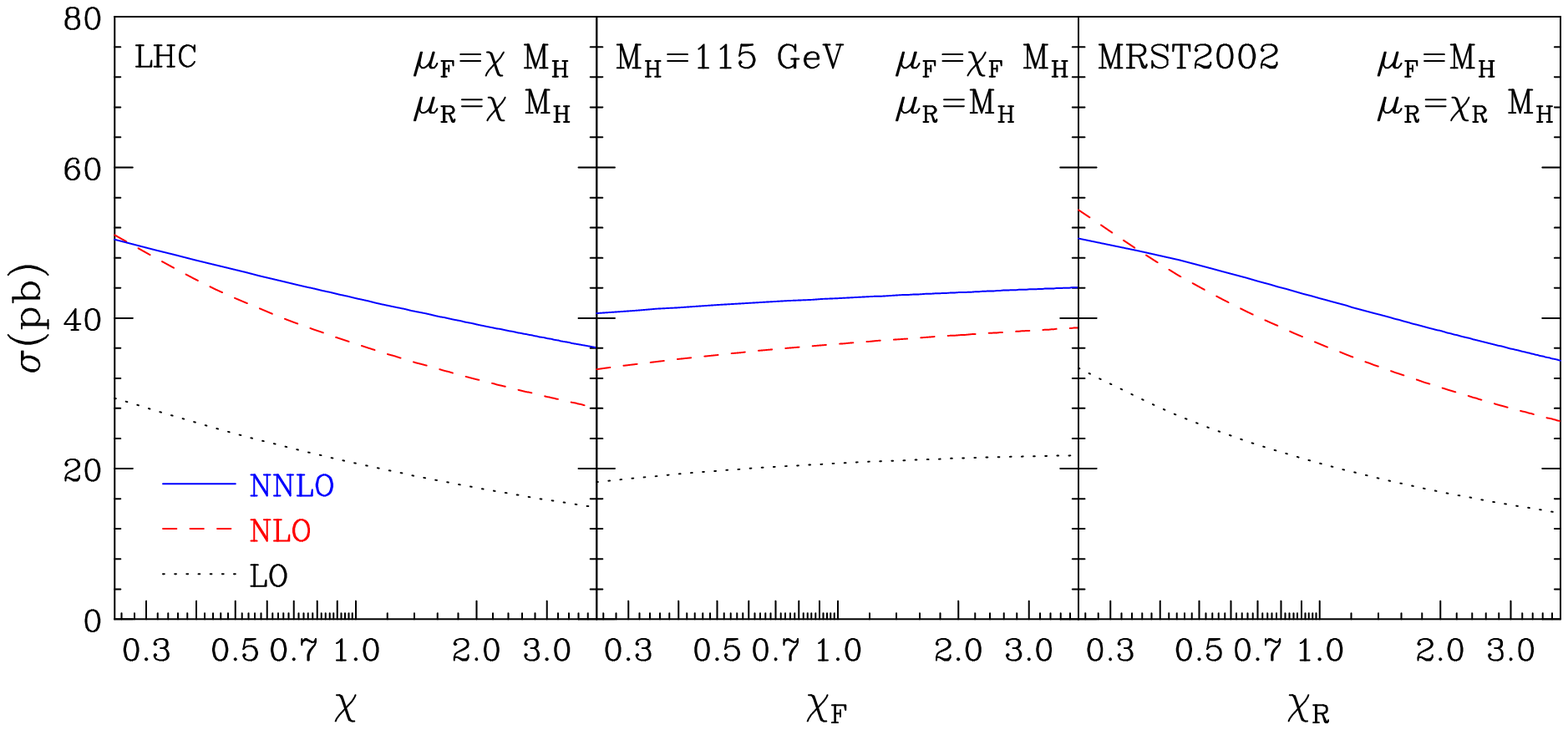}\\
\epsfxsize=15truecm
\epsffile{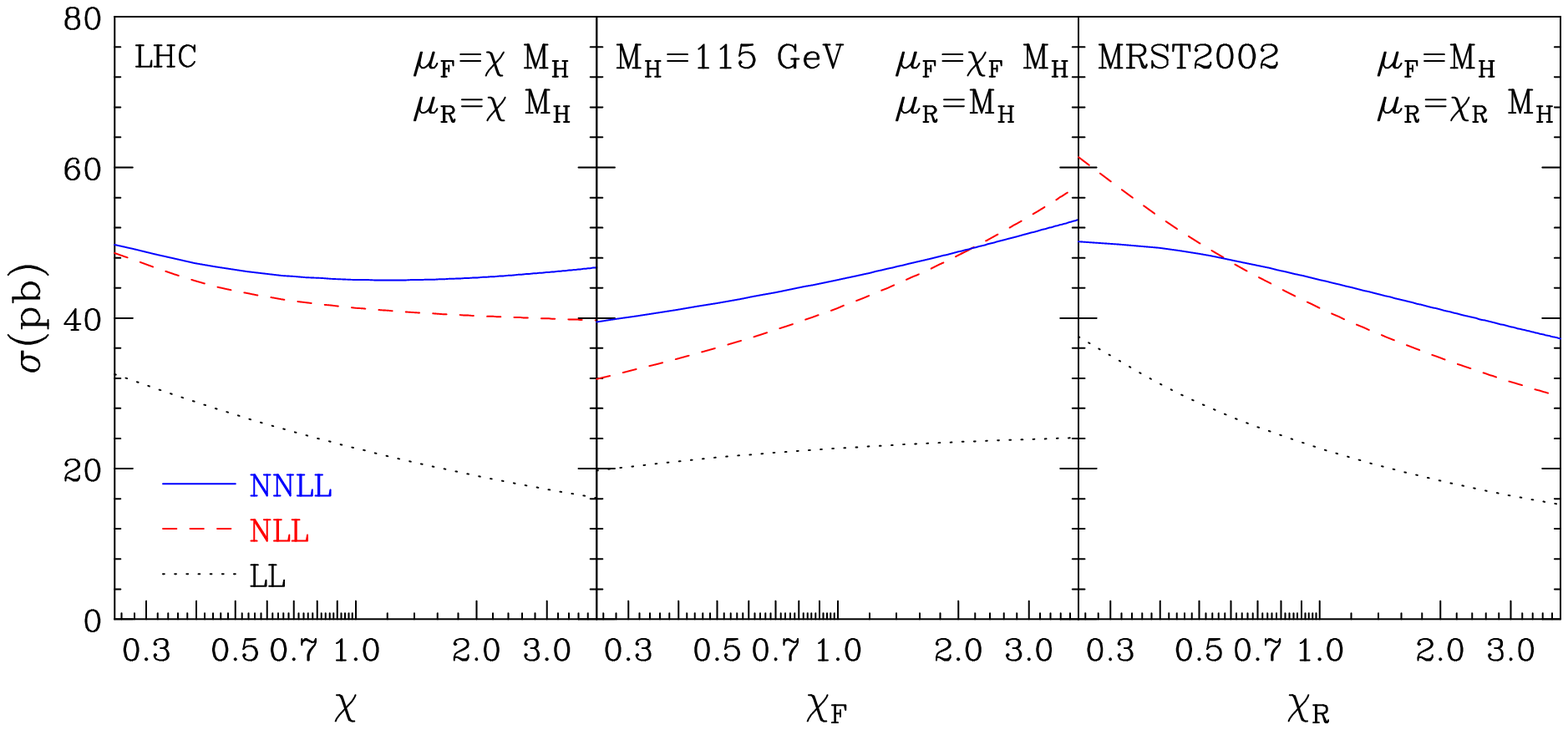}
\end{tabular}
\end{center}
\caption{\label{fig:lhcscale}{\em Scale dependence of the Higgs production
cross section at the LHC for $M_H=115$ GeV at a) (upper) LO, NLO, NNLO 
and b) (lower) LL, NLL, NNLL accuracy.}}
\end{figure}

In this subsection we study the phenomenological impact of soft-gluon
resummation on the production of the SM Higgs boson at the LHC.  
The cross sections are computed using the MRST2002
set of parton distributions \cite{mrst2002}, with densities and QCD coupling
evaluated at each corresponding order,  as stated in Sect.~\ref{sec:resxs}
and done in Sect.~\ref{sec:fonsp}.

We begin the presentation of our results by showing in Fig.~\ref{fig:lhcscale}
the scale dependence of the cross section for the production of a Higgs boson
with $M_H=115$~GeV. The scale dependence is analysed by varying the
factorization and renormalization scales around
the default value $M_H$.  The plot on the left corresponds to the simultaneous
variation of both scales, $\mu_F=\mu_R=\chi \, M_H$, whereas the plot in the
centre (right) corresponds to the
variation of the factorization (renormalization) scale 
$\mu_F=\chi_F \, M_H$ ($\mu_R=\chi_R \, M_H$) by fixing the other scale
at the default value $M_H$.

As expected from the QCD running of $\as$, the cross sections typically
decrease when $\mu_R$ increases around the characteristic hard scale $M_H$,
at fixed $\mu_F=M_H$.
In the case of variations of $\mu_F$ at fixed $\mu_R=M_H$, we observe the opposite behaviour. In
fact, when $M_H=115$ GeV, the cross sections are mainly sensitive to partons
with momentum fraction $x \sim 10^{-2}$, and in this $x$-range scaling
violation of the parton densities is (moderately) positive. 
Varying the two scales simultaneously ($\mu_F=\mu_R$) leads to a compensation
of the two different behaviours.
As a result, the
scale dependence is mostly driven by the renormalization scale, because the
lowest-order contribution to the process is proportional to $\as^2$, a
(relatively) high power of $\as$.

\begin{figure}[htb]
\begin{center}
\begin{tabular}{c}
\epsfxsize=15truecm
\epsffile{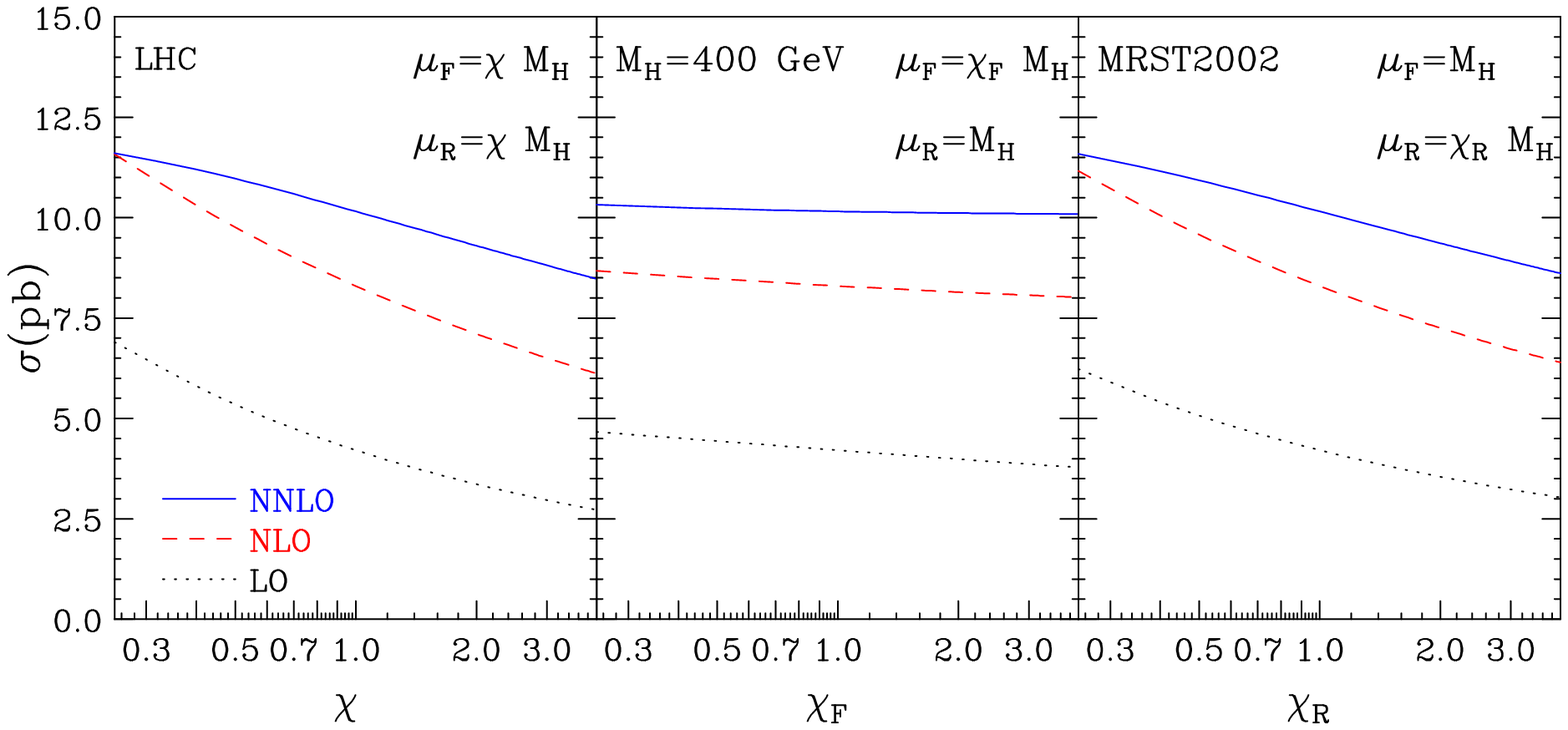}\\
\epsfxsize=15truecm
\epsffile{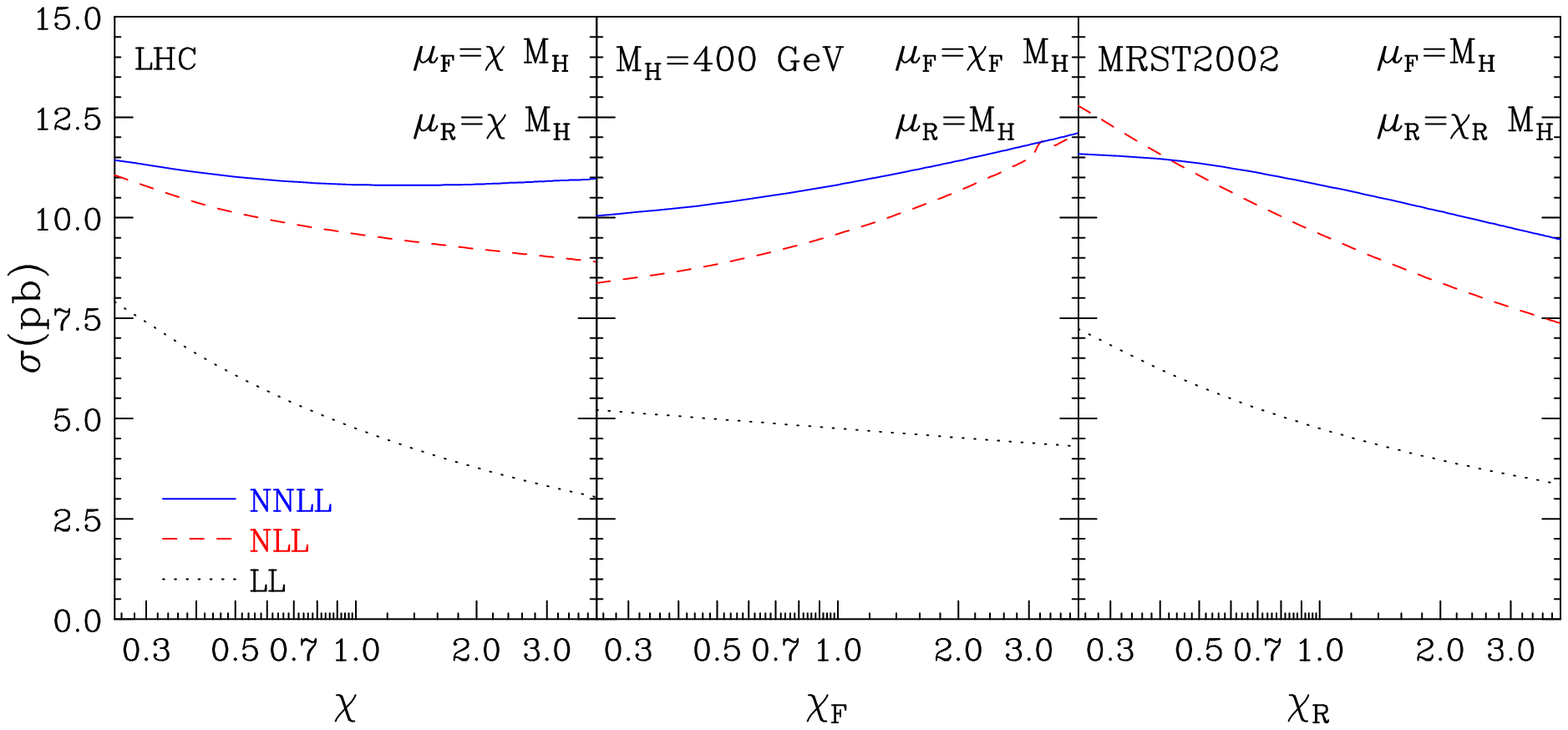}
\end{tabular}
\end{center}
\caption{\label{fig:lhc400scale}{\em Scale dependence of the Higgs production 
cross section at the LHC for $M_H=400$ GeV at a) (upper) LO, NLO, NNLO 
and b) (lower) LL, NLL, NNLL accuracy.}}
\end{figure}

Figure~\ref{fig:lhcscale}a shows that
the scale dependence is reduced when higher-order corrections
are included.
When resummation effects are implemented (Fig.~\ref{fig:lhcscale}b),
we typically observe a further (slight) reduction of the scale dependence,
with the exception of the factorization-scale dependence at fixed $\mu_R=M_H$
that is marginally stronger after resummation. This 
suggests that
the rather flat dependence on $\mu_F$ at NNLO can be an accidental effect,
as also suggested by the fact that the $\mu_F$ dependence is much weaker
than the $\mu_R$ dependence at each fixed order (LO, NLO, NNLO).

In Fig.~\ref{fig:lhc400scale},
analogous results are plotted at a higher value, $M_H=400$~GeV, of the Higgs
boson mass.
The overall features of Figs.~\ref{fig:lhcscale} and \ref{fig:lhc400scale}
are similar, although we notice that the 
improvement in the scale dependence when higher-order contributions 
are included is slightly better in Fig.~\ref{fig:lhc400scale} than in 
Fig.~\ref{fig:lhcscale}.
An interesting difference between these two figures regards
the $\mu_F$ dependence at fixed $\mu_R$.
The LO, NLO, NNLO and LL results in Fig.~\ref{fig:lhc400scale} show that the
corresponding cross sections (very) slightly decrease as $\mu_F$ 
increases around $M_H$. This is because, when increasing $M_H$ from 
$M_H=115$~GeV to $M_H=400$~GeV, the cross section is sensitive to partons 
with higher values of the momentum fraction $x$, so that scaling
violation of the parton densities can become slightly negative.
The fact that the parton densities are evaluated in an $x$-range where scaling
violation changes sign is also suggested by the change in the slope of the 
$\mu_F$ dependence when going from LL to NLL and NNLL order.

\begin{figure}[htb]
\begin{center}
\begin{tabular}{c}
\epsfxsize=12truecm
\epsffile{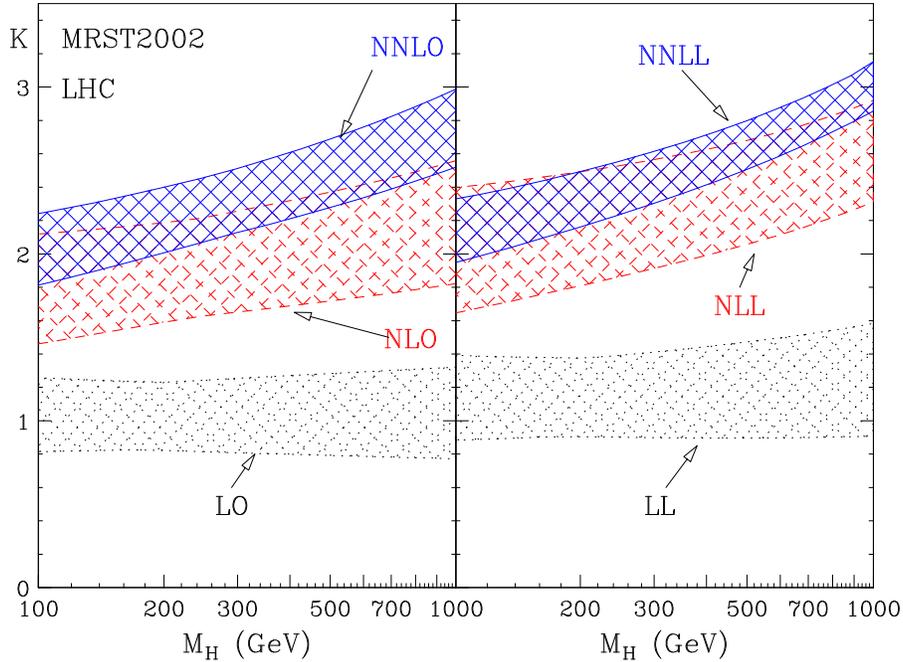}\\
\end{tabular}
\end{center}
\caption{\label{fig:lhcbands}{\em  Fixed-order and resummed K-factors for Higgs production at the LHC.}}
\end{figure}

The impact of higher-order corrections is sometimes presented through the
$K$-factors, defined as the ratio of the cross section evaluated at each
corresponding order over the LO result.
The $K$-factors are shown in
Fig.~\ref{fig:lhcbands}, where the bands are obtained,
as in Sect.~\ref{sec:fonsp}, 
by varying the scales $\mu_R$ and $\mu_F$
(simultaneously and independently) 
in the range $0.5 M_H\le \mu_F,\mu_R \le 2M_H$, with the constraint
$0.5\le \mu_F/\mu_R\le 2$.  
The LO result that normalizes the $K$-factors is computed at the default scale $M_H$ in all
cases.
We see that the effect of the higher-order corrections 
increases with $M_H$.
We also see that the soft-gluon resummation effects are more 
important at higher values of $M_H$. This is expected, since by increasing $M_H$
we are closer to the hadronic threshold, where soft-gluon effects are larger.
When $M_H$ increases, the scale dependence
after resummation is smaller than at the corresponding fixed orders.
In the case of a light Higgs boson ($M_H \ltap 200$~GeV), the NNLO K-factor is about 2.1--2.2,
which corresponds to an increase of about $20\%$ with respect to the NLO K-factor.
In this low-mass range,  the
effects of resummation are also moderate: at NNLL accuracy the central value of the cross section increases
by about $6\%$ with respect to NNLO.

\begin{figure}[htb]
\begin{center}
\begin{tabular}{c}
\epsfxsize=12truecm
\epsffile{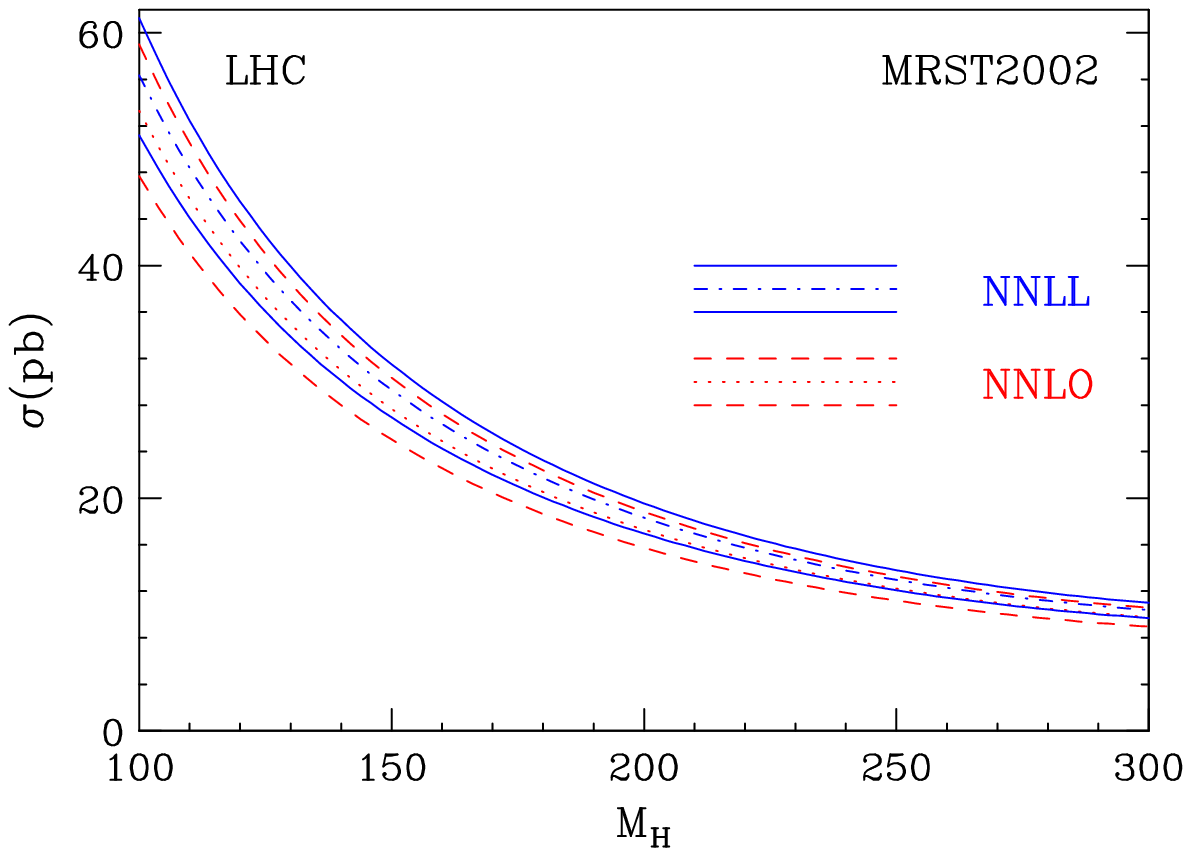}\\
\end{tabular}
\end{center}
\caption{\label{fig:xslhc}{\em NNLL and NNLO cross sections at the LHC,
using MRST2002 parton densities.}}
\end{figure}

In Fig.~\ref{fig:xslhc} we plot the
NNLO and NNLL cross sections, with the corresponding scale-dependence
bands (computed as in Fig.~\ref{fig:lhcbands}),
in the range $M_H=$100--300~GeV. 
The corresponding numerical results are given in Table~\ref{tab:lhc}, where
$\sigma_{\rm min}$, $\sigma_{\rm max}$ and $\sigma_{\rm ref}$ correspond to the minimum,
maximum and central values in the bands.

{\renewcommand{\arraystretch}{1.8} 
\begin{table}
\begin{center}
\begin{tabular}{c|ccc|ccc}
\hline
$M_H$ & $\sigma^{NNLO}_{\rm min}$ & $\sigma^{NNLO}_{\rm ref}$ & $\sigma^{NNLO}_{\rm max}$ & $\sigma^{NNLL}_{\rm min}$ & $\sigma^{NNLL}_{\rm ref}$ & $\sigma^{NNLL}_{\rm max}$\\
\hline
100 & 47.71 & 53.30 & 59.02  & 51.22 & 56.36 & 61.28\\

110 & 41.08 & 45.77 & 50.55  & 44.12 & 48.41 & 52.47\\ 

120 & 35.81 & 39.80 & 43.85  & 38.46 & 42.10 & 45.52\\  

130 & 31.52 & 34.96 & 38.45  & 33.87 & 37.00 & 39.92\\

140 & 27.99 & 30.99 & 34.02  & 30.09 & 32.80 & 35.32\\

150 & 25.05 & 27.69 & 30.34 & 26.94 & 29.31 & 31.51\\

160 & 22.56 & 24.90 & 27.25 & 24.28 & 26.37 & 28.31\\

170 & 20.45 & 22.54 & 24.64 & 22.02 & 23.88 & 25.59\\

180 & 18.65 & 20.52 & 22.40 & 20.08 & 21.74 & 23.27\\

190 & 17.09 & 18.79 & 20.48 & 18.39 & 19.90 & 21.27\\

200 & 15.74 & 17.28 & 18.83 & 16.96 & 18.32 & 19.55\\

210 & 14.57 & 15.98 & 17.39 & 15.70 & 16.94 & 18.07\\

220 & 13.55 & 14.85 & 16.14 & 14.60 & 15.74 & 16.78\\

230 & 12.65 & 13.86 & 15.05 & 13.65 & 14.70 & 15.65\\

240 & 11.87 & 12.99 & 14.10 & 12.81 & 13.78 & 14.67\\

250 & 11.19 & 12.24 & 13.28 & 12.08 & 12.99 & 13.81\\

260 & 10.60 & 11.58 & 12.55 & 11.45 & 12.30 & 13.06\\

270 & 10.09 & 11.01 & 11.93 & 10.90 & 11.70 & 12.42\\

280 & 9.648 & 10.53 & 11.39 & 10.42 & 11.18 & 11.86\\

290 & 9.270 & 10.11 & 10.94 & 10.02 & 10.74 & 11.39\\

300 & 8.960 & 9.773 & 10.57 & 9.696 & 10.38 & 11.00\\

\hline
\end{tabular}                                                                 
\ccaption{}{\label{tab:lhc}{\em NNLO and NNLL cross sections (in pb) at the LHC, 
using MRST2002 parton densities.}}
\end{center}
\end{table}}

\subsection{Tevatron}

\begin{figure}[htb]
\begin{center}
\begin{tabular}{c}
\epsfxsize=15truecm
\epsffile{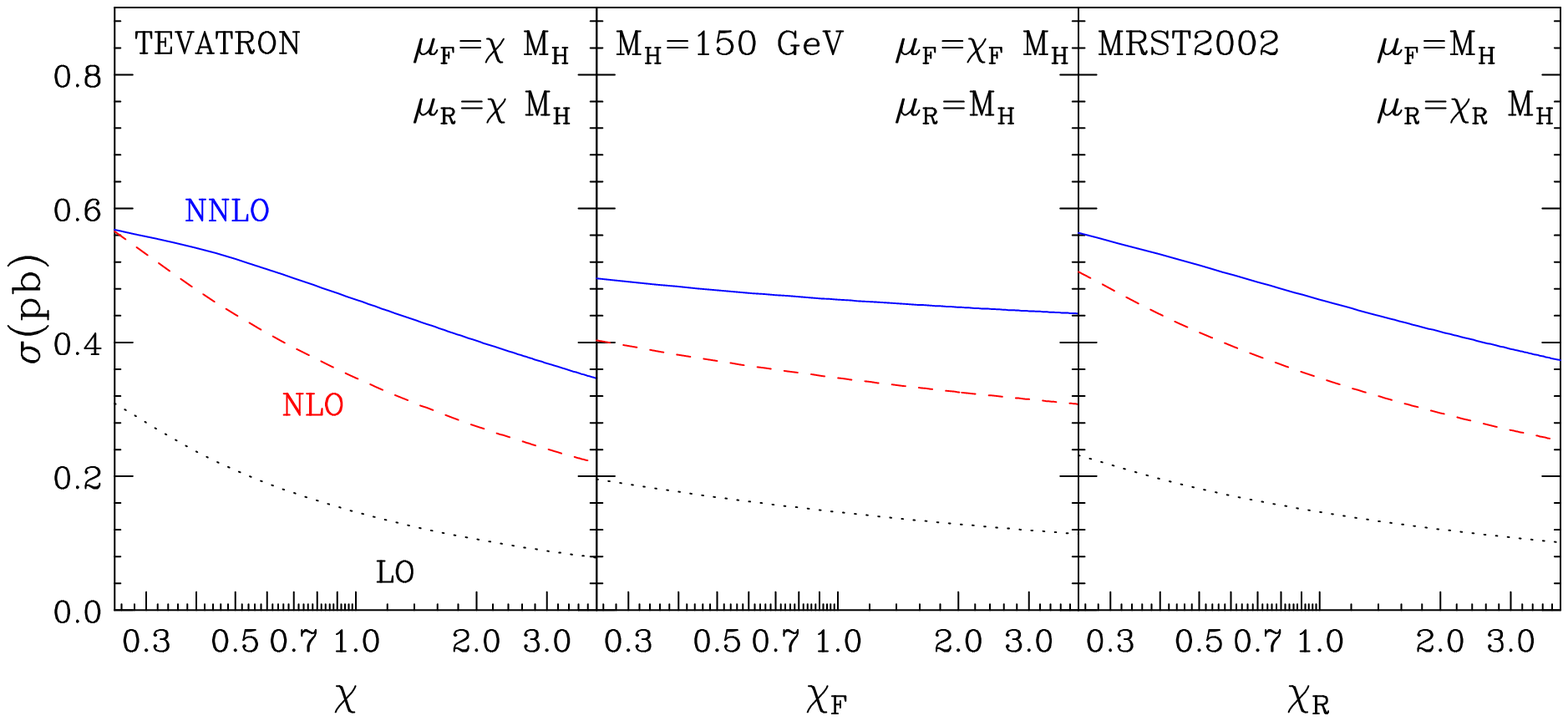}\\
\epsfxsize=15truecm
\epsffile{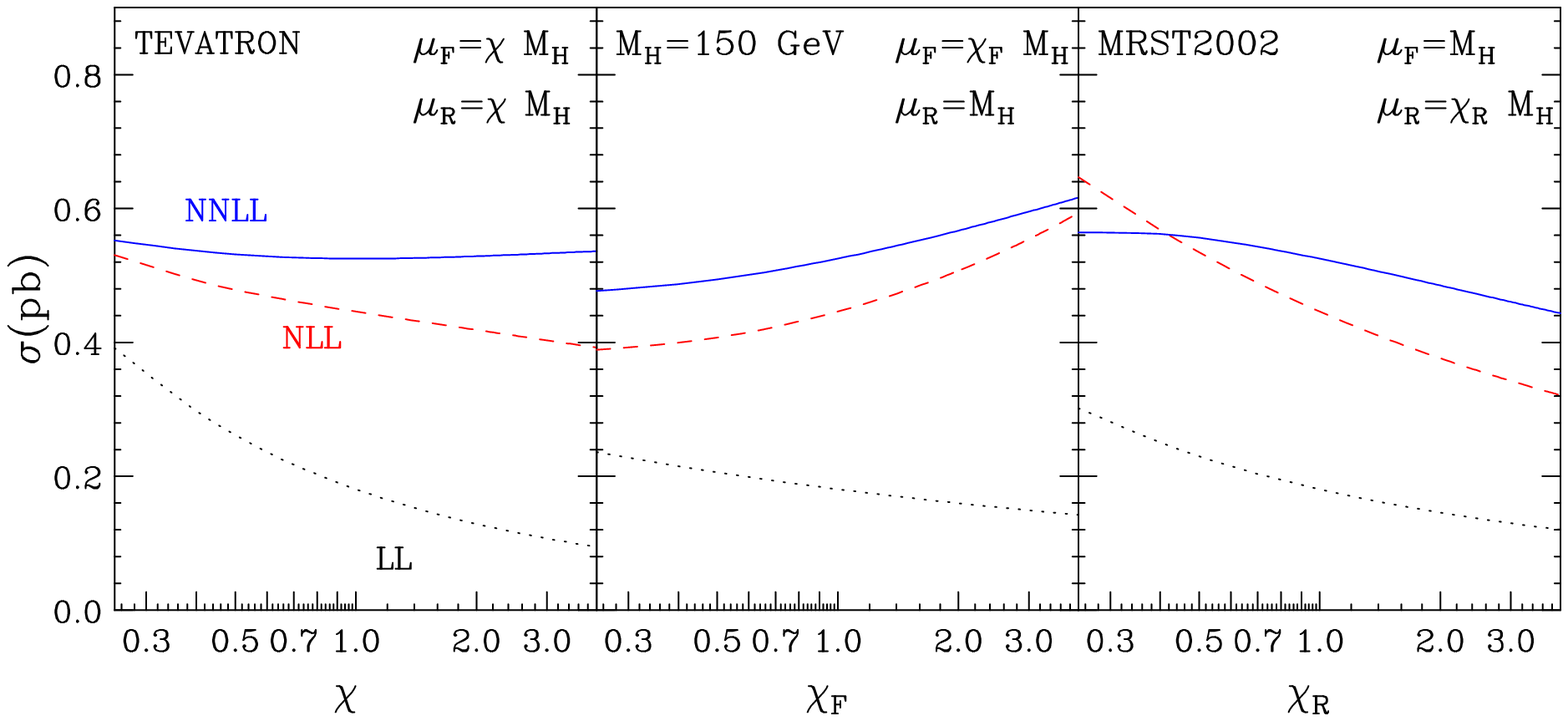}
\end{tabular}
\end{center}
\caption{\label{fig:tevscale}{\em Scale dependence of the Higgs production
cross section at the Tevatron for $M_H=150$ GeV at
a) (upper) LO, NLO, NNLO 
and b) (lower) LL, NLL, NNLL accuracy.}}
\end{figure}

Here we study the phenomenological
impact of soft-gluon resummation on the production
of the SM Higgs boson at the Tevatron Run~II.

As in the previous subsection, we show in Fig.~\ref{fig:tevscale} the scale
dependence of the fixed-order and resummed results. We use $M_H=150$~GeV. As in the LHC case, the cross
sections typically decrease when $\mu_R$ increases around the characteristic
hard scale $M_H$.  Figure~\ref{fig:tevscale}a shows that the fixed-order cross
section decreases when $\mu_F$ increases at fixed $\mu_R$.  
This is not unexpected: at the Tevatron,
the cross section is mainly sensitive to partons with $x\sim 0.05$--0.1, where
the scaling violation is slightly negative.  As in the LHC case, the $\mu_F$
dependence of the resummed results appears to be stronger 
than the $\mu_F$ dependence of the fixed-order results. The slope of the 
$\mu_F$ dependence changes sign in going from LL to  
NLL and NNLL order, as in the case of Fig.~\ref{fig:lhc400scale}.

\begin{figure}[htb]
\begin{center}
\begin{tabular}{c}
\epsfxsize=12truecm
\epsffile{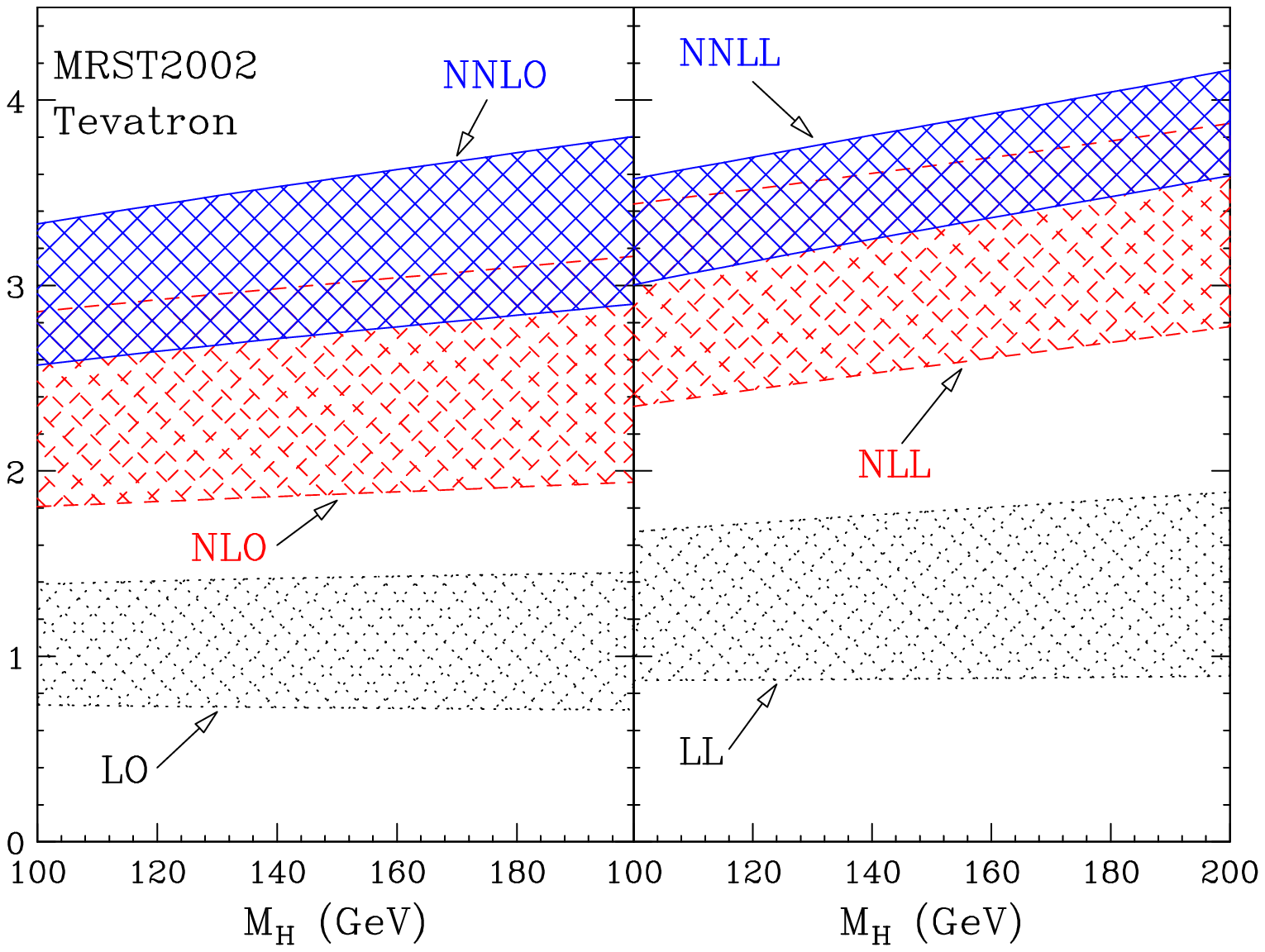}\\
\end{tabular}
\end{center}
\caption{\label{fig:tevbands}{\em Fixed order and resummed K-factors for Higgs
production at the Tevatron Run II ($\sqrt{s}=1.96$ TeV).}}
\end{figure}

In Fig.~\ref{fig:tevbands} we plot the K-factor bands defined as in
Fig.~\ref{fig:lhcbands}. We see  that the impact of
higher-order corrections at fixed $M_H$ is larger at the Tevatron than at the
LHC. This is not unexpected: at the Tevatron, the Higgs boson is 
produced closer to the hadronic threshold and
soft-gluon effects are therefore more sizeable.
The NNLO K-factor is about 3--3.2, with
an increase of about $40\%$ with respect to NLO. Correspondingly, also the
resummation effects are larger: at NNLL they increase the 
NNLO 
cross section by
about 12--15$\%$.
The scale dependence of the NNLL result is smaller than at NNLO.

\begin{figure}[htb]
\begin{center}
\begin{tabular}{c}
\epsfxsize=12truecm
\epsffile{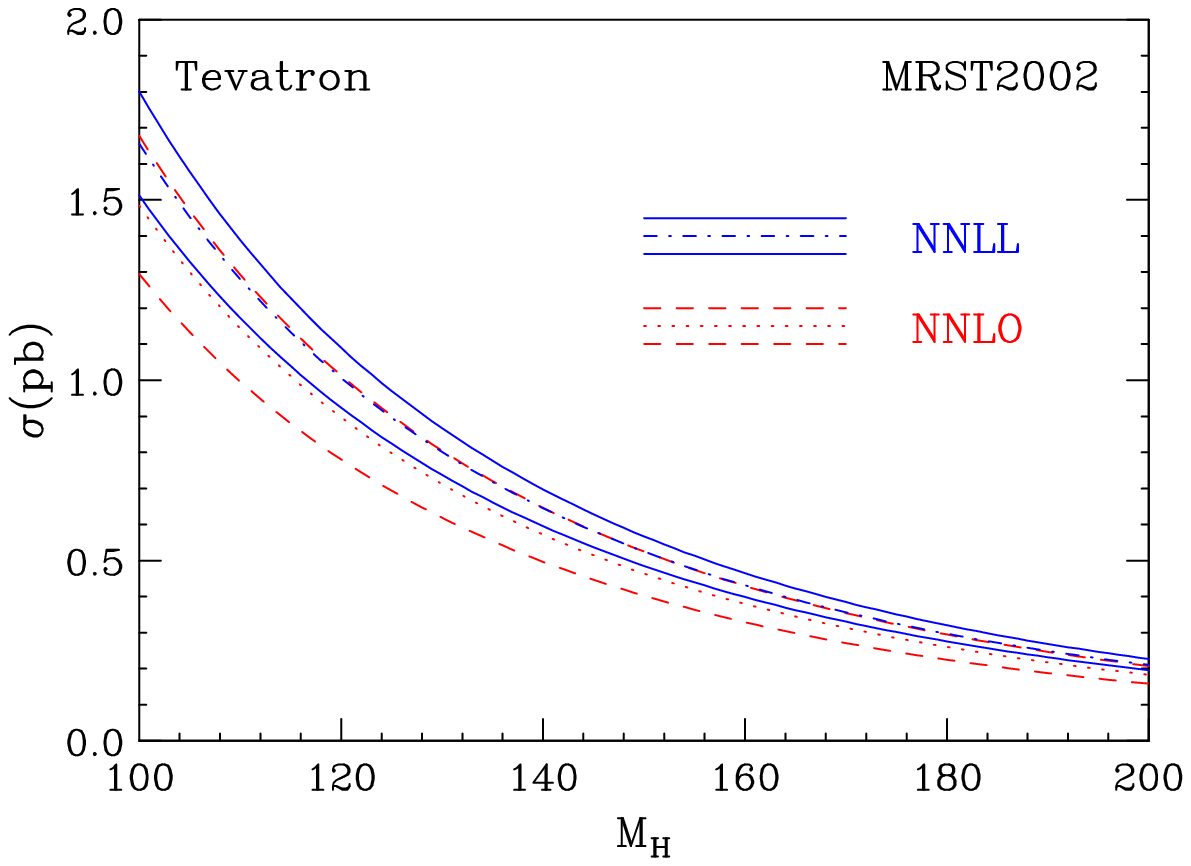}\\
\end{tabular}
\end{center}
\caption{\label{fig:xstev}{\em NNLL and NNLO cross sections at the Tevatron
($\sqrt{s}=1.96$~TeV), using MRST2002 parton densities.}}
\end{figure}

In Fig.~\ref{fig:xstev} we plot the
NNLO and NNLL cross-section bands, obtained as
in Figs.~\ref{fig:lhcbands} and \ref{fig:tevbands},
in the mass range $M_H=100$--200~GeV.
The corresponding numerical results are given in Table~\ref{tab:tev}.

{\renewcommand{\arraystretch}{1.8} 
\begin{table}
\begin{center}
\begin{tabular}{c|ccc|ccc}
\hline
$M_H$ & $\sigma^{NNLO}_{\rm min}$ & $\sigma^{NNLO}_{\rm ref}$ & $\sigma^{NNLO}_{\rm max}$ & $\sigma^{NNLL}_{\rm min}$ & $\sigma^{NNLL}_{\rm ref}$ & $\sigma^{NNLL}_{\rm max}$\\
\hline

 100 &   1.2946  &    1.4846  &  1.6783  &  1.5129  &  1.6568  &  1.8009 \\

 105 &   1.1335  &    1.3006  &  1.4701  &  1.3292  &  1.4535  &  1.5784 \\

 110 &   0.9970  &   1.1445  &  1.2936  &  1.1728  &  1.2808  &  1.3894 \\

 115 &   0.8804  &   1.0112  &  1.1429  &  1.0389  &  1.1331  &  1.2281 \\

 120 &   0.7803  &   0.8967  &  1.0135  &  0.9237  &  1.0062  &  1.0898 \\

 125 &   0.6939  &   0.7978  &  0.9017  &  0.8240  &  0.8965  &  0.9704 \\

 130 &   0.6190  &   0.7120  &  0.8048  &  0.7373  &  0.8013  &  0.8668 \\

 135 &   0.5537  &   0.6373  &  0.7204  &  0.6615  &  0.7182  &  0.7766 \\

 140 &   0.4967  &   0.5719  &  0.6465  &  0.5951  &  0.6455  &  0.6976 \\

 145 &   0.4466  &   0.5145  &  0.5817  &  0.5367  &  0.5815  &  0.6282 \\

 150 &   0.4025  &   0.4639  &  0.5246  &  0.4850  &  0.5251  &  0.5670 \\

 155 &   0.3635  &   0.4193  &  0.4742  &  0.4393  &  0.4752  &  0.5130 \\

 160 &   0.3290  &   0.3797  &  0.4295  &  0.3988  &  0.4310  &  0.4651 \\

 165 &   0.2984  &   0.3445  &  0.3898  &  0.3627  &  0.3917  &  0.4226 \\

 170 &   0.2712  &   0.3132  &  0.3544  &  0.3305  &  0.3566  &  0.3847 \\

 175 &   0.2469  &   0.2852  &  0.3228  &  0.3017  &  0.3253  &  0.3509 \\

 180 &   0.2251  &   0.2602  &  0.2946  &  0.2758  &  0.2972  &  0.3206 \\

 185 &   0.2056  &   0.2378  &  0.2693  &  0.2526  &  0.2720  &  0.2933 \\

 190 &   0.1881  &   0.2177  &  0.2465  &  0.2317  &  0.2493  &  0.2688 \\

 195 &   0.1724  &   0.1996  &  0.2261  &  0.2129  &  0.2289  &  0.2469 \\

 200 &   0.1582  &   0.1832  &  0.2076  &  0.1959  &  0.2105  &  0.2270 \\

\hline
\end{tabular}                                                                 
\ccaption{}{\label{tab:tev}{\em NNLO and NNLL cross sections (in pb) at the
Tevatron ($\sqrt{s}=1.96$~TeV), using MRST2002 parton densities.}}
\end{center}
\end{table}}

\subsection{Uncertainties on Higgs production cross section}

In this subsection we would like to discuss the various sources
of QCD uncertainty that
still affect the Higgs production cross section,
focusing on the low-$M_H$ region ($M_H\ltap 200$ GeV).
The uncertainty basically has two origins:
the one coming from still unknown perturbative QCD contributions 
to the coefficient function $G_{ab}$ in Eq.~(\ref{had}),
and the one originating from our limited knowledge of the parton distributions.

Uncalculated higher-order QCD radiative corrections are 
the most important source of uncertainty on the coefficients $G_{ab}$.
A method, which is customarily used in perturbative QCD calculations,
to estimate their size is to vary
the renormalization and factorization scales
around the hard scale $M_H$.
In general, this procedure can only give a lower limit on the `true'
uncertainty. This is well demonstrated by Figs.~\ref{fig:lhcbands} and 
\ref{fig:tevbands}, which show no overlap between the LO and NLO 
(or, LL and NLL) bands. However, the NLO and NNLO bands and, also,
the NNLO and NNLL bands do overlap. Furthermore, the central value of the
NNLL bands lies inside the corresponding NNLO bands.
This gives us confidence in using scale
variations to estimate the uncertainty at NNLO and at NNLL order.

Performing scale variations as in Figs.~\ref{fig:lhcbands},~\ref{fig:xslhc},
\ref{fig:tevbands} and \ref{fig:xstev}, 
from the numerical results in Tables~\ref{tab:lhc} and \ref{tab:tev},
we find the following results.
At the LHC, the NNLO scale dependence ranges from about $\pm 10 \%$
when $M_H=120$~GeV, to about $\pm 9 \%$ when $M_H=200$~GeV.
At NNLL order, it is about $\pm 8\%$ when $M_H\ltap 200$~GeV.
At the Tevatron, when $M_H\ltap 200$~GeV,
the NNLO scale dependence is about $\pm 13 \%$, whereas
the NNLL scale dependence is about $\pm 8\%$.

Another method to estimate the size of higher-order corrections is
to compare the results at the highest available order
with those at the previous order. Considering the differences between the
NNLO and NNLL cross sections, we obtain results that are consistent with
the uncertainty estimated from scale variations.

We have also considered other possible sources of higher-order uncertainty.
The NNLL coefficient $A^{(3)}$ is not yet exactly known (see Eq.~(\ref{A3coef})
and the accompanying comment).
One can thus wonder which is the numerical impact of $A^{(3)}$ in the calculation.
We have investigated the numerical effect of $A^{(3)}$
by comparing the full NNLL result with the same result
obtained by setting $A^{(3)}=0$. The differences are below $1\%$,
allowing us to conclude that the uncertainty from $A^{(3)}$ can safely be neglected.
Similar conclusions can be drawn about the inclusion of the dominant collinear
contributions (see Eq.~(\ref{lcollpres})) in the resummed formula.
At NNLL order, it gives an effect
below $1\%$, in agreement with the fact that collinear logarithmic 
contributions, which are formally suppressed by powers of $1/N$,
give only a small correction to the dominant terms due to soft and virtual
contributions (see Sect.~\ref{sec:SV}).

A different and relevant source of perturbative QCD uncertainty
comes from 
the use of the
large-$M_t$ approximation in the computation of $G_{ab}$ beyond the LO.
The comparison \cite{Spira:1995rr,Kramer:1998iq}
between the exact NLO cross section and the one obtained in
the large-$M_t$ approximation  
(but rescaled with the full Born result, including its
exact dependence on $M_t$ and $M_b$) shows
that the approximation works well also for $M_H\gtap M_t$.
This is not accidental. In fact,
the higher-order contributions to the cross section are dominated
by soft radiation, which is weakly sensitive to the mass of the heavy quark in the 
loop at the Born level. In other words, as for the size of the QCD radiative
corrections, what matters is that the heavy-quark mass, $M_t$ is actually
larger than the soft-gluon scale, $M_H/N$, rather than the Higgs boson scale
$M_H$.
This
feature, i.e. the dominance of soft-gluon effects, 
persists at NNLO (see Sect.~\ref{sec:SV})
and it is thus natural to assume that,
having normalized our cross sections with the exact Born result,
the uncertainty ensuing from the large-$M_t$ approximation should be
of order of few per cent for $M_H\ltap 200$~GeV, as it is at NLO.

Besides QCD radiative corrections, electroweak corrections 
also have to be considered.
For a light Higgs, the ${\cal O}(G_F M_t^2)$
dominant corrections in the large-$M_t$ limit
have been computed
and found to give
a very small effect (well below $1\%$) \cite{ewc}.

The other independent and important source of theoretical uncertainty in the cross section
is the one coming from parton distributions.

We start our discussion by considering NLO results.
We have compared the MRST2002 NLO cross sections with
the ones computed with CTEQ6 \cite{Pumplin:2002vw}
and Alekhin \cite{Alekhin:2002fv} distributions.
All three sets include a study of the effect of the experimental uncertainties
in the extraction of the parton densities.
At the LHC, we find that the CTEQ6M results are slightly larger than the MRST2002 ones,
the differences decreasing from about $2\%$ at $M_H=100$~GeV to below $1\%$
at $M_H=200$~GeV. Alekhin's results are instead slightly smaller than the MSRT2002 ones,
the difference being below $3\%$ for $M_H\ltap 200$ GeV.
At the Tevatron, CTEQ6 (Alekhin) cross sections are smaller than the MRST2002 ones,
the differences increasing from $6\%$ ($7\%$) to $10\%$ ($9\%$) when $M_H$ increases from
$100$~GeV to $200$~GeV.
These results are not unexpected,
since in Higgs production at the Tevatron the gluon distribution is
probed at larger values of $x$,
where its (experimental) uncertainty is definitely larger.
The cross section differences that we find at NLO
are compatible with the experimental uncertainty
on the NLO gluon luminosity quoted by the three groups,
which is below about $\pm 5\%$ at the LHC
and about $\pm 10\%$ at the Tevatron 
\cite{mrst2002,Pumplin:2002vw,Alekhin:2002fv}.

Throughout the paper we have used the MRST2002 set \cite{mrst2002}
as the reference set of parton distributions.
The main motivation for this choice
is that this set includes (approximated) NNLO densities,
allowing a consistent study at NNLL (NNLO) accuracy.
The CTEQ collaboration does not provide a NNLO set,
and a complete, consistent comparison with MRST can therefore not be performed.
In Ref.~\cite{Alekhin:2002fv},
a NNLO set of parton distributions (set A02 from here on) has been released. 
This set also includes an estimate of the corresponding uncertainties.
We can thus perform a comparison with the MRST2002 results.
{\renewcommand{\arraystretch}{2.0}  \renewcommand{\tabcolsep}{1mm}
\begin{table}
\begin{center}
\begin{minipage}{0.31\textwidth}
\begin{tabular}{c|c|c}
\hline
$M_H$ & $\sigma^{NNLO}$ & $\sigma^{NNLL}$\\
\hline
$100$ & $57.52^{+0.93}_{-0.93}$ & $60.95^{+1.00}_{-1.00}$\\

$110$ & $48.97^{+0.77}_{-0.77}$ & $51.81^{+0.82}_{-0.82}$\\

$120$ & $42.29^{+0.64}_{-0.64}$ & $44.82^{+0.68}_{-0.69}$\\

$130$ & $36.99^{+0.54}_{-0.55}$ & $39.17^{+0.58}_{-0.59}$\\

$140$ & $32.59^{+0.46}_{-0.47}$ & $34.55^{+0.50}_{-0.50}$\\

$150$ & $28.94^{+0.40}_{-0.40}$ & $30.74^{+0.43}_{-0.43}$\\

$160$ & $25.91^{+0.35}_{-0.35}$ & $27.52^{+0.38}_{-0.38}$\\
\hline
\end{tabular}                                                                 
\end{minipage}
\begin{minipage}{0.31\textwidth}
\begin{tabular}{c|c|c}
\hline
$M_H$ & $\sigma^{NNLO}$ & $\sigma^{NNLL}$\\
\hline
$170$ & $23.35^{+0.31}_{-0.31}$ & $24.75^{+0.34}_{-0.33}$\\

$180$ & $21.17^{+0.28}_{-0.28}$ & $22.42^{+0.30}_{-0.30}$\\

$190$ & $19.29^{+0.25}_{-0.25}$ & $20.44^{+0.27}_{-0.27}$\\

$200$ & $17.65^{+0.23}_{-0.23}$ & $18.76^{+0.25}_{-0.25}$\\

$210$ & $16.25^{+0.21}_{-0.21}$ & $17.28^{+0.23}_{-0.23}$\\

$220$ & $15.03^{+0.20}_{-0.20}$ & $15.98^{+0.21}_{-0.21}$\\

$230$ & $13.96^{+0.18}_{-0.18}$ & $14.84^{+0.19}_{-0.19}$\\
\hline
\end{tabular}                                                                 
\end{minipage}
\begin{minipage}{0.31\textwidth}
\begin{tabular}{c|c|c}
\hline
$M_H$ & $\sigma^{NNLO}$ & $\sigma^{NNLL}$\\
\hline

$240$ & $13.04^{+0.17}_{-0.17}$ & $13.86^{+0.18}_{-0.18}$\\

$250$ & $12.23^{+0.16}_{-0.16}$ & $13.00^{+0.17}_{-0.17}$\\

$260$ & $11.53^{+0.16}_{-0.16}$ & $12.26^{+0.16}_{-0.16}$\\

$270$ & $10.92^{+0.15}_{-0.15}$ & $11.62^{+0.16}_{-0.16}$\\

$280$ & $10.39^{+0.15}_{-0.15}$ & $11.05^{+0.15}_{-0.15}$\\

$290$ & $9.939^{+0.14}_{-0.14}$ & $10.57^{+0.15}_{-0.15}$\\

$300$ & $9.562^{+0.14}_{-0.14}$ & $10.16^{+0.15}_{-0.15}$\\

\hline
\end{tabular}                                                                 
\end{minipage}
\ccaption{}{\label{tab:aleklhc}{\em NNLO and NNLL cross sections (in pb)
at the LHC ($\mu_F=\mu_R=M_H$)
using the parton distributions of Ref.~\cite{Alekhin:2002fv}.}}
\end{center}
\end{table}}
{\renewcommand{\arraystretch}{2.0} \renewcommand{\tabcolsep}{1mm}
\begin{table}
\begin{center}
{\small
\begin{minipage}{0.32\textwidth}
\begin{tabular}{c|c|c}
\hline
$M_H$ & $\sigma^{NNLO}$ & $\sigma^{NNLL}$\\
\hline
$100$ & $1.378^{+0.041}_{-0.041}$ & $1.548^{+0.043}_{-0.042}$\\

$105$ & $1.203^{+0.038}_{-0.038}$ & $1.347^{+0.039}_{-0.039}$\\

$110$ & $1.054^{+0.035}_{-0.035}$ & $1.181^{+0.037}_{-0.036}$\\

$115$ & $0.9280^{+0.033}_{-0.032}$ & $1.042^{+0.034}_{-0.034}$\\

$120$ & $0.8194^{+0.030}_{-0.030}$ & $0.9227^{+0.031}_{-0.032}$\\

$125$ & $0.7259^{+0.028}_{-0.028}$ & $0.8191^{+0.029}_{-0.029}$\\

$130$ & $0.6445^{+0.026}_{-0.026}$ & $0.7292^{+0.027}_{-0.027}$\\

\hline
\end{tabular}                                                                 
\end{minipage}
\begin{minipage}{0.32\textwidth}
\begin{tabular}{c|c|c}
\hline
$M_H$ & $\sigma^{NNLO}$ & $\sigma^{NNLL}$\\
\hline
$135$ & $0.5740^{+0.024}_{-0.024}$ & $0.6503^{+0.026}_{-0.025}$\\

$140$ & $0.5135^{+0.023}_{-0.023}$ & $0.5816^{+0.024}_{-0.024}$\\

$145$ & $0.4604^{+0.021}_{-0.021}$ & $0.5210^{+0.023}_{-0.022}$\\

$150$ & $0.4136^{+0.020}_{-0.020}$ & $0.4686^{+0.021}_{-0.021}$\\

$155$ & $0.3723^{+0.019}_{-0.019}$ & $0.4231^{+0.020}_{-0.020}$\\

$160$ & $0.3357^{+0.018}_{-0.017}$ & $0.3831^{+0.019}_{-0.019}$\\

$165$ & $0.3032^{+0.017}_{-0.016}$ & $0.3471^{+0.018}_{-0.017}$\\

\hline
\end{tabular}                                                                 
\end{minipage}
\begin{minipage}{0.32\textwidth}
\begin{tabular}{c|c|c}
\hline
$M_H$ & $\sigma^{NNLO}$ & $\sigma^{NNLL}$\\
\hline
$170$ & $0.2741^{+0.015}_{-0.015}$ & $0.3142^{+0.016}_{-0.016}$\\

$175$ & $0.2483^{+0.014}_{-0.014}$ & $0.2841^{+0.015}_{-0.015}$\\

$180$ & $0.2253^{+0.014}_{-0.014}$ & $0.2577^{+0.015}_{-0.015}$\\

$185$ & $0.2054^{+0.013}_{-0.013}$ & $0.2353^{+0.014}_{-0.014}$\\

$190$ & $0.1877^{+0.012}_{-0.012}$ & $0.2157^{+0.013}_{-0.013}$\\

$195$ & $0.1716^{+0.011}_{-0.011}$ & $0.1981^{+0.012}_{-0.012}$\\

$200$ & $0.1572^{+0.011}_{-0.011}$ & $0.1818^{+0.012}_{-0.012}$\\
\hline
\end{tabular}                                                                 
\end{minipage}}
\ccaption{}{\label{tab:alektev}{\em NNLO and NNLL cross sections (in pb)
at the Tevatron ($\mu_F=\mu_R=M_H$)
using the parton distributions of Ref.~\cite{Alekhin:2002fv}.}}
\end{center}
\end{table}}

In Table~\ref{tab:aleklhc} (Table~\ref{tab:alektev}) we report the NNLO and NNLL
cross sections (computed with $\mu_F=\mu_R=M_H$)
obtained at the LHC (Tevatron) by using the set A02 of parton distributions.
We also report the corresponding errors,
computed by using the dispersion array\footnote{More precisely,
30 parton distributions are generated according to
$f_i(x,k)=f_i(x)\pm \Delta_i(x,k),~k=1,\dots, 15$,
and the cross section for each $k$ is evaluated. The positive (negative) deviations
from the central value are then summed in quadrature to obtain
the upper (lower) error.}
of
the A02 sets.
Comparing Tables~\ref{tab:aleklhc} and~\ref{tab:alektev}
with the corresponding central results
in Tables~\ref{tab:lhc} and~\ref{tab:tev},
we see that there are relatively large differences that cannot be
accounted for by the errors provided in
the A02 set.
At the LHC the cross section is larger than the one computed
with the MRST2002 set, and the difference goes from about $8\%$
at low masses to about $2\%$ at $M_H=200$~GeV.
At the Tevatron the cross section is lower than that using MRST2002,
with a difference that ranges from about $7\%$ at low $M_H$ to about $14\%$
at $M_H=200$~GeV.

\begin{figure}[htb]
\begin{center}
\begin{tabular}{c}
\epsfxsize=9truecm
\epsffile{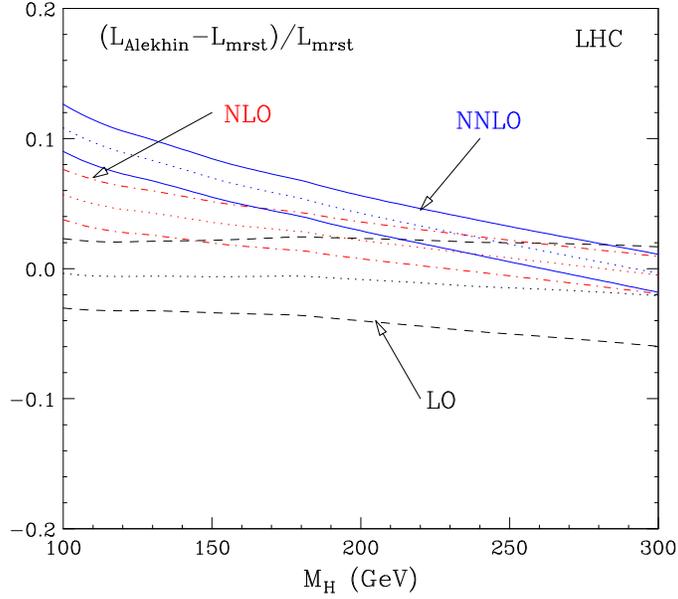}\\
\end{tabular}
\end{center}
\caption{\label{fig:lumilhc}{\em Comparison of A02 and MRST2002 gg luminosities at the LHC.}}
\end{figure}

\begin{figure}[htb]
\begin{center}
\begin{tabular}{c}
\epsfxsize=9truecm
\epsffile{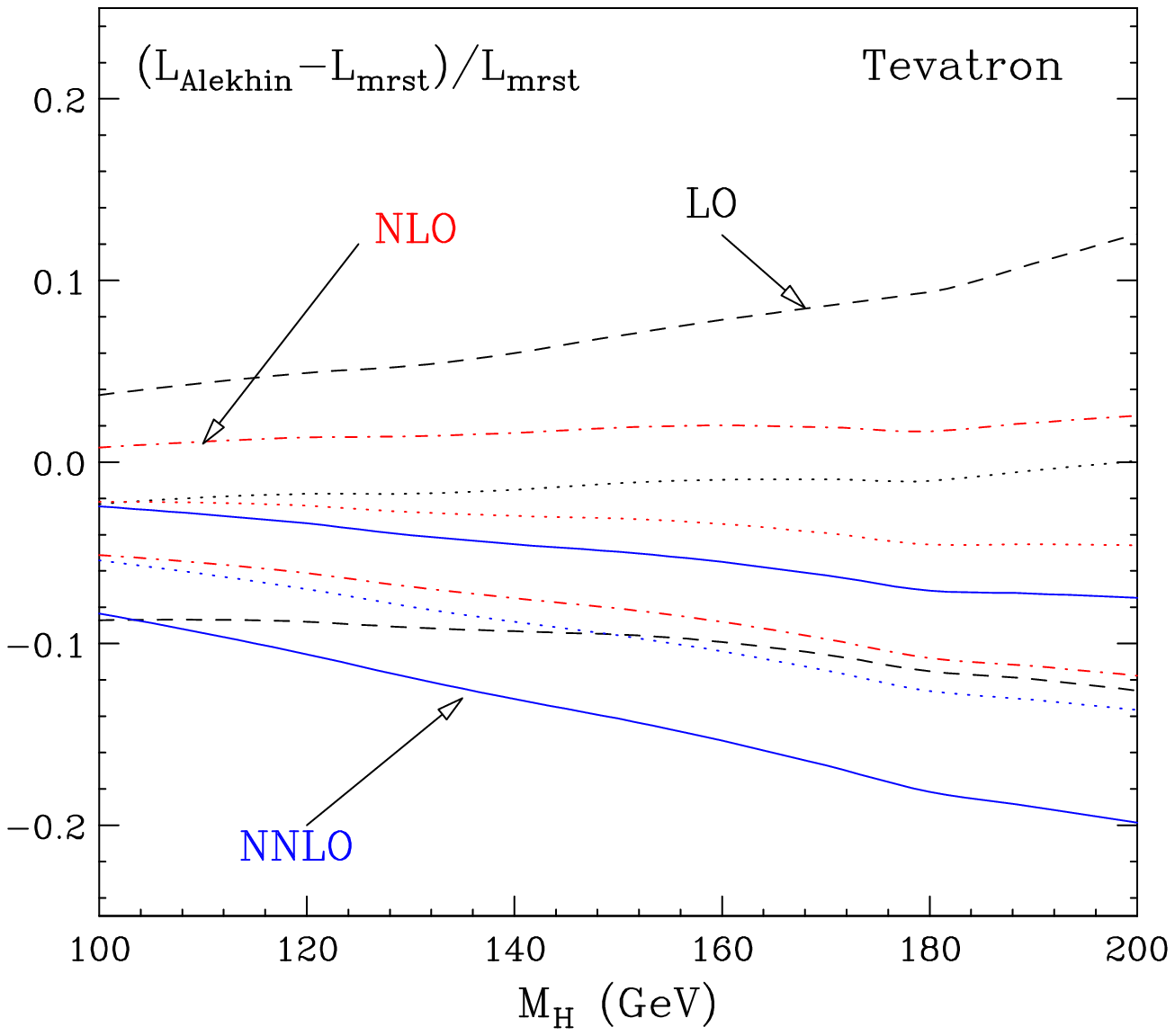}\\
\end{tabular}
\end{center}
\caption{\label{fig:lumitev}{\em Comparison of A02 and MRST2002 gg luminosities at the Tevatron.}}
\end{figure}

These differences can be better understood by looking
at Figs.~\ref{fig:lumilhc} and \ref{fig:lumitev},
where a comparison of Alekhin (with errors)
and MSRT2002 luminosities is
presented.
Comparing Tables~\ref{tab:lhc}, \ref{tab:tev}, \ref{tab:aleklhc} 
and~\ref{tab:alektev} with Figs.~\ref{fig:lumilhc} and \ref{fig:lumitev},
we see that the differences in the cross sections are basically
due to differences in the NNLO gluon--gluon luminosity.
Moreover, Figs.~\ref{fig:lumilhc} and \ref{fig:lumitev} show that 
the differences between the luminosities are typically larger than
the estimated uncertainty of experimental origin.
In particular, the differences between the $gg$ luminosities appear to
increase with the perturbative order (i.e. going from LO to NLO and to NNLO).


We are not able to pin down the origin of these differences.
References~\cite{mrst2002} and \cite{Alekhin:2002fv} use the same
(though approximated) NNLO evolution kernels, but
the MRST2002 set is based on a fit of deep-inelastic scattering (DIS), DY 
and Tevatron jet data, whereas 
the A02 set is obtained through a fit to DIS data only.

From the above discussion, 
we conclude
 that the theoretical uncertainties of perturbative origin in the calculation
of the Higgs production cross section, after inclusion of both NNLO corrections
and soft-gluon resummation
at the NNLL level, are below 10\% 
in the low-mass range.
However, it is also apparent that there are uncertainties in the
(available) parton densities alone that can reach values larger than 10\%, and that are not fully understood
at
the moment.
Improvements of these aspects may only come from better understanding of parton
density
determinations, and are
therefore
totally unrelated to the calculation presented in this work.

\section{Conclusions}
\label{sec:conclusions}

In this work we have presented a next-to-next-to-leading resummation of soft-gluon effects
in the gluon fusion cross section for Higgs production. Furthermore, we have
supplemented
our resummed result with available fixed-order (up to NNLO) calculations. We have presented a
phenomenological study of the Higgs cross sections at the LHC and the Tevatron,
and showed that, once problems with parton density
determinations
are overcome,
a theoretical precision of about 10\% in the predicted cross section can be attained.

The impact of the NNLL corrections included in this work was found to be 
modest in the case of light-Higgs production at the LHC. The NNLL corrections
are more significant at the Tevatron, and allow us to reduce the theoretical
error band at the same precision level as at the LHC.
We believe that our results add much confidence in the predicted cross sections
for the following reason. The pattern of radiative corrections from the NNLL resummation, when truncated at
fixed order, reproduces quite well the pattern from the exact fixed-order 
calculations. This adds confidence that the higher-order 
(beyond the NNLO level) radiative corrections
arising from the NNLL treatment reflect the behaviour of the
exact theory.

As a side product of this work, we have shown that the $N$-space formulation 
of the soft-gluon approximation
is superior to the $x$-space one. In Ref.~\cite{Catani:1996yz} it was
shown
that the $N$-space formulation of soft-gluon resummation should be used, since it avoids a gross violation of
momentum conservation, which  in turn leads to an unphysically large asymptotic growth
of the coefficients of the perturbative expansion. In the present work, the better
behaviour of the $N$-space approach has also been shown to hold in practice,
when comparing known fixed-order results with the fixed-order expansion of
the resummation formulae, up to the NNLO level.

\appendix
\section{Appendix: $N$-moments of soft-gluon contributions}
\label{appa}

In this appendix we present a simple method to evaluate $N$-moments
of soft-gluon contributions to arbitrary logarithmic accuracy.

We consider the $N$-moments $I_n(N)$ of the singular distributions
${\cal D}_n(z)$ defined in Eq.~(\ref{singd}): 
\begin{equation}
\label{inmom}
I_n(N) \equiv \int_0^1 dz \;z^{N-1} {\cal D}_n(z) =
\int_0^1 dz \; \frac{z^{N-1} - 1}{1-z} \;\ln^n(1-z) \;\;.
\end{equation}
These $N$-moments were computed in Ref.~\cite{Catani:ne}. It was shown that, in
the large-$N$ limit, the leading and next-to-leading logarithmic terms 
($\ln^{n+1} N$ and $\ln^{n} N$) arising from the integration over $z$
can straightforwardly be obtained by a very simple prescription.
It is indeed sufficient \cite{Catani:ne}
to replace the integrand weight $(z^{N-1} - 1)$ by the 
approximation
\begin{equation}
\label{nllpres}
z^{N-1} - 1 \simeq - \; \Theta(1 - z - N_0/N) \;\;,
\end{equation}
where $N_0= e^{- \gamma_E}$. 
In the following we show how this NLL prescription
can be generalized to any logarithmic accuracy.
The all-order generalization of Eq.~(\ref{nllpres}) is given in 
Eq.~(\ref{npres}) or, equivalently, in Eq.~(\ref{nzeropres}).

The $N$-moments in Eq.~(\ref{inmom})
can be evaluated as described in Ref.~\cite{Catani:ne}.
Using the following identity
\begin{equation}
\ln^n(1-z) = \lim_{\epsilon \to 0} 
\left( \frac{\partial}{\partial \epsilon} \right)^n
(1-z)^\epsilon
\end{equation}
to replace the logarithmic term in the integrand on the right-hand side of
Eq.~(\ref{inmom}), we obtain
\begin{equation}
\label{inanyn}
I_n(N) = \lim_{\epsilon \to 0}
\left( \frac{\partial}{\partial \epsilon} \right)^n
\left\{ \frac{1}{\epsilon} 
\left[ \frac{\Gamma(N) \;\Gamma(1+\epsilon)}{\Gamma(N+\epsilon)} - 1 \right]
\right\} \;\;,
\end{equation}
where $\Gamma(x)$ is the Euler $\Gamma$-function. The expression (\ref{inanyn})
can be used to evaluate $I_n(N)$ for any given value of $n$ without
approximations of the dependence on $N$ \cite{Catani:ne}.

We are interested in the large-$N$ behaviour of $I_n(N)$. Neglecting terms
that are suppressed by powers of $1/N$ in the large-$N$ limit, we can use the
approximation
\begin{equation}
\frac{\Gamma(N)}{\Gamma(N+\epsilon)} = e^{- \epsilon \ln N} 
\left[ 1 + {\cal O}\left(\frac{1}{N} \right) \right] \;\;,
\end{equation}
and we can rewrite Eq.~(\ref{inanyn}) as
\begin{equation}
\label{inepln}
I_n(N) = \lim_{\epsilon \to 0}
\left( \frac{\partial}{\partial \epsilon} \right)^n
\left\{ \frac{1}{\epsilon} 
\left[ e^{- \epsilon \ln N} \Gamma(1+\epsilon) - 1 \right]
\right\} +  {\cal O}\left(\frac{1}{N} \right) \;\;.
\end{equation}
Using the expression
\begin{equation}
\label{gamexp}
\Gamma(1+\epsilon) = \exp \left\{ -\gamma_E \;\epsilon 
+ \sum_{n=2}^{+\infty} \left(-1 \right)^n
\frac{\zeta(n)}{n} \;  \epsilon^n \right\} \;\;,
\end{equation}
the term in the curly bracket of Eq.~(\ref{inepln}) can easily be expanded
in powers of $\epsilon$. The result for $I_n(N)$ is thus a polynomial 
of degree $n+1$ in the large logarithm $\ln N$:
\begin{align}
\label{inres}
I_n(N) &= \frac{(-1)^{n+1}}{n+1} \,(\ln N + \gamma_E)^{n+1}
+ \frac{(-1)^{n-1}}{2} \;n \,\zeta(2) \; (\ln N + \gamma_E)^{n-1} \nonumber \\
&+ \sum_{k=0}^{n-2} \;a_{nk} \; (\ln N + \gamma_E)^{k} 
+ {\cal O}\!\left(\frac{1}{N} \right) \;\;,
\end{align}
where the coefficients $a_{nk}$ are combinations of powers of
$\zeta(2), \zeta(3), \dots, \zeta(n+1)$ (see e.g. Table~1 
in Ref.~\cite{Catani:ne}).

As shown by Eq.~(\ref{inres}) and pointed out in Ref.~\cite{Catani:ne},
the LL and NLL
terms ($\ln^{n+1} N$ and 
$\ln^{n} N$) 
can straightforwardly be obtained from the definition of the 
$N$-moments in Eq.~(\ref{inmom}) by simply implementing the prescription in
Eq.~(\ref{nllpres}). To show how this NLL prescription
can be generalized to any logarithmic accuracy,
we consider the right-hand side of Eq.~(\ref{inepln}) and
use the following formal identity:
\begin{equation}
e^{- \epsilon \ln N} \; \Gamma(1+\epsilon)  = 
\Gamma\!\left(1-\frac{\partial}{\partial \ln N}\right) \;e^{- \epsilon \ln N}
\;\;.
\end{equation}
Then we can perform the $n$-th derivative with respect to $\epsilon$, and we
obtain
\begin{equation}
\label{inln}
I_n(N) = \Gamma\!\left(1-\frac{\partial}{\partial \ln N}\right)
\frac{\left( - \ln N \right)^{n+1}}{n+1} + {\cal O}\left(\frac{1}{N} \right)
\;\;.
\end{equation}
This expression can be regarded as a replacement of Eq.~(\ref{inepln})
to compute the polynomial coefficients $a_{nk}$ in Eq.~(\ref{inres}). Moreover,
by noting that
\begin{equation}
\frac{\left( - \ln N \right)^{n+1}}{n+1} =
- \int_0^{1-1/N} dz \; \frac{\ln^n(1-z)}{1-z} \;\;,
\end{equation}
we shortly get the all-order generalization of Eq.~(\ref{nllpres}).
It is:
\begin{align}
\label{npres}
z^{N-1} - 1 &= - \; \Gamma\!\left(1-\frac{\partial}{\partial \ln N}\right)
\; \Theta(1 - z - 1/N) + {\cal O}(1/N) \\
\label{nzeropres}
&= - \; {\widetilde \Gamma}\!\left(1-\frac{\partial}{\partial \ln N}\right)
\; \Theta(1 - z - N_0/N) + {\cal O}(1/N) \;\;,
\end{align}
where we have also introduced the function ${\widetilde \Gamma}$ defined by
\begin{equation}
\label{gtildedef}
{\widetilde \Gamma}(1+\epsilon) \equiv 
e^{\gamma_E \,\epsilon}\; \Gamma(1+\epsilon) = \exp \left\{ 
\sum_{n=2}^{+\infty} \left(-1 \right)^n
\frac{\zeta(n)}{n} \;  \epsilon^n \right\} \;\;.
\end{equation}
Note that the equality between Eqs.~(\ref{npres}) and (\ref{nzeropres}) simply
follows from 
$\Gamma= {\widetilde \Gamma} \exp[\gamma_E N\partial/(\partial N)]$ and
from the fact that the exponential of the differential operator acts as the rescaling
$N \to \exp[\gamma_E]N=N/N_0$.

Although we have derived it by starting from Eq.~(\ref{inmom}), it is
straightforward to show that
the prescription (\ref{npres}) (or (\ref{nzeropres})) can be applied as follows
\begin{align}
\int_0^1 dz \, \frac{z^{N-1} - 1}{1-z} \,F(\as , \ln (1-z) )
\!\!&= - \,\Gamma\!\left(1-\frac{\partial}{\partial \ln N}\right)
\int_0^{1-1/N} \; \frac{dz}{1-z} \; F(\as , \ln (1-z) ) +
{\cal O}\!\left(\frac{1}{N}\right) \nonumber \\
\!\!&= 
- \,{\widetilde \Gamma}\!\left(1-\frac{\partial}{\partial \ln N}\right)
\int_0^{1-N_0/N} \!\!\frac{dz}{1-z} \;F(\as , \ln (1-z) ) +
{\cal O}\!\left(\frac{1}{N}\right) \nonumber 
\end{align}
to evaluate the $\ln N$-contributions arising from the integration of any
soft-gluon function $F$ that has a generic perturbative expansion of the type
\begin{equation}
F(\as, \ln (1-z) ) = \sum_{k=1}^{+\infty} \as^k \;
\sum_{n=0}^{2k-1} F_{kn} \;\ln^n(1-z) \;\;.
\end{equation}

The all-order differential operator $\Gamma$ (${\widetilde \Gamma}$) in
Eq.~(\ref{npres}) (Eq.~(\ref{nzeropres})) is obviously defined 
only in a formal sense through Eq.~(\ref{gamexp}) (Eq.~(\ref{gtildedef})).
However, such a definition is sufficient for any
formal manipulation to all logarithmic orders. Moreover, it is also sufficient
for any practical use at fixed and arbitrary logarithmic accuracy.
Since each additional
power of $\partial/(\partial \ln N)$ leads to terms that are suppressed by a
power of $\ln N$, we can obtain all the terms up to a given (and arbitrary)
logarithmic order by simply truncating the power series expansion of $\Gamma$
at the corresponding order in the logarithmic derivative.
For example, using the truncation
\begin{align}
\Gamma\!\left(1-\frac{\partial}{\partial \ln N}\right) &= 1 +
\gamma_E \frac{\partial}{\partial \ln N}
+ \frac{1}{2} \left( \gamma_E^2 + \zeta(2) \right) 
\left(\frac{\partial}{\partial \ln N}\right)^2 \nonumber \\
&+
\frac{1}{6} \left( \gamma_E^3 + 3 \gamma_E \,\zeta(2) + 2 \zeta(3) \right)
\left(\frac{\partial}{\partial \ln N}\right)^3 + \dots \;\;,
\end{align}
the first, second, third and fourth terms on the right-hand side lead
to the LL, NLL, NNLL and N$^3$LL contributions, respectively.

\section{Appendix: Equivalence between resummation formulae}
\label{appb}

Here we prove the equivalence between the two all-order
representations (\ref{deltarep}) and (\ref{deltarepn0}) of the Sudakov 
radiative factor $\Delta_N^{H}$. In particular, we derive the relations
between the functions $A(\as)$ and $D(\as)$ in Eq.~(\ref{deltarep})
and the functions ${\widetilde D}(\as)$ and ${\widetilde C}_{gg}(\as)$
in Eq.~(\ref{deltarepn0}). These relations are:
\begin{align}
\label{wtild}
{\widetilde D}(\as) = D(\as) 
+ 4 \;\partial_{\as} \;\Gamma_2(\partial_{\as}) 
\left[ A(\as) - \f{1}{4} \;\partial_{\as} \;D(\as) \right] \;\;,
\end{align}
\begin{align}
\label{wtilc}
{\widetilde C}_{gg}(\as(M_H^2),M_H^2/\mu^2_R=1) =
\exp \left\{ - 4 \;\Gamma_2(\partial_{\as})
\left[ A(\as) - \f{1}{4} \;\partial_{\as} \;D(\as) \right]
\right\} \;\;,
\end{align}
where the differential operator $\partial_{\as}$ is defined by
\begin{equation}
\label{das}
\partial_{\as} \equiv -2 \beta(\as) \;\as \;\f{\partial}{\partial\as} \;\;,
\end{equation}
$\beta(\as)$ is the QCD $\beta$-function,
\begin{equation}
\label{rge}
\f{d \ln \as(\mu^2)}{d \ln \mu^2} = \beta(\as(\mu^2)) = 
- \sum_{n=0}^{+\infty} b_n \as^{n+1}\;\;,
\end{equation}
whose first three coefficients, $b_0, b_1, b_2$, are given in
Eq.~(\ref{bcoef}),
and the function $\Gamma_2(\epsilon)$ is defined by its power series expansion
in $\epsilon$:
\begin{equation}
\label{ga2def}
\Gamma_2(\epsilon) \equiv \f{1}{\epsilon^2} \left[ 1 - e^{-\gamma_E \epsilon} \,
\Gamma(1-\epsilon) \right] = \f{1}{\epsilon^2} \left\{ 1 - \exp \left[
\sum_{n=2}^{+\infty} \frac{\zeta(n)}{n} \;  \epsilon^n \right] \right\} \;\;.
\end{equation}
Note that Eq.~(\ref{wtilc}) gives 
${\widetilde C}_{gg}(\as(\mu^2_R),M_H^2/\mu^2_R)$ at the scale $\mu^2_R=M_H^2$
as a function of $\as(M_H^2)$. The dependence on $\mu^2_R$ is simply recovered
by using renormalization-group invariance, i.e. by expressing $\as(M_H^2)$
as a function of $\as(\mu^2_R)$ and $M_H^2/\mu^2_R$ through the solution of
the renormalization-group equation (\ref{rge}) (see also Eq.~(\ref{rgesol})).

Inserting the following expansion
\begin{equation}
\Gamma_2(\epsilon) =  - \f{1}{2} \;\zeta(2) - \f{1}{3} \; \zeta(3) \;\epsilon 
+{\cal O}(\epsilon^2) \;\;,
\end{equation}
in Eqs.~(\ref{wtild}) and (\ref{wtilc}), we can straightforwardly obtain
the expansions of the functions ${\widetilde D}$ and ${\widetilde C}_{gg}$
up to NLL accuracy:
\begin{equation}
\label{wtildex}
{\widetilde D}(\as) - D(\as) = - \left( \f{\as}{\pi} \right)^2 4 \zeta(2)
\pi \b0 A^{(1)} + {\cal O}(\as^3)
\end{equation}
\begin{align}
{\widetilde C}_{gg}(\as(\mu^2_R),M_H^2/\mu^2_R) = 1 
&+\f{\as(\mu^2_R)}{\pi} \;2 \zeta(2) A^{(1)} 
+\left( \f{\as(\mu^2_R)}{\pi} \right)^2 
\left[ 2 \zeta(2) \left( A^{(2)} - \pi \b0 A^{(1)} \ln \f{M_H^2}{\mu^2_R}
\right) \right. \nn \\
\label{wtilcex}
& \left. + 2 \left( \zeta(2) A^{(1)} \right)^2
+ \frac{8}{3} \zeta(3) \pi \b0 A^{(1)}
\right] + {\cal O}(\as^3) \;\;.
\end{align}

To derive Eqs.~(\ref{wtild}) and (\ref{wtilc}), we start from 
Eq.~(\ref{deltarep}), we replace the integrand weight $(z^{N-1} - 1)$
by the all-order logarithmic prescription in Eq.~(\ref{nzeropres}), and 
we change the integration variable $z$ as $z \to y=1-z$.
We obtain:
\begin{align}
\label{lndrep}
\ln \Delta_N^{H}\!\left(\as(\mu^2_R),\f{M_H^2}{\mu^2_R};
\f{M_H^2}{\mu_F^2}\right) 
= - \; {\widetilde \Gamma}\!\left(1-\frac{\partial}{\partial \ln N}\right)
\int_{N_0/N}^1 \;\frac{dy}{y}  &\Bigl[ 2 \int_{\mu_F^2}^{y^2M_H^2}
\; \frac{dq^2}{q^2} \; A(\as(q^2)) \Bigr. \\
& \Bigl. \f{}{} + D(\as(y^2M_H^2))
\Bigr]  + {\cal O}(1/N) \;\;. \nn
\end{align}
Then we use the definitions in Eqs.~(\ref{gtildedef}) and (\ref{ga2def}) to get
\begin{equation}
\label{gtilga2}
{\widetilde \Gamma}\!\left(1-\frac{\partial}{\partial \ln N}\right)
= 1 - \Gamma_2\!\left(\frac{\partial}{\partial \ln N}\right) 
\left(\frac{\partial}{\partial \ln N}\right)^2 \;\;.
\end{equation}
Inserting Eq.~(\ref{gtilga2}) in Eq.~(\ref{lndrep}) and comparing the latter
with Eq.~(\ref{deltarepn0}), we get
\begin{align}
\int_{N_0/N}^1 &\;\frac{dy}{y} \;{\widetilde D}(\as(y^2M_H^2))
- \ln {\widetilde C}_{gg}(\as(M_H^2),M_H^2/\mu^2_R=1) 
= \int_{N_0/N}^1 \;\frac{dy}{y} \;D(\as(y^2M_H^2)) \nn \\
\label{deltarel}
&- \Gamma_2\!\left(\frac{\partial}{\partial \ln N}\right) 
\left(\frac{\partial}{\partial \ln N}\right)^2
\int_{N_0/N}^1 \;\frac{dy}{y}  \;\Bigl[ \,2 \int_{\mu_F^2}^{y^2M_H^2}
\; \frac{dq^2}{q^2} \; A(\as(q^2)) 
\Bigl. \f{}{} + D(\as(y^2M_H^2))
\Bigr] \\
&= \int_{N_0/N}^1 \;\frac{dy}{y} \;D(\as(y^2M_H^2)) \nn \\
\label{deltarel1}
&- \Gamma_2\!\left(\frac{\partial}{\partial \ln N}\right)
\;\Bigl[ \,- 4 \; A(\as(N_0^2M_H^2/N^2)) + 
\left(\frac{\partial}{\partial \ln N}\right) D(\as(N_0^2M_H^2/N^2))
\Bigr] \;\;,
\end{align}
where Eq.~(\ref{deltarel1}) is obtained from Eq.~(\ref{deltarel})
by acting with the operator 
$\left(\frac{\partial}{\partial \ln N}\right)^2$ onto the $y$-integral.
Using the renormalization-group equation (\ref{rge}),
we now observe that the relation
\begin{equation}
\label{dlndas}
\left(\frac{\partial}{\partial \ln N}\right) f(\as(k^2/N^2))
= -2 \left(\frac{\partial}{\partial \ln k^2}\right)
f(\as(k^2/N^2))
=\partial_{\as} f(\as(k^2/N^2)) 
\end{equation}
is valid for any arbitrary function $f(\as)$ (the operator $\partial_{\as}$
is defined in Eq.~(\ref{das})). 
Therefore, we can perform the replacement
$\frac{\partial}{\partial \ln N} \to \partial_{\as}$ in Eq.~(\ref{deltarel1}),
and we obtain
\begin{align}
\int_{N_0/N}^1 &\;\frac{dy}{y} \;{\widetilde D}(\as(y^2M_H^2))
- \ln {\widetilde C}_{gg}(\as(M_H^2),M_H^2/\mu^2_R=1) 
= \int_{N_0/N}^1 \;\frac{dy}{y} \;D(\as(y^2M_H^2)) \nn \\
\label{deltarel2}
&+ \left\{ \Gamma_2\left(\partial_{\as}\right)
\;\Bigl[ \;4 A(\as) -  \partial_{\as}
 D(\as) \Bigr] \right\}_{\as=\as(N_0^2M_H^2/N^2)} \;\;.
\end{align}
Note that this equation holds for any value of $N$. Thus, setting $N=N_0$
the $y$-integrals vanish, and we immediately get Eq.~(\ref{wtilc}).
Moreover, we can apply the operator 
$\frac{\partial}{\partial \ln N}$ on both sides of Eq.~(\ref{deltarel2})
and, since ${\widetilde C}_{gg}$ is $N$-independent, we obtain the relation
\begin{align}
{\widetilde D}(\as(N_0^2M_H^2/N^2))
&= D(\as(N_0^2M_H^2/N^2)) \nn \\
&+ \left(\frac{\partial}{\partial \ln N}\right)
\left\{ \Gamma_2\left(\partial_{\as}\right)
\;\Bigl[ \;4 A(\as) -  \partial_{\as}
 D(\as) \Bigr] \right\}_{\as=\as(N_0^2M_H^2/N^2)}\; ,
\end{align}
which, owing to Eq.~(\ref{dlndas}), gives exactly Eq.~(\ref{wtild}).

\section{Appendix: Resummation formulae at NNLL accuracy and beyond}
\label{appc}

In this appendix we sketch the calculation of the logarithmic functions
$g_H^{(n)}$ in Eq.~(\ref{gexpan}). These functions originate from the Sudakov
radiative factor $\Delta_{N}^{H}$ in Eq.~(\ref{resfdelta}). Thus we write
\begin{align}
\label{logexp}
\ln \Delta_N^{H}\!\left(\as(\mu^2_R),\f{M_H^2}{\mu^2_R};\f{M_H^2}{\mu_F^2}\right) 
&= \ln N \; g_H^{(1)}(\b0 \as(\mu^2_R) \ln N) + 
\\
&+ \sum_{n=2}^{+\infty} \left[ \as(\mu^2_R)\right]^{n-2}
\; g_H^{(n)}(\b0 \as(\mu^2_R)\ln N,M_H^2/\mu^2_R;M_H^2/\mu_F^2 ) 
+ {\cal O}(1) \,\,, \nonumber
\end{align} 
where the term ${\cal O}(1)$ stands for contributions that are constant
in the large-$N$ limit. This $N$-independent term eventually contributes to 
the function $C_{gg}(\as)$ in Eq.~(\ref{resformula}).

The logarithmic expansion on the right-hand side of Eq.~(\ref{logexp})
is most easily computed by starting from the integral representation in
Eq.~(\ref{deltarepn0}) and by using the method described in 
Refs.~\cite{Catani:ne,Catani:1992ua}. We first use the renormalization group
equation (\ref{rge}) to change the integration variables $\{q^2,y\}$ in
$\{\as(q^2)=\alpha, \as(y^2M_H^2)=\alpha'\}$, so Eq.~(\ref{deltarepn0})
becomes
\begin{align}
\label{deltarepalpha}
\ln \Delta_N^{H}\!\left(\as(\mu^2_R),\f{M_H^2}{\mu^2_R};\f{M_H^2}{\mu_F^2}\right) 
&= - \int_{\as(N_0^2M_H^2/N^2)}^{\as(M_H^2)} 
\;\frac{d\alpha'}{\alpha'} \frac{1}{\beta(\alpha')} 
\left[ \,\int_{\as(\mu_F^2)}^{\alpha'}
\; \frac{d\alpha}{\alpha} \;\frac{A(\alpha)}{\beta(\alpha)} 
+ \frac{1}{2} \,{\widetilde D}(\alpha)
\right] + {\cal O}(1) \;.
\end{align}
Now, the integrand can be expanded (see Eqs.~(\ref{Afun}), (\ref{Dfun}), 
(\ref{wtild}) and (\ref{rge})) in power series of 
$\alpha$ and $\alpha'$, and the expansion can be truncated 
to the required (and arbitrary) logarithmic accuracy. This procedure leads to
simple (logarithmic and polynomial) integrals, and the result is expressed
in terms of elementary functions of the perturbative coefficients $A^{(n)},
D^{(n)}, b_n$ and of $\as(k^2)$ with $k^2=N_0^2M_H^2/N^2, M_H^2$ or $\mu_F^2$.
To obtain the functions 
$g_H^{(n)}(\b0 \as(\mu^2_R)\ln N,M_H^2/\mu^2_R;M_H^2/\mu_F^2 )$, it is 
sufficient to express $\as(k^2)$
in terms of $\as(\mu_R^2)$ and $\ln(k^2/\mu_R^2)$ 
(i.e. $\ln N,$$\ln (M_H^2/\mu_R^2),$$\ln (M_H^2/\mu_F^2),$$\ln N_0= -\gamma_E$)
according to the perturbative solution of the renormalization group
equation (\ref{rge}), and then to compare the results with the right-hand side
of Eq.~(\ref{logexp}).
For instance, to extract the LL, NLL and NNLL functions 
$g_H^{(1)}, g_H^{(2)}$ and $g_H^{(3)}$ in Eqs.~(\ref{g1fun}),(\ref{g2fun})
and (\ref{g3fun}), it is sufficient to use the NNLO solution of
Eq.~(\ref{rge}):
\begin{align}
\label{rgesol}
\as(k^2) &= \frac{\as(\mu_R^2)}{\ell} 
\Bigl\{ 1 - \frac{\as(\mu_R^2)}{\ell} \,\frac{b_1}{b_0}\, 
\ln \ell + \left( \frac{\as(\mu_R^2)}{\ell} \right)^2
\left[ \, \frac{b_1^2}{b_0^2}  \left(
\ln^2 \ell - \ln \ell + \ell - 1 \right) -  \frac{b_2}{b_0} ( \ell -1 )
\, \right] \nn
\\
&+ {\cal O}(\as^3(\mu_R^2) (\as(\mu_R^2) \ln (k^2/\mu_R^2))^n) \Bigr\} \;, 
\end{align}
where we have defined ${\ell}= 1+ b_0 \,\as(\mu_R^2) \ln (k^2/\mu_R^2)$.
The extension to arbitrarily higher logarithmic accuracy is straightforward.

\section{Appendix: Convergence of the fixed-order\\ soft-gluon expansion}
\label{appd}

We check here the numerical convergence of the fixed-order soft-gluon expansion 
toward the resummed result obtained by using the Minimal Prescription.

The rapid convergence of the higher-order soft-gluon corrections is displayed in
Table~\ref{tab:pert}, for the case of Higgs production at the LHC and the
Tevatron. All the entries reported in Table~\ref{tab:pert} are obtained
by using the MRST2002 NNLO parton densities.
The first three columns show the contributions to the total cross
section from the first three orders in perturbation theory. The sum of the first 
three columns thus gives the exact NNLO result.
The next two columns show the contributions from the next two fixed-order
terms, as obtained by the (truncated)
fixed-order expansion of our NNLL resummed formula. 
The sum of all entries in each row corresponds to the full NNLL resummed result
as obtained with the Minimal
Prescription,
while each fixed-order term has no ambiguity due to the choice
of the contour for the Mellin transformation in Eq.~(\ref{hadnres}).
The results in Table~\ref{tab:pert}
support the validity of the Minimal Prescription,
since the truncated resummed expansion converges to it very rapidly.

{\renewcommand{\arraystretch}{1.8}
\begin{table}
\begin{center}
\begin{tabular}{|c|c|c|c|c|c|c|} \hline
& $\alpha_s^2$ & $\alpha_s^3$ & $\alpha_s^4$ & $\alpha_s^5$ &
$\alpha_s^6$ & $\alpha_s^{\ge 7}$  \\ \hline \hline
LHC & 11.67&     14.72 &   8.61  &  1.65  &    0.30 &    0.08   \\
Tevatron    & 0.228 &   0.291  &  0.194  &  0.065  &  0.017  &  0.007\\
\hline
\end{tabular}
\ccaption{}{\label{tab:pert}
{\em Contributions to the total cross sections
 (in pb) at the LHC and the Tevatron
from higher orders in the expansion of
the NNLL resummed result, with
$\mu_R=\mu_F=M_H$, $M_H=130$ GeV and MRST2002 NNLO parton densities.
The sum of the first three columns gives the exact NNLO result.}}
\end{center}
\end{table} }

\newpage
\section{Appendix: Soft-gluon expansion at N$^3$LO}
\label{appe}

In this final appendix we present the soft-gluon approximation,
$G_{gg,\,N}^{(3) \,\mlbl{SV-$N$}}$, in $N$ space of the N$^3$LO contribution,
$G_{gg,\,N}^{(3)}$, to the coefficient function in Eq.~(\ref{alloexpansion}).
The analytic expression of $G_{gg,\,N}^{(3) \,\mlbl{SV-$N$}}$,
obtained by performing the expansion of the all-order
resummation formula in Eq.~(\ref{resformula}), is 
\begin{align}
\label{ggSV-N_NNNLO}
G_{gg}^{(3) \,\mlbl{SV-$N$}}(& M_H^2/\mu_R^2,
M_H^2/\mu_F^2)= \frac{4 \, \left(A^{(1)}\right)^3}{3}\,\ln^6 N\nn\\
 +\ln^5 N  &\Bigg[
\frac{8  \,\left(A^{(1)}\right)^2 \, b_0 \pi}{3} + 8 \,{\gamma_E} \,
\left(A^{(1)}\right)^3 -4\,\left(A^{(1)}\right)^3  \ln\f{M_H^2}{\mu_F^2}
\Bigg] \nn \\
+ \ln^4 N & \Bigg[ 4\, A^{(1)} A^{(2)} + \frac{4  A^{(1)} b_0^2 \pi^2}{3} +2
\left(A^{(1)}\right)^2 C_{gg}^{(1)} +\frac{40 \left(A^{(1)}\right)^2 b_0 \pi
\,{\gamma_E}}{3}
+ 16 \left(A^{(1)}\right)^3 \,{\gamma_E^2}  \nn\\
& -4 \left(A^{(1)}\right)^2\, b_0 \pi  \ln\f{M_H^2}{\mu_R^2} +
4 \left(A^{(1)}\right)^3\, \ln^2\f{M_H^2}{\mu_F^2}
-\ln\f{M_H^2}{\mu_F^2} \Bigg( \frac{8\,\left(A^{(1)}\right)^2 b_0 \pi}{3} +
16\left(A^{(1)}\right)^3
\,{\gamma_E} \Bigg)
\Bigg] \nn\\
+ \ln^3 N & \Bigg[ \frac{8\,A^{(2)}\, b_0 \pi}{3} +
 \frac{4\,A^{(1)}\, b_1 \pi^2}{3}  +
 \frac{4\,A^{(1)}\, b_0\pi \,C_{gg}^{(1)}  }{3} - 2\, A^{(1)}\, D^{(2)}
+ 8\,\left(A^{(1)}\right)^2\, b_0 \pi\, \zeta(2)\nn\\
&+ \frac{32\,\left(A^{(1)}\right)^3 \gamma_E^3}{3}
 + 24\, \left(A^{(1)}\right)^2\, b_0 \pi\,  \gamma_E^2 + \gamma_E
 \Big(  16\, A^{(1)} A^{(2)} + \frac{16\, A^{(1)} \, b_0^2 \pi^2 }{3} + 8
\left(A^{(1)}\right)^2 \,C_{gg}^{(1)} \Big)\nn\\
& -\frac{4\, \left(A^{(1)}\right)^3}{3} \ln^3\f{M_H^2}{\mu_F^2}
+\ln^2\f{M_H^2}{\mu_F^2} \Big( 8\, \left(A^{(1)}\right)^3 \gamma_E
-2\, \left(A^{(1)}\right)^2\, b_0\pi \Big)
- \ln\f{M_H^2}{\mu_R^2} \Big( \frac{8\,A^{(1)}\, b_0^2 \pi^2}{3}\nn\\
&+16\,\left(A^{(1)}\right)^2\, b_0 \pi\, \gamma_E  \Big)
 -\ln\f{M_H^2}{\mu_F^2} \Big( 8\,A^{(1)}\, A^{(2)} +
4\,\left(A^{(1)}\right)^2 \,  C_{gg}^{(1)} +8\, \left(A^{(1)}\right)^2 \,
b_0 \pi\,  \gamma_E\nn\\
&+16\, \left(A^{(1)}\right)^3\,\gamma_E^2 \Big)
+8\, \left(A^{(1)}\right)^2\, b_0 \pi\, \ln\f{M_H^2}{\mu_F^2}\,
\ln\f{M_H^2}{\mu_R^2}
\Bigg]\nn\\
+ \ln^2 N & \Bigg[   2\, A^{(3)} +   2\, A^{(2)}\, C_{gg}^{(1)}+ 2\,
A^{(1)}\, C_{gg}^{(2)} -2\, b_0 \pi\, D^{(2)}+   8\, A^{(1)}\,b_0^2 \pi^2\, \zeta(2)
+16\,\left(A^{(1)}\right)^2 \, b_0 \pi\, \gamma_E^3  \nn \\
& + \gamma_E^2 \Big(8\, \left(A^{(1)}\right)^2\,
C_{gg}^{(1)}+16\,  A^{(1)}\, A^{(2)} +8 \, A^{(1)}\, b_0^2 \pi^2
 \Big) + \gamma_E \Big( 8\,  A^{(2)}\,b_0 \pi +4\,  A^{(1)}\,b_1 \pi^2\nn\\
& +4\,  A^{(1)}\,b_0 \pi\,  C_{gg}^{(1)}
 - 4\,  A^{(1)}\,  D^{(2)}
+16\,  \left(A^{(1)}\right)^2\, b_0 \pi\, \zeta(2) \Big)
+2\, A^{(1)}\, b_0^2 \pi^2\, \ln^2\f{M_H^2}{\mu_R^2}\nn\\
& -4\, \left(A^{(1)}\right)^2\, b_0\pi\,
\ln^2\f{M_H^2}{\mu_F^2}\,\ln\f{M_H^2}{\mu_R^2}
+ 16 \left(A^{(1)}\right)^2\, b_0\pi\,\gamma_E
\ln\f{M_H^2}{\mu_F^2}\,\ln\f{M_H^2}{\mu_R^2}
+2\, \left(A^{(1)}\right)^2\, b_0\pi\, \ln^3\f{M_H^2}{\mu_F^2}\nn\\
& + \ln^2\f{M_H^2}{\mu_F^2} \Big(4\, A^{(1)}\,  A^{(2)}
 + 2\, \left(A^{(1)}\right)^2 C_{gg}^{(1)} -4\,  \left(A^{(1)}\right)^2\,
b_0\pi\,\gamma_E  \Big)
+ \ln\f{M_H^2}{\mu_F^2} \Big( 2\, A^{(1)}\, D^{(2)}\nn\\
& -16 \, A^{(1)}\,A^{(2)}\,\gamma_E 
 -8\, \left(A^{(1)}\right)^2\, C_{gg}^{(1)} \,\gamma_E
-8\, \left(A^{(1)}\right)^2\,b_0\pi\,\gamma_E^2-8\,
\left(A^{(1)}\right)^2\,b_0\pi\,\zeta(2)
\Big) \nn \\
& 
- \ln\f{M_H^2}{\mu_R^2} \Big( 4\, A^{(2)}\, b_0\pi + 2\, A^{(1)}\, b_1\pi^2 +
 2\, A^{(1)}\, b_0\pi \, C_{gg}^{(1)} +8\,  A^{(1)}\, b_0^2\pi^2\,\gamma_E
+16\, \left(A^{(1)}\right)^2\,b_0\pi
\Big)
\Bigg] \nn
\end{align}
\begin{align}
+ \ln N & \Bigg[   -D^{(3)} -  C_{gg}^{(1)} \, D^{(2)}
+\frac{32\,A^{(1)}\, b_0^2\pi^2\,\zeta(3)}{3}
+8\, A^{(2)}\, b_0\pi\,\zeta(2) + 4 \, A^{(1)}\, b_1\pi^2\,\zeta(2)
\nn\\
& + 4 \, A^{(1)}\, b_0\pi\,C_{gg}^{(1)} \, \zeta(2)
+\frac{16 \,A^{(1)}\, b_0^2\pi^2\,\gamma_E^3}{3} + \gamma_E^2 \Big(
8\, A^{(2)}\, b_0\pi +
4\, A^{(1)}\, b_1\pi^2 + 4\, A^{(1)}\, C_{gg}^{(1)} \, b_0\pi \Big)
 \nn\\
&  +\gamma_E \Big(4\, A^{(3)} + 4\, A^{(2)}\, C_{gg}^{(1)} + 4\,
A^{(1)}\, C_{gg}^{(2)}
-4\, b_0\pi \, D^{(2)} +16 \,A^{(1)}\, b_0^2\pi^2\,\zeta(2)\Big)
-\frac{2\,A^{(1)}\, b_0^2\pi^2}{3}
\, \ln^3\f{M_H^2}{\mu_F^2}\nn\\
& 
+ 2\,A^{(1)}\, b_0^2\pi^2\,\ln^2\f{M_H^2}{\mu_F^2}\, \ln\f{M_H^2}{\mu_R^2} -
2\,A^{(1)}\, b_0^2\pi^2\,\ln\f{M_H^2}{\mu_F^2}\, \ln^2\f{M_H^2}{\mu_R^2}
 + 4\,A^{(1)}\, b_0^2\pi^2\, \gamma_E\,\ln^2\f{M_H^2}{\mu_R^2}
\nn\\
&
-\ln^2\f{M_H^2}{\mu_F^2} \Big(  2\,A^{(2)}\, b_0\pi + A^{(1)}\, b_1\pi^2
+ A^{(1)}\, b_0\pi\, C_{gg}^{(1)} \Big)
-\ln\f{M_H^2}{\mu_F^2} \Big( 2\,A^{(3)} + 2\, A^{(2)}\, C_{gg}^{(1)}+ 2\,
A^{(1)}\, C_{gg}^{(2)}\Big)
 \nn \\
& 
+\ln\f{M_H^2}{\mu_F^2}\, \ln\f{M_H^2}{\mu_R^2} \Big( 4\,A^{(2)}\, b_0\pi +
2\,A^{(1)}\, b_1\pi^2 + 2\, A^{(1)}\, b_0\pi\, C_{gg}^{(1)}\Big)
 \nn \\
&
- \ln\f{M_H^2}{\mu_R^2} \Big( 8\,A^{(2)}\, b_0\pi\, \gamma_E +  4\,A^{(1)}\,
b_1\pi^2\, \gamma_E +
4\, A^{(1)}\, C_{gg}^{(1)}\, b_0\pi\, \gamma_E +8 A^{(1)}\, b_0^2\pi^2\,\gamma_E^2
+ 8 A^{(1)}\, b_0^2\pi^2\,\zeta(2) \Big)
\Bigg] \nn \\
+ \; C_{gg}^{(3)} &\;\;.
\end{align}
Note that with the present knowledge (see Sect.~\ref{sec:resNNLL}) of the
perturbative coefficients of the
functions $A(\as)$, $D(\as)$ and $C_{gg}(\as)$, it is
possible to predict the numerical values of the coefficients in 
the expansion (\ref{ggSV-N_NNNLO}) only up to ${\cal O}(\ln^2 N)$ terms.
The term proportional to $\ln N$ and the $N$-independent term depend on the 
unknown coefficients $D^{(3)}$ and
$C_{gg}^{(3)}$, respectively\footnote{Note that the results in the 
the fourth column of Table~\ref{tab:pert}
correspond to using the N$^3$LO coefficient in 
Eq.~(\ref{ggSV-N_NNNLO}) with $C_{gg}^{(3)}=0$ and with an approximate
form 
(which follows from setting $g_H^{(4)}=0$ in Eq.~(\ref{gexpan}))
of the coefficient of the $\ln N$ term. 
}.
The same results apply to the DY process,
by simply replacing the $gg$-channel coefficients $A^{(n)}, D^{(n)}$ and
$C_{gg}^{(n)}$ with the corresponding  $q{\bar q}$-channel coefficients. 
The expression (\ref{ggSV-N_NNNLO}) can be used to check future perturbative
calculations at N$^3$LO.

\end{document}